\documentclass{wiley-article}
\usepackage{graphicx}
\usepackage[space]{grffile}
\usepackage{latexsym}
\usepackage{textcomp}
\usepackage{longtable}
\usepackage{tabulary}
\usepackage{booktabs,array,multirow}
\usepackage{amsfonts,amsmath,amssymb}
\usepackage{natbib}
\usepackage{url}
\usepackage{etoolbox}
\usepackage{anyfontsize} 
\usepackage{silence} 
\makeatletter

\graphicspath{{figures}}

\usepackage[utf8]{inputenc}
\usepackage[english]{babel}

\usepackage{subcaption}
\usepackage{placeins}

\newcommand{\indep}{\perp \!\!\! \perp}
\newcommand{\med}{\mbox{median}}

\newcommand{\cov}{\mbox{Cov}}
\newcommand{\var}{\mbox{Var}}
\newcommand{\dCov}{\mbox{dCov}} 
\newcommand{\dVar}{\mbox{dVar}}
\newcommand{\dStd}{\mbox{dStd}}
\newcommand{\dCor}{\mbox{dCor}}
\newcommand{\E}{\mathbb{E}}
\newcommand{\IF}{\mbox{IF}}

\newcommand{\bX}{\boldsymbol{X}}
\newcommand{\bZ}{\boldsymbol{Z}}

\newcommand{\biloop}{\mathbin{\rotatebox[origin=c]{-90}{$\scriptscriptstyle0$}\rotatebox[origin=c]{-90}{$\scriptscriptstyle0$}}}

\newcommand{\Xe}{X_{\varepsilon}}
\newcommand{\Ye}{Y_{\varepsilon}}
\newcommand{\eps}{\varepsilon}

\usepackage{xcolor}
\definecolor{blue}{RGB}{0,0,255}
\definecolor{red}{RGB}{255,0,0}

\papertype{Original Article}

\title{Robust Distance Covariance}

\author[1]{Sarah Leyder}
\author[1]{Jakob Raymaekers}
\author[2]{\mbox{Peter J. Rousseeuw}}
\affil[1]{Department of Mathematics, University of Antwerp, Antwerp, Belgium}
\affil[2]{Department of Mathematics, University of Leuven, Leuven, Belgium}

\corraddress{Peter J. Rousseeuw, Department of Mathematics, University of Leuven, Celestijnenlaan 220B, BE-3001 Heverlee, Belgium}
\corremail{peter.rousseeuw@kuleuven.be}
\runningauthor{Leyder, Raymaekers, Rousseeuw}
\fundinginfo{The first author was supported by Fonds Wetenschappelijk onderzoek - Vlaanderen 
(FWO) as a PhD fellow Fundamental Research (PhD fellowship 11K5523N).
The second author was supported by the European Union’s Horizon 2022 research and 
innovation program under the Marie Sklodowska Curie grant agreement 
No 101103017.}

\begin{document}

\begin{frontmatter}
\maketitle

\begin{abstract}
Distance covariance is a popular measure of 
dependence between random variables. 
It has some robustness properties, but not all.
We prove that the influence function of the usual
distance covariance is bounded, but that its breakdown
value is zero. Moreover, it has an unbounded sensitivity 
function converging to the bounded influence function for
increasing sample size.
To address this sensitivity to outliers we construct
a more robust version of distance covariance and
distance correlation, based on a new data transformation.
Simulations indicate that the resulting method is quite
robust, and has good power in the presence of outliers.
We illustrate the method on genetic data. Comparing the 
classical distance correlation with its more robust 
version provides additional insight.

\end{abstract}
\end{frontmatter}

\section{Introduction}\label{sec:motivation}

Distance covariance ($\dCov$), proposed by 
\citet{szekely2007}, is a relatively recent 
measure of dependence between real random 
variables $X$ and $Y$.
Its popularity stems from the fact that 
$\dCov(X,Y)$ is zero if and only if $X$ and 
$Y$ are independent, and positive otherwise.
Therefore {\it any} type of dependence makes it 
nonzero, unlike the product-moment covariance 
$\cov(X,Y)$ which aims specifically for linear 
relations.

Another feature of $\dCov$ is its simple 
definition. Assume that X has a finite first 
moment. Then \citet{szekely2007} compute the 
{\it doubly centered} interpoint distances of
X, given by
\begin{align} \label{eq:DC}
     \Delta(X,X') = |X-X'| - E_{X''}[|X-X''|] 
  -E_{X''}[|X''-X'|] + E_{X'',X'''}E[|X''-X'''|] 
\end{align}
where $X'$, $X''$ and $X'''$ are independent copies of $X$.
Note that $\Delta(X,X') = \Delta(X',X)$ is not
a distance itself, since it also takes on negative values.
Moreover, $E_{X}[\Delta(X,X')]$ is zero, and the same holds
for $E_{X'}[\Delta(X,X')]$ and $E_{X,X'}[\Delta(X,X')]$.
This explains the name `doubly centered'. It turns out that 
the {\it second} moments of $\Delta(X,X')$ exist as well.

If also $E[|Y|]$ is finite, \citet{szekely2007}
introduced the distance covariance defined as
\begin{equation} \label{eq:dCov}
  \dCov(X,Y) := \cov(\Delta(X,X'),\Delta(Y,Y')).
\end{equation}
(In fact they took the square root of the right hand
side, but we prefer not to because the units 
of~\eqref{eq:dCov} are those of $X$ times $Y$.)
They proved the amazing result that when the first 
moments of $X$ and $Y$ exist, it holds that
\begin{equation} \label{eq:dCovzero}
  X \indep Y \Longleftrightarrow \dCov(X,Y)=0.
\end{equation}
This yields a necessary {\it and sufficient} 
condition for independence. 
The implication $\Longleftarrow$ is not obvious at 
all, and was proven by complex analysis.

In the same paper also another expression 
for $\dCov$ was derived. Working out the covariance 
in~\eqref{eq:dCov} yields
$4 \times 4 = 16$ terms, that exist when $X$ and $Y$ 
also have second moments. These can be reduced to 
four and even three terms:
\begin{align*} 
  \dCov(X, Y) &=
   \E[|X-X'||Y-Y'|]+\E[|X-X'|] \E[|Y-Y'|]
  -\E[|X-X'||Y-Y''|]-\E[|X-X''||Y-Y'|] \nonumber\\
  &=\, \E[|X-X'||Y-Y'|]+\E[|X-X'|]\E[|Y-Y'|] 
     -2\E[|X-X'||Y-Y''|]. \nonumber 
\end{align*}
In view of the occurrence of terms containing 
$X-X'$ and $Y-Y'$, \cite{RR_dCov} investigated various 
relations between the properties $X \indep Y$, 
$(X - X') \indep Y$, and
$(X - X') \indep (Y - Y')$.
 
Based on the distance covariance one also defines
the {\it distance variance} of a variable as
\begin{equation} \label{eq:dVar}
  \dVar(X) := \dCov(X,X)\;,
\end{equation}
which has the units of $X^2$. The {\it distance 
standard deviation} is its square root
\begin{equation} \label{eq:dsd}
   \dStd(X) := \sqrt{\dVar(X)}
\end{equation}
with the units of $X$. The {\it distance 
correlation} is given by
\begin{equation} \label{eq:dCor}
   \dCor(X,Y) = \frac{\dCov(X,Y)}
   {\sqrt{\dVar(X)\,\dVar(Y)}}\;.
\end{equation}
The distance correlation has no units and lies between
zero and one, and finite sample versions of it are 
often used to test for independence.

In this paper $X$ and $Y$ will typically be univariate
random variables, but the above definitions can also
be used for multivariate variables, in which case
$|\cdot|$ stands for the \mbox{Euclidean} norm. 
\citet{szekely2012} generalized these notions to 
{\it $\alpha$-distance dependence measures} where
$0<\alpha<2$. If the moments of order $2\alpha$ of
$|X|$ and $|Y|$ are finite, $\dCov(X,Y;\alpha)$ is 
defined as
\begin{equation*}
\dCov(X, Y ; \alpha) =
  \E[|X-X^{\prime}|^\alpha|Y-Y^{\prime}|^\alpha] +
  \E[|X-X^{\prime}|^\alpha] \E[|Y-Y^{\prime}|^\alpha]
  - 2 \E[|X-X^{\prime}|^\alpha|Y-Y^{\prime \prime}|^\alpha]
\end{equation*}
for i.i.d. $(X,Y)$, $(X^{\prime},Y^{\prime})$, and
$(X^{\prime\prime},Y^{\prime\prime})$. It satisfies
the equivariance property
\begin{equation*}
   \dCov(a_1+b_1 X, a_2+b_2 Y; \alpha)^{1/\alpha}
   =b_1 b_2 \dCov(X, Y; \alpha)^{1/\alpha} 
\end{equation*}
for all $a_1$\,, $a_2$\,, $b_1>0$, and $b_2 > 0$. For
appropriate measurement units we should thus work with 
$\dCov(\cdot,\cdot,\alpha)^{1/\alpha}$ when $\alpha \neq 1$.

Distance covariance is a general method
with interesting properties and wide ranging
applications, for instance in feature 
selection \citep{song2012feature}, 
sparse contingency tables \citep{Zhang2019}, 
and time series 
\citep{zhou2012measuring,Davis2018}.
Fast algorithms for its computation have been
constructed \citep{Huo2016,chaudhuri2019fast}.
\cite{sejdinovic2013} and \cite{Edelmann2022} 
found interesting connections with other 
dependence measures, such as the Hilbert-Schmidt 
Independence Criterion of 
\cite{gretton2005kernel}.

The distance covariance and distance correlation have
sometimes been credited with natural robustness properties,
see e.g. \cite{matteson2017independent}, \cite{CHEN2018118},
\cite{PhysRevLett.125.122001} and \cite{ugwu2023distance},
but to the best of our knowledge they have not yet been 
formally investigated from this perspective.  

In Sections~\ref{sec:IFs} and~\ref{sec:bdv} we study
the robustness of $\dCov$ and $\dVar$ against outliers
by deriving influence functions and breakdown values.
Section~\ref{sec:transf} then constructs a new 
approach to make $\dCov$ and $\dCor$ more robust.
The new method is compared with alternatives 
in Section~\ref{sec:simulation}, and it is illustrated
on a real data example in Section~\ref{sec:appl}.
Section~\ref{sec:outlook} provides an outlook on
potential applications, and Section~\ref{sec:concl} concludes.

\section{Influence functions}\label{sec:IFs}

Let $T$ be a statistical functional that maps a
bivariate distribution $F$ to a scalar.
Following \cite{Hampel:IFapproach}, the
{\it influence function} (IF) of $T$ at $F$ is 
defined as 
\begin{equation} \label{eq:IF}
   \IF((s,t),T,F) = \lim_{\eps \to 0}
   \frac{T((1-\eps)F + \eps
   \Delta_{(s,t)}) - T(F)}{\eps}
	=\frac{\partial}{\partial \eps}
	T((1-\eps)F_\rho + \eps
	\Delta_{(s,t)})\Big|_{\eps = 0+}
\end{equation}
for any $(s,t)$, where $\Delta_{(s,t)}$ is 
the probability distribution that puts all 
its mass in the point $(s,t)$.
The IF quantifies\linebreak 
the effect of a small amount 
of contamination in $(s,t)$ on $T$\;. The 
supremum of the influence function\linebreak
$\gamma^*(T) \coloneqq \sup_{(s,t)\in 
\mathbb{R}^2}{|\IF((s,t),T,F)|}$
is called the \textit{gross-error sensitivity}.
One goal of robust estimation is to have
a finite gross-error sensitivity, or 
equivalently, to have a bounded
influence function.

Let $(X,Y)$ be a bivariate random
vector with distribution $F$, with 
marginal distributions $F_X$ and $F_Y$.
We can then consider the distance
covariance and distance correlation
as statistical functionals that map
$F$ to the population
value of their respective quantities. 
In this section we will derive the
influence functions of the distance 
covariance and the distance correlation, 
and study their behavior.

\subsection{Distance covariance}\label{subsec:dcov}

\noindent
The influence function of the $\alpha$-distance covariance is
given in the following result:

\begin{proposition}\label{prop:IFdcov}
    Assume $E[|X|^{2\alpha}] < \infty$ 
    and $E[|Y|^{2\alpha}] < \infty$.
    The influence function of the $\alpha$-distance
    covariance between $X$ and $Y$ is given by
     \begin{equation*}
   \IF((s,t),\dCov(X,Y;\alpha),F) =
  -2\,\dCov(X,Y;\alpha) + 2\,\eta(s,t,X,Y,\alpha)\, ,
\end{equation*}
where
\begin{align*}
\eta(s,t,X,Y,\alpha) \coloneqq 
\cov(|X-s|^{\alpha}-|X-X^{\prime}|^{\alpha},
|Y-t|^{\alpha}-|Y-Y^{\prime\prime}|^{\alpha})\,.
\end{align*} 
\end{proposition}
We have introduced the notation 
$\eta(s,t,X,Y,\alpha)$ for the second term of 
the IF. This is the quantity that depends on 
the position of the contamination $(s,t)$, 
and it governs the behavior of most 
influence functions that we will consider.
The derivations of all our influence function
results can be found in Section~A of the 
Supplementary Material.

Figure \ref{fig:IFdCov} shows the 
IF of the distance covariance
for various values of $\alpha$,
when $(X,Y)$ follows a bivariate normal 
distribution with mean zero, unit 
variances, and correlation $\rho = 0.6$.
The contamination is placed at
$(s,t) = (s,s)$ in the left panel,
and at $(s,t) = (s,-s)$
in the right panel.
We compare the influence functions of
$c_\alpha\dCov(\cdot,\cdot,\alpha)
 ^{1/\alpha}$ where 
$c_\alpha := \dCov(X,Y;1)/
\dCov(X,Y;\alpha)^{1/\alpha}$.
The exponent $1/\alpha$ gives the
quantities under comparison 
the same units. Moreover, the factor 
$c_\alpha$ ensures that all measures
estimate the same population quantity
on uncontaminated data.
Without these adjustments, the 
influence functions would not be 
directly comparable.

\begin{figure}[!ht]
\centering
\includegraphics[width=0.42\textwidth]
{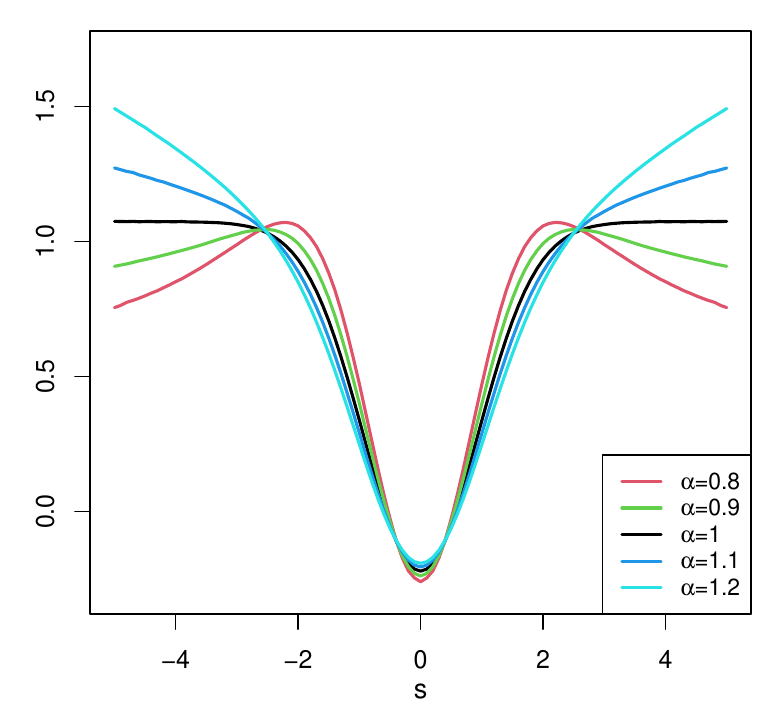}
\includegraphics[width=0.42\textwidth] 
{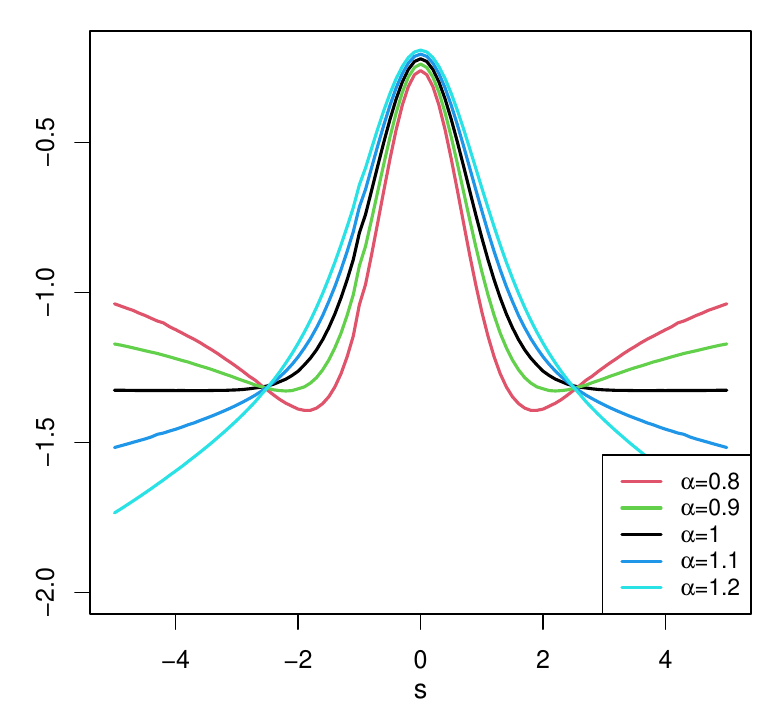}
\caption{\centering Influence function of 
$c_\alpha\dCov(\cdot,\cdot;\alpha)^{1/\alpha}$ 
for different values of $\alpha$. The left panel 
shows the IF in $(s,t) = (s,s)$, whereas right 
panel shows it in $(s,t) = (s, -s)$. Here $(X,Y)$ 
follows a bivariate normal distribution with 
correlation $\rho=0.6$.}
\label{fig:IFdCov}
\end{figure}

For small and intermediate values of $s$,
the influence functions are roughly
quadratic in $|s|$ and behave very similarly.
However, the behavior for large values of 
$|s|$ depends strongly on $\alpha$.
We see that for $\alpha < 1$ the IF is not 
only bounded but redescends towards zero.
For $\alpha > 1$ the IF becomes unbounded. 
This is not unexpected because 
$c_\alpha\dCov(\cdot,\cdot;\alpha)^{1/\alpha}$ 
becomes proportional to the absolute value 
of the classical covariance between $X$ 
and $Y$ when $\alpha \to 2$. 
The standard choice $\alpha = 1$ plays a special 
role, as the highest value of $\alpha$
for which the IF is bounded.
We formalize this behavior in
the proposition below.

\begin{proposition}\label{prop:boundeddCov}
The IF of the $\alpha$-distance 
covariance has the following properties:

\vspace{-6mm}
\begin{itemize}
\item If $\alpha > 1$ then 
  $\IF((s,t),\dCov(X,Y;\alpha),F)$ can be unbounded.
\item  If $\alpha = 1$ then 
  $\IF((s,t),\dCov(X,Y;\alpha),F)$ is bounded.        
\item  If $\alpha < 1$ then 
  $\IF((s,t),\dCov(X,Y;\alpha),F)$ is bounded and 
  redescending.
\end{itemize}
\end{proposition}

The dependency structure between $X$ and $Y$
affects the behavior for $\alpha > 1$. 
If $X$ and $Y$ are independent, then 
$\eta(s,t,X,Y,\alpha)=0$ so the IF remains bounded.
At the other extreme, if $X=Y$ they are perfectly 
dependent, and 
$\IF((s,s),\dCov(X,Y;\alpha),F) = 
\IF(s,\dVar(X;\alpha),F_X)$ is unbounded 
for $\alpha>1$ as we will see in the next section. 

\subsection{Distance variance and standard deviation}
\label{subsec:dvardsd}

From the IF of the distance covariance we can derive
those of $\dVar$ and $\dStd$. 

\begin{corollary}\label{cor:IFdvardsd}
The influence functions of $\dVar$ and $\dStd$ are given by
\begin{align*}
   \IF(s,\dVar(X;\alpha),F) =& -2\dVar(X;\alpha)
   + 2 \eta(s, s, X,X, \alpha))\\
   \IF(s,\dStd(X;\alpha),F) =&\; \frac{-2 \dVar(X;\alpha) 
   + 2 \eta(s, s, X,X, \alpha)}{2 \dStd(X;\alpha)}\;.
 \end{align*}
\end{corollary}

\begin{figure}[!ht]
\centering
\includegraphics[width=0.42\textwidth]
  {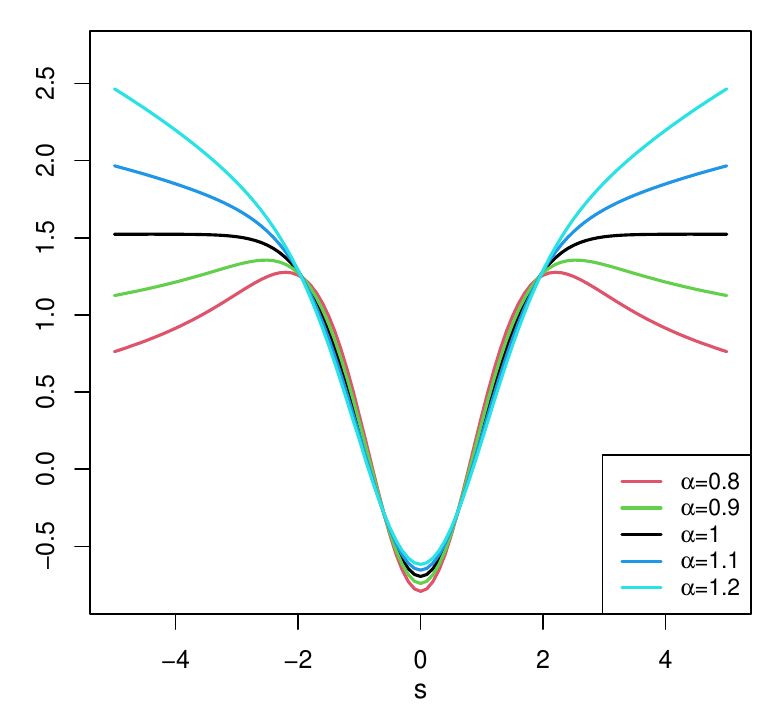}
\includegraphics[width=0.42\textwidth]
  {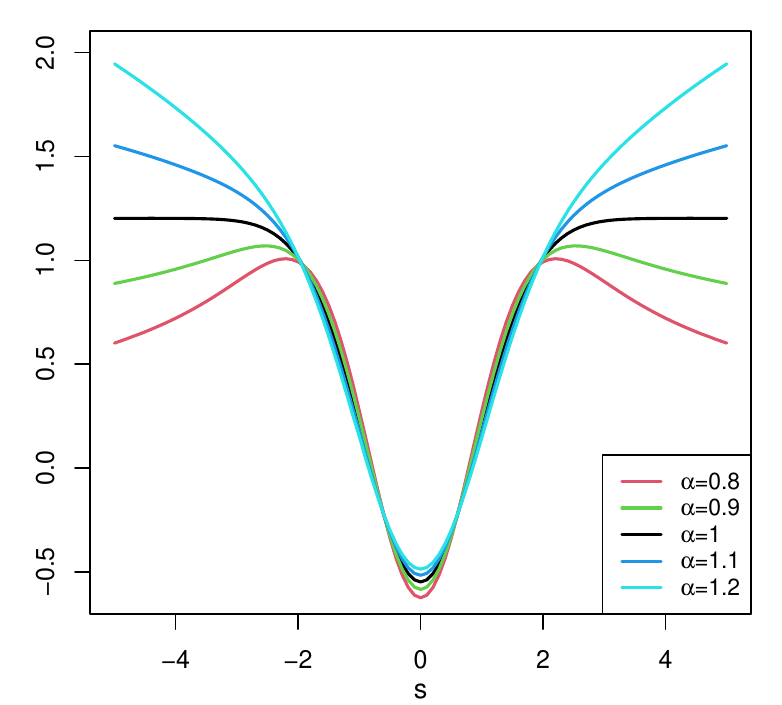}
\caption{ \centering Influence function of 
  $v_\alpha\dVar(\cdot;\alpha)^{1/\alpha}$ (left) and 
  $v_\alpha^{1/2}\dStd(\cdot;\alpha)^{1/\alpha}$ 
  (right) for different values of $\alpha$,\\ for 
  standard normal data.}
\label{fig:IFdVar_and_IFdStd}
\end{figure}

The left panel of Figure \ref{fig:IFdVar_and_IFdStd} 
plots the IF of the quantities 
$v_\alpha\dVar(\cdot;\alpha)^{1/\alpha}$ with 
$v_\alpha:=\dVar(X;1)/\dVar(X;\alpha)^{1/\alpha}$
that are comparable across $\alpha$.
Here $F$ is the standard normal distribution. 
The behavior of the IF again depends on $\alpha$. 
In particular, $\alpha = 1$ is the highest value
for which the IF is bounded.
Larger $\alpha$ yield unbounded influence functions, 
whereas smaller $\alpha$ lead to redescending curves. 

\begin{proposition}\label{prop:boundeddvar}
The IF of the $\alpha$-distance variance has the 
following properties:

\vspace{-5mm}
\begin{itemize}
\item  If $\alpha >  1$ then 
       $\IF(s,\dVar(X;\alpha),F)$ is unbounded.
\item  If $\alpha = 1$ then 
       $\IF(s,\dVar(X;\alpha),F)$ is bounded.
\item  If $\alpha < 1$ then 
       $\IF(s,\dVar(X;\alpha),F)$ is bounded 
       and redescending.
\end{itemize}
\end{proposition}

\noindent
As $\dVar$ and $\dStd$ are scale estimators,
we can compare them with popular alternatives.
We need to be careful though, since different
scale estimators may be estimating different
population quantities. For instance, if we want 
to estimate the variance of a Gaussian 
distribution we need a consistency factor
for $\dVar$. For $\alpha=1$, \citet{szekely2007} 
find at the standard normal distribution that
\begin{align*}
    \dVar(X) &= \dCov(X,X) = 
    \frac{4}{3\pi}(\pi -3\sqrt{3}+3)\,.
\end{align*}
For $X \sim \mathcal{N}(0,\sigma^2)$ and 
$c := 3\pi/(4(\pi -3\sqrt{3}+3))$
we thus obtain $c\,\dVar(X)=\sigma^2$ and 
$\sqrt{c}\,\dStd(X)=\sigma$. For other $\alpha$ we 
use $v_\alpha c\, \dVar^{1/\alpha}(X)$ with
$v_\alpha:=\dVar(X;1)/\dVar(X;\alpha)^{1/\alpha}$.

\begin{figure}[!ht] 
\centering
\includegraphics[width = 0.45\textwidth]
   {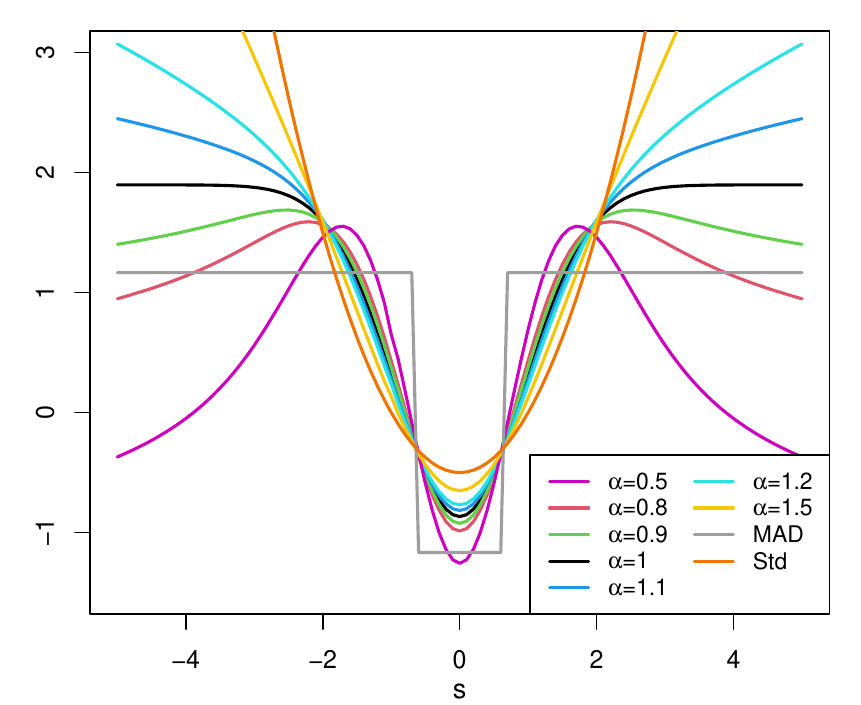}
\caption{ \centering Influence function of different 
  scale estimators, including 
  $\sqrt{v_\alpha c}\,\dStd(X;\alpha)^{1/\alpha}$ 
  for various $\alpha$, the MAD, 
  and the classical standard deviation Std.}
\label{fig:IFscaleestimators}
\end{figure}

Figure \ref{fig:IFscaleestimators} plots 
the influence function of 
$\dStd(X;\alpha)^{1/\alpha}$ with its 
consistency factor. It also contains the
IF of the classical standard deviation 
Std, and the MAD given by
$1.483\, \med(X - \med(X))$. We know 
$\dStd(X;\alpha)^{1/\alpha}$ 
becomes proportional to Std for 
$\alpha \to 2$, which explains why its 
influence function tends to a parabola.

A further investigation of the $\alpha$-distance 
standard deviation as a scale estimator can be 
found in Section~B of the Supplementary Material.

\subsection{Distance correlation}\label{subsec:dcor}

Based on Proposition \ref{prop:IFdcov}
and Corollary \ref{cor:IFdvardsd} we now
obtain the IF of the distance correlation.
\begin{corollary}\label{cor:IFdcor}
The influence function of $\dCor$ is given by
\begin{align*}
 \IF\big((s,t),\dCor(X,Y;\alpha),F\big)
  = \frac{2 \eta(s, t, X,Y,\alpha)}
     {\dStd(X;\alpha)\,\dStd(Y;\alpha)}
 - \dCor(X,Y;\alpha)\Bigg(
     \frac{\eta(s, s, X,X, \alpha)}{\dVar(X;\alpha)}+
     \frac{\eta(t, t, Y,Y,\alpha)}{\dVar(Y;\alpha)}\Bigg).
\end{align*}
\end{corollary}

\begin{figure}[!ht]
\centering
\includegraphics[width=0.42\textwidth]
  {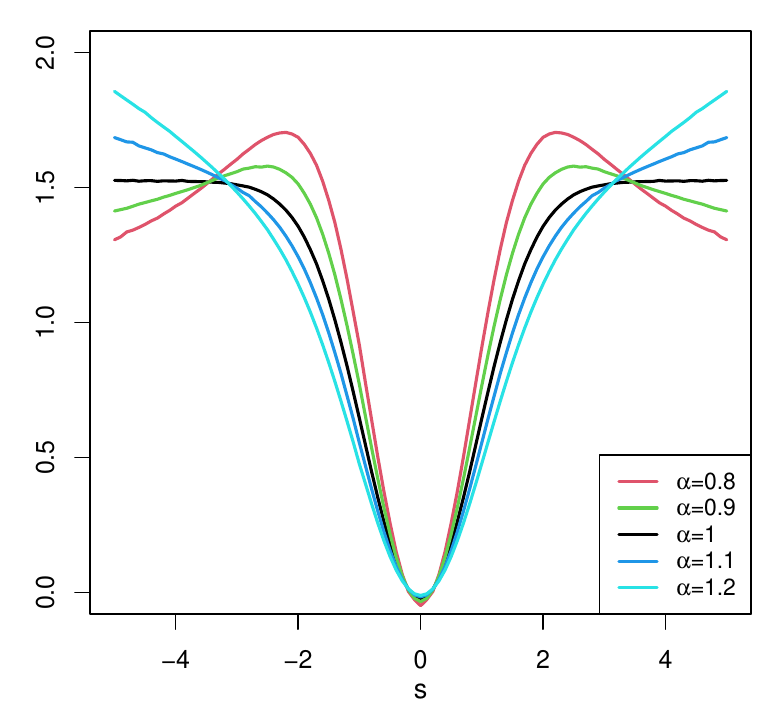}
\includegraphics[width=0.42\textwidth]
  {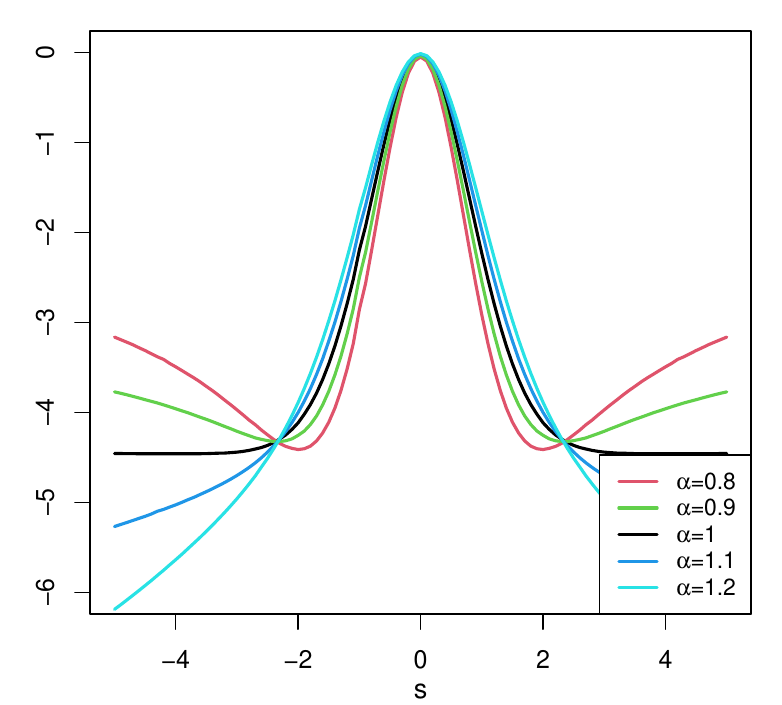}
\caption{\centering Influence function of 
$r_\alpha\dCor(\cdot,\cdot;\alpha)^{1/\alpha}$ 
for various $\alpha$. The left panel shows 
$(s,t) = (s,s)$, whereas the right panel shows 
$(s,t) = (s, -s)$. Here $(X,Y)$ follows a 
bivariate normal distribution with 
$\rho=0.6$.}
\label{fig:IFdCor}
\end{figure}

Figure \ref{fig:IFdCor} shows the IF 
of $r_\alpha\dCor(\cdot,\cdot;\alpha)^{1/\alpha}$ on 
bivariate normal data with correlation $\rho = 0.6$, 
with\linebreak $r_\alpha := \dCor(\cdot,\cdot;1)/
\dCor(\cdot,\cdot;\alpha)^{1/\alpha}$. 
The behavior of the IF 
is similar to that in Figure~\ref{fig:IFdCov}.

\begin{proposition}\label{prop:boundeddcor}
The influence function of the $\alpha$-distance 
correlation satisfies:

\vspace{-5mm}
\begin{itemize}
\item If $\alpha > 1$ then 
   $\IF\big((s,t),\dCor(X,Y;\alpha),F\big)$ 
   can be unbounded.
\item If $\alpha = 1$ then 
   $\IF\big((s,t),\dCor(X,Y;\alpha),F\big)$ is bounded.
\item If $\alpha < 1$ then 
   $\IF\big((s,t),\dCor(X,Y;\alpha),F\big)$ is bounded 
   and redescending.
\end{itemize}
\end{proposition}

\section{Breakdown values} \label{sec:bdv}

From the study of the influence functions in 
Section \ref{sec:IFs} one may conclude that
$\dCov$ and $\dCor$ indeed have natural robustness 
properties when $\alpha \leq 1$, and even a 
redescending nature when $\alpha < 1$. 
However, the IF only tells part of the story. 
Here we complement the analysis with 
a discussion of the breakdown value, a popular
and intuitive measure of robustness.
It quantifies how much contamination an estimator 
can take before it becomes completely 
uninformative, i.e., it carries no information 
about the uncontaminated data.

The {\it finite-sample breakdown value} 
\citep{donoho1983notion}
of an estimator $T$ at a dataset $\bZ$ is the 
smallest fraction of observations that needs 
to be replaced to make the estimate useless. 
Here $\bZ$ is a dataset of size $n$, 
and we denote by $\bZ^m$ any corrupted dataset 
obtained by replacing at most $m$ cases of 
$\bZ$ by arbitrary cases. Then the finite-sample 
breakdown value of $T$ at $\bZ$ is 
defined as
\begin{equation*} 
  \varepsilon^*_n(T, \bZ)=
  \min \left\{\frac{m}{n}:\;
	\sup_{\bZ^m}{\left|T(\bZ^m) - 
	T(\bZ)\right|} = \infty\right\}.
\end{equation*}

We now investigate the finite-sample breakdown 
value of the distance variance. For a
univariate sample $X_n$ we denote
$d_{ij} := |x_i-x_j|^{\alpha}$ for $\alpha>0$.
(In higher dimensions $|\cdot|$ would be
interpreted as the Euclidean norm.)
Double centering yields the values
$$\Delta_{ij} := d_{ij} - \overline{d_{i.}} - 
  \overline{d_{.j}} + \overline{d_{..}}$$
so  
$\sum_{j=1}^n \Delta_{ij} = 0$ for all $i$ and
$\sum_{i=1}^n \Delta_{ij} = 0$ for all $j$.

Now we replace the observation $x_1$ by a large number
$s>0$ that we will let go to infinity afterward.
[For $s<0$ we would write $|s|$ below, and in higher
dimensions we could replace $x_1$ by e.g.
$(s,0,\ldots,0)$.]
When $s$ grows, the $n-1$ values $d_{1j} = |s-x_j|^\alpha$
for $j \neq 1$ are of the order $s^\alpha$. We can think
of the $x_j$ as small relative to $s$, and positioned
around zero. We will write the quantities of interest 
in terms of the variable $u := s^{\alpha/2}$\,. 
This yields
$$d_{1j} = |s-x_j|^\alpha = ((s-x_j)^2)^{\alpha/2}
         = u^2 + O(u)\;.$$
From this we derive expressions for 
$\overline{d_{i.}}$,
$\overline{d_{.j}}$, 
$\overline{d_{..}}$,
as well as $\Delta_{ij}$ in 
Section~C.1 of the Supplementary Material. 
Combining these, the distance variance of $X_n$ 
becomes
\begin{equation*}
  \dVar(X_n; \alpha) = 4\frac{(n-1)^2}{n^4}u^4
  + O(u^3)\;.
\end{equation*}
For large $s$ and sample size $n$ we thus obtain
\begin{equation} \label{eq:dVara}
  \dVar(X_n; \alpha) \approx \frac{4}{n^2}s^{2\alpha}
  + O(s^{3\alpha/2})
\end{equation}  
which goes to infinity with $s$ (for any $\alpha > 0$), 
so the breakdown value of $\dVar$ is only 
$\frac{1}{n} \approx 0$. This says that a single 
outlier can destroy it.

Interestingly, this does not contradict the fact that the
influence function is bounded for $\alpha \leqslant 1$.
This is because the denominator of the
leading term of the right hand side
of~\eqref{eq:dVara} contains $n^2$ instead of the
usual $n$. To see the effect of this, note that the
contamination mass $\varepsilon$ in the definition of
the influence function corresponds to $\frac{1}{n}$, hence
\begin{align*}
   \IF(s,\dVar,F) &= 
   \frac{\partial}{\partial \varepsilon}\left[
    \dVar(F_\varepsilon) 
    \right]_{\varepsilon=0}
   = \lim_{\varepsilon \rightarrow 0+}
     \frac{\dVar(F_\varepsilon) -
     \dVar(F)}{\varepsilon}\\
   &= \lim_{\varepsilon \rightarrow 0+}
     \frac{4 \varepsilon^2 s^{2\alpha}}{\varepsilon}
     + \lim_{\varepsilon \rightarrow 0+}
       \frac{O(s^{3\alpha/2})}{\varepsilon}
    = \lim_{\varepsilon \rightarrow 0+}
      4 \varepsilon s^{2\alpha}
         + \lim_{\varepsilon \rightarrow 0+}
        \frac{O(s^{3\alpha/2})}{\varepsilon}  
\end{align*}
so the highest order term in $s$ vanishes due to 
the remaining factor $\varepsilon$, whereas
the next term corresponds to the IF.

This might be the first occurrence of a
natural statistic 
with a bounded IF and a zero breakdown value. 
The opposite situation was known before,
for instance the normal scores R-estimator of
location has an unbounded IF and a positive 
breakdown value of $23.9\%$, see page 112 of
\cite{Hampel:IFapproach}. Also, the normal
scores correlation coefficient has unbounded
IF and breakdown value 12.4\% 
\citep{boudt2012gaussian}. 
But a combination of a bounded IF with a zero 
breakdown value appears to be new.
In retrospect we can construct artificial but 
simpler estimators with these properties,
such as a modified trimmed mean. 
The usual 10\% trimmed mean is given by
$$T(X_n) = \frac{\sum_{i=1}^n w_i x_{i:n}}
  {\sum_{i=1}^n w_i}$$
where $x_{1:n} \leqslant \ldots \leqslant x_{n:n}$
and the $w_i$ are 1, except for the first and last 10\% 
of them for which $w_i=0$. If we instead put the first
and last 10\% of them equal to $w_i = 1/n^2$ then
the breakdown value will go down from 10\% to 
$1/n \approx 0$, but the influence function will 
remain the same as that of the usual 10\% trimmed 
mean.

\begin{figure}[!ht]
\centering
\includegraphics[height=4.5cm] {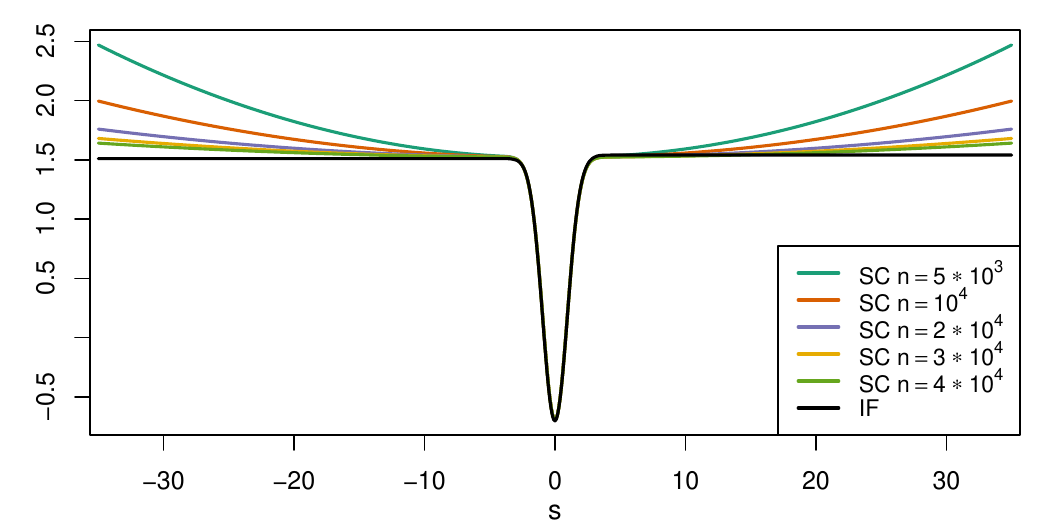}
\caption{\centering Unbounded sensitivity curve 
  of $\dVar$ for 
  different sample sizes $n$, which converges for 
  $n \rightarrow \infty$ to the bounded influence 
  function.}
\label{fig:SCdVar}
\end{figure}

The {\it sensitivity curve} of an estimator $T$
is a finite-sample version of the IF, given by
\begin{align}
\mbox{SC}\big(s, &T, (x_1,\ldots,x_n)\big) 
  :=n\,\big(T(s,x_1,\ldots,x_n) -
  T(x_1,\ldots,x_n)\big)
\end{align}
at a dataset $(x_1,\ldots,x_n)$. Here we set
$x_i = \Phi^{-1}((i-\frac{1}{2})/(n+\frac{1}{2}))$ 
for illustration purposes.
Note that~\eqref{eq:dVara} implies that the 
sensitivity curve of the distance variance for
$\alpha=1$ is unbounded, since it becomes 
arbitrarily high for large $s$.
This also seems to be at odds with the fact that  
the IF is bounded. However, the 
unbounded sensitivity curve does converge to the 
bounded IF for increasing sample
size $n$. This is is illustrated in 
Figure~\ref{fig:SCdVar}
which shows the sensitivity curve for a
wide range of $s$, for different
sample sizes $n$.

For the breakdown of $\dCov$ we start from a 
bivariate dataset $Z = [X_n\,;\,Y_n] = 
\{(x_i,y_i);\;i=1,\ldots,n\}$
and replace $(x_1,y_1)$ by some $(s,t)$ for large 
positive $s$ and $t$. In Section~C.1 of the 
Supplementary Material we then obtain
\begin{equation}\label{eq:dCovab}
  \dCov(X_n,Y_n;\alpha)
  \approx \frac{4}{n^2} s^\alpha t^\alpha
  + O(s^{3\alpha/4}t^{3\alpha/4})\,,
\end{equation}
so the breakdown value of $\dCov$ is 
$\frac{1}{n} \rightarrow 0$ as well.

Note that~\eqref{eq:dCovab} implies that the 
sensitivity surface of the distance covariance is 
unbounded, since it grows without bound for large 
arguments $s$ and $t$. But when $\alpha \leqslant 1$
the unbounded sensitivity surface does converge to 
the bounded influence function for increasing 
sample size $n$. The situation is analogous to the
illustration in Figure~\ref{fig:SCdVar}.

For $\dCor$ we can for instance take $t=s$. Then we
obtain
\begin{equation} \label{eq:dCor+}
  \dCor(X_n,Y_n) = 
  \frac{\dCov(X_n,Y_n)}
  {\sqrt{\dVar(X_n)}\sqrt{\dVar(Y_n)}}
 \longrightarrow 1
\end{equation}
for $s \rightarrow \infty$. So replacing a single data
point can bring $\dCor$ arbitrarily close to 1, even 
if the variables $X$ and $Y$ are independent. This can
also be seen as breakdown.

So far we have looked at the finite-sample breakdown
value, but we can also compute the asymptotic version.
For the distance variance we consider the distribution
$F$ of $X$. We can construct contaminated distributions
$F_\varepsilon = (1-\varepsilon)F + \varepsilon H$
where $H$ may be any distribution.
The asymptotic breakdown value is then defined as
$$\varepsilon^*(\dVar) = \inf \{\varepsilon;\; 
  \sup_H\;\dVar\big( (1-\varepsilon)F + 
  \varepsilon H \big) = \infty\}\;.$$
Section~C.2 of the Supplementary Material 
shows that $\varepsilon^*(\dVar) = 0$ by 
inserting distributions $H = \Delta_s$ for 
$|s| \rightarrow \infty$.

For the asymptotic breakdown value of $\dCov$
we contaminate the bivariate distribution $F$ of $(X,Y)$,
and define
$$\varepsilon^*(\dCov) = \inf \{\varepsilon;\; 
  \sup_H\;\dCov \big( (1-\varepsilon)F + 
  \varepsilon H \big) = \infty\}\;.$$
Section~C.3 of the Supplementary Material 
shows that $\varepsilon^*(\dCov) = 0$
by employing $H = \Delta_{(s,s)}$ for 
$|s| \rightarrow \infty$. 
One can verify that~\eqref{eq:dCor+}
also holds in the asymptotic setting.

Whether finite-sample or asymptotic, the only way to 
prevent breakdown of $\dCov$ and $\dCor$ 
is to ensure that $s$ and $t$ remain bounded.

\section{Robustness by transformation} 
\label{sec:transf}

From the results of Sections \ref{sec:IFs} and \ref{sec:bdv}, it is clear that if we want a 
strictly positive breakdown value we cannot use 
$\dCov$ directly, no matter the value of $\alpha$. 
One could think of applying a robust covariance
measure to the doubly centered distances, but that 
may lose the crucial property that a population 
result of zero implies $X \indep Y$. 

However, if we first apply a bounded 
transformation to $X$ and $Y$ prior to computing $\dCov$ 
and $\dCor$, the breakdown value would be strictly larger 
than 0 and the influence function would be bounded for 
any $\alpha$. We will consider 
dependence measures of the type
\begin{equation} \label{eq_transfo}
  \dCor(\psi_1(X), \psi_2(Y); 1)
\end{equation}
where the $\psi_j$ are bounded functions transforming 
$X$ and $Y$. The $\psi_j$ have to be \mbox{invertible}, 
so that independence of $\psi_1(X)$ and $\psi_2(Y)$ 
implies independence of
$\psi_1^{-1}(\psi_1(X)) = X$ and 
$\psi_2^{-1}(\psi_2(Y)) = Y$. 
Some suggestions in this direction have been made 
before. For instance, \cite{szekelyrizzo2023} suggest 
computing the distance correlation after transforming 
the data to ranks, which corresponds to using 
$\psi_1(x) = F_X(x)$ and $\psi_2(y) = F_Y(y)$,
the cumulative distribution functions of $X$ and $Y$.
\cite{Mai2023} apply $\dCor$ after transforming $X$ 
and $Y$ to normal scores (also called Gaussian ranks), 
by the unbounded transformation 
$\psi_j(t) = \Phi^{-1}(F_j(t))$ with 
$\Phi$ the standard Gaussian cdf.
This approach of computing a classical estimator after 
marginal transformation of the variables has also been 
used successfully in the context of correlation 
estimation in high dimensions \citep{RR_FROC}. 
In that study, transformations with a continuous $\psi$
that is redescending in the sense that 
$\lim_{x\to \pm \infty}|\psi(x)| = 0$ turned out to be 
most effective, in line with the success of such
$\psi$-functions for M-estimation of location
in \cite{andrews}.

\begin{figure}[!ht]
\centering
\includegraphics[width = 0.75\textwidth]{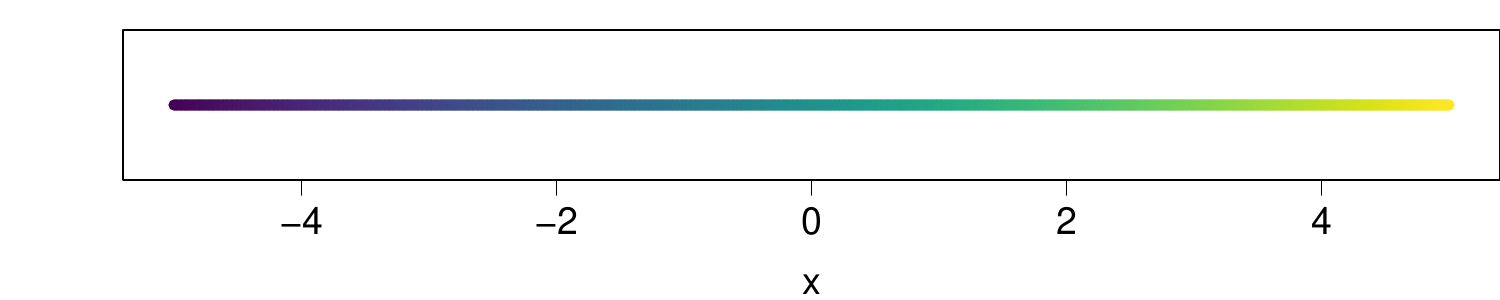}
\includegraphics[width = 0.75\textwidth]{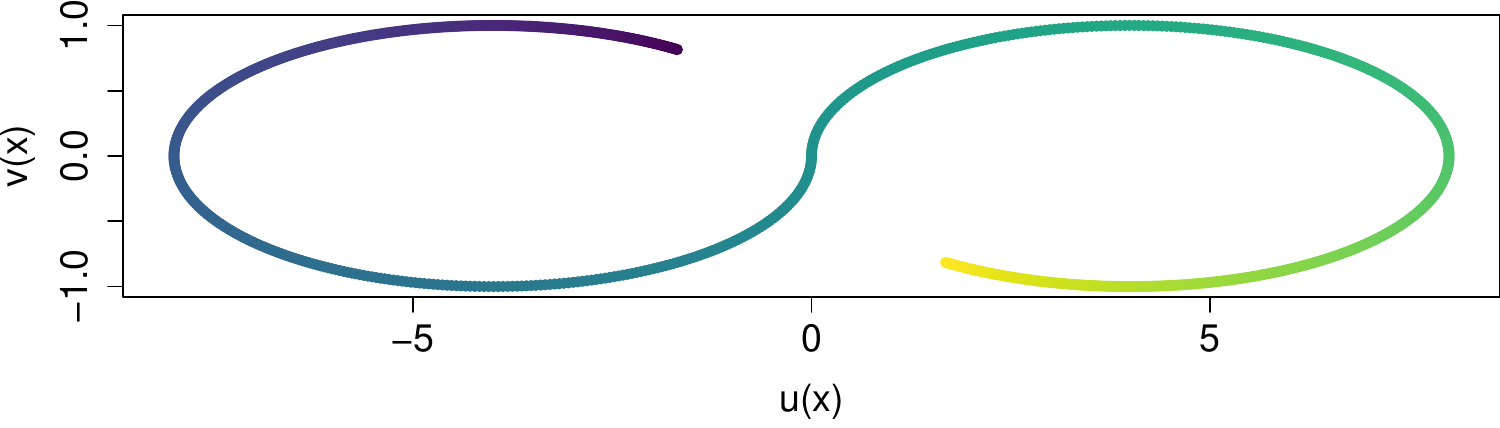}
\caption{\centering Illustration of the biloop 
         transformation.}
\label{fig:psibiloop}
\end{figure}

This raises a question: is it possible to design a 
bounded transformation $\psi$ which (i) is invertible
so we keep the independence property, and (ii) is 
continuous and redescending? 
At first sight this seems impossible: any invertible 
continuous function 
$\psi: \mathbb{R} \mapsto \mathbb{R}$ has to be 
strictly monotone, so it cannot be redescending. 
However, if the goal is to quantify (in)dependence of 
random variables, we need not stay in one dimension, 
and can in fact allow the image of $\psi$ to be in 
$\mathbb{R}^2$. This unlocks the possibility of 
creating a $\psi$-function which is simultaneously 
continuous, invertible, and redescending. 
We propose to use the function
$\psi_{\biloop}: \mathbb{R} \to \mathbb{R}^2: 
x \mapsto (u(x),v(x))$ with
\begin{align*}
u(x) &= 
\begin{cases}
   \;\;\;c\, ( 1 + \cos{(2 \pi \tanh{(x/c)}+ \pi)}) 
   & \text{if } x\geqslant 0\\
   -c\, ( 1 + \cos{(2 \pi \tanh{(x/c)}- \pi)})  
   & \text{if } x < 0
\end{cases}\\
v(x) &= \sin(2 \pi \tanh{(x/c)})
\end{align*}
where $c>0$ is a tuning constant with default 
$c = 4$ that gave good results in simulations. 
We apply this function to  
robustly standardized data variables, that is, 
their median is set to 0 and their MAD to 1.
Note that the image of $\psi_{\biloop}$ is a combination 
of two ellipses around $(c,0)$ and $(-c,0)$, since 
$\psi_{\biloop}(\mathbb{R}) = \{(u,v) \in \mathbb{R}^2;
\;||(u-c,cv)|| = c \;\mbox{ or }\; ||(u+c,cv)|| = c \}$. 
Figure \ref{fig:psibiloop} illustrates the function 
$\psi_{\biloop}$. As the argument increases from $x=0$, 
$\psi_{\biloop}(x)$ first moves to the right away from 
the origin, but for 
$x/c > 4 \tanh^{-1}(\frac{1}{2}) \approx 2.2$ it 
returns on the ellipse, and   
$\lim_{x\to \pm \infty}|\psi_{\biloop}(x)| = 0$.
We call $\psi_{\biloop}$ a \textit{biloop function}. 
Here `loop' refers to an ellipse, and `bi' alludes to 
the number of ellipses, the bivariate nature of
$\psi_{\biloop}$\,, and the redescending nature of 
the biweight function \citep{Tukey1974}. 

\begin{figure}[!hb]
\centering
\includegraphics[height=57mm]
  {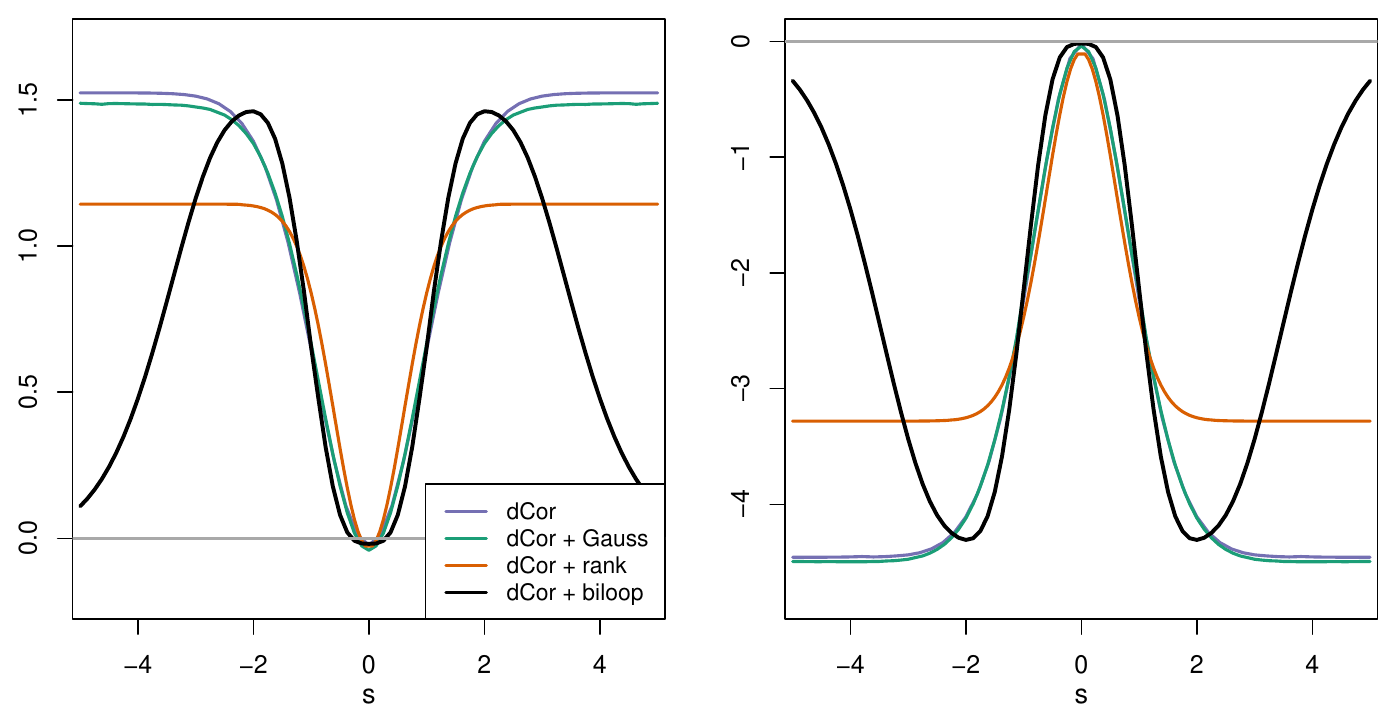}
\caption{\centering IF of $\dCor(\psi(X),\psi(Y);1)$ 
where $\psi$ is the identity, the normal score 
(Gaussian rank), the rank transform, 
or the biloop. The left panel 
shows $(s,t) = (s,s)$, whereas the right panel 
shows $(s,t) = (s, -s)$.\\ Here $(X,Y)$ is bivariate 
normal with correlation $\rho=0.6$.}
\label{fig:IFpsibiloop}
\end{figure}

We now consider the biloop $\dCor$, i.e., the distance 
correlation after the $\psi_{\biloop}$ transformation,
as in~\eqref{eq_transfo}. 
Due to the bounded and redescending nature of the 
biloop transformation, the biloop $\dCor$ has a 
bounded and redescending influence function. 
Figure \ref{fig:IFpsibiloop} shows this influence 
function and compares it to the influence functions 
of the classical $\dCor$ with $\alpha=1$, as well 
as to the $\dCor$ after rank transform 
\citep{szekelyrizzo2023} 
and after normal scores \citep{Mai2023}. 
For the expressions of these
influence functions we refer to Section~D 
of the Supplementary Material.

We see that the biloop $\dCor$ indeed has a 
redescending influence function. 
Using the rank transform does lead to 
a lower gross-error sensitivity, but does 
not yield a redescending influence function. 
The difference between the classical $\dCor$ and 
$\dCor$ after normal scores is tiny. This similarity 
is to be expected since the model distribution in 
this plot is Gaussian too. 

We end with a remark on the utility of a 
highly robust measure of independence. 
The robustness properties of the biloop $\dCor$ 
make it less sensitive to observations 
in the tails of the distribution, far away 
from the center.
In contrast, the classical $\dCor$ picks up
dependence in these tails very easily.
In some situations we may want to focus on 
dependence in the tails, and in other
situations we may prefer not to.
In general, comparing the classical $\dCor$
with the robust version 
helps to identify whether the dependence was 
mainly in the tails or rather in the center of 
the data.

\section{Empirical results} \label{sec:simulation}

\subsection{Power simulation}
\label{sec:powersim}

In this simulation we study the power of $\dCor$ 
after applying the transformations discussed in 
Section \ref{sec:transf}. More precisely, we 
compare the original $\dCor(X,Y)$ with the proposed 
biloop $\dCor$ given by 
$\dCor(\psi_{\biloop}(X),\psi_{\biloop}(Y))$, with 
$\dCor(F_X(X),F_Y(Y))$ by transforming the data to 
ranks, and with 
$\dCor(\Phi^{-1}(F_X(X)),\Phi^{-1}(F_Y(Y)))$ using 
coordinatewise normal scores.

\begin{figure*}[!ht]
\centering
\includegraphics[width = 0.99\textwidth]
  {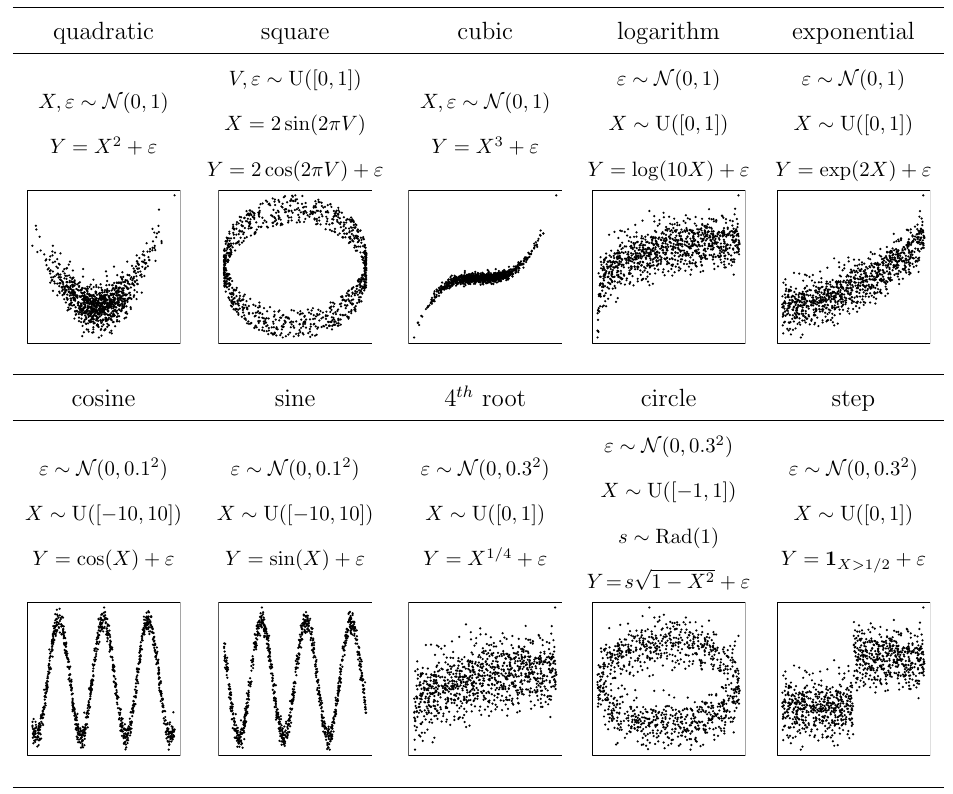} 
\caption{\centering The bivariate settings 
        for the power simulation.}
\label{fig:simsettings_bivariate}
\end{figure*}

To thoroughly compare the methods we use 16 data settings. We consider 10 bivariate settings, based on \cite{chaudhuri2019fast} and \cite{Reshef2018}. In addition we study 6 multivariate settings, following the simulation in \cite{szekely2007}. An overview of the bivariate simulation settings is given in Figure~\ref{fig:simsettings_bivariate}. For each setting we generate 2000 samples according to the specified distributions. On each of these samples, we then execute a permutation test with $\lfloor(200+5000/n)\rfloor$ permutations and a significance level of $0.1$ as in \cite{szekely2007}.

\begin{figure}[!ht]
\begin{subfigure}{0.245\textwidth}
  \includegraphics[width=\linewidth]
  {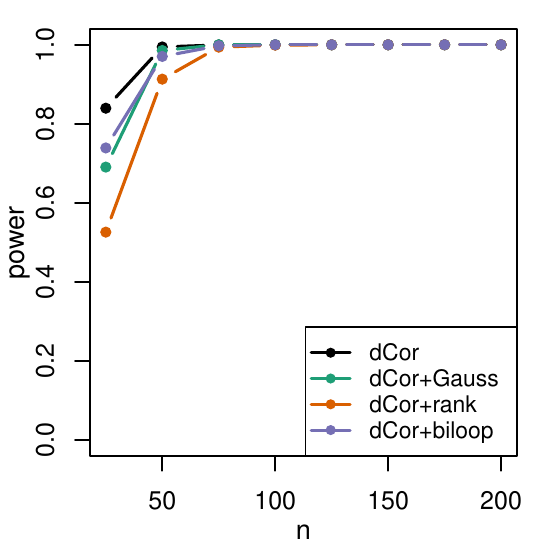}
  \caption{\centering quadratic}
\end{subfigure}
\begin{subfigure}{0.245\textwidth}
  \includegraphics[width=\linewidth]
  {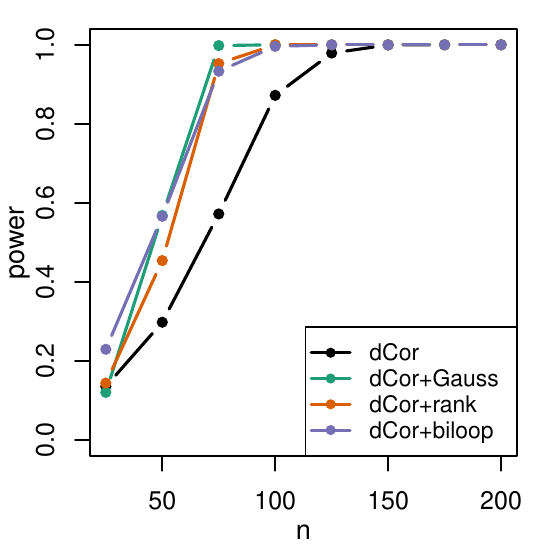}
  \caption{\centering square}
\end{subfigure}
\begin{subfigure}{0.245\textwidth}
  \includegraphics[width=\linewidth]
  {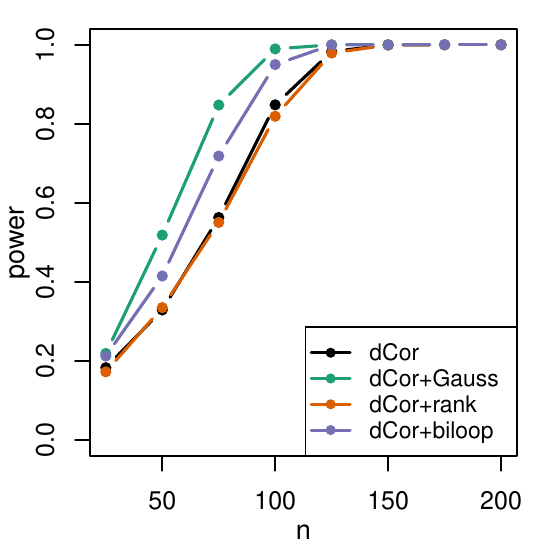}
  \caption{\centering cosine}
\end{subfigure}
\begin{subfigure}{0.245\textwidth}
  \includegraphics[width=\linewidth]
  {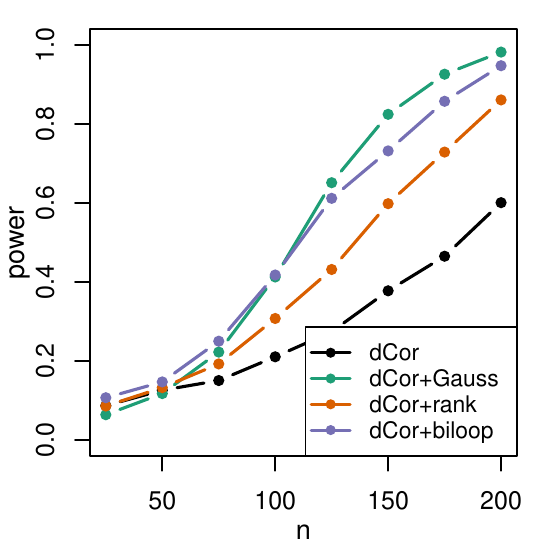}
  \caption{\centering circle}
\end{subfigure}
\caption{\centering Results of the power simulation 
on bivariate $(X,Y)$ for different sample sizes $n$.}
\label{fig:powersim_univariate}
\end{figure}

We first discuss the results of the bivariate settings. We present the most interesting results here, and refer to Section~E of the Supplementary Material 
for the results on the other settings, which gave more moderate and qualitatively similar results. Figure \ref{fig:powersim_univariate} shows the results of the power simulation on the quadratic, square, cosine and circle setups. We see that all methods perform reasonably well and usually achieve a high power for sample sizes of $n = 100$ and upwards. An exception is the circle setup, where all methods require a higher sample size to achieve a satisfactory power level. Interestingly, the classical $\dCor$ performs worst here, with the $\dCor$ after rank transform as a second. The biloop $\dCor$ and $\dCor$ after normal scores perform similarly and are on top. 

\begin{figure}[!ht]
\centering
\includegraphics[width = 0.62\textwidth]
  {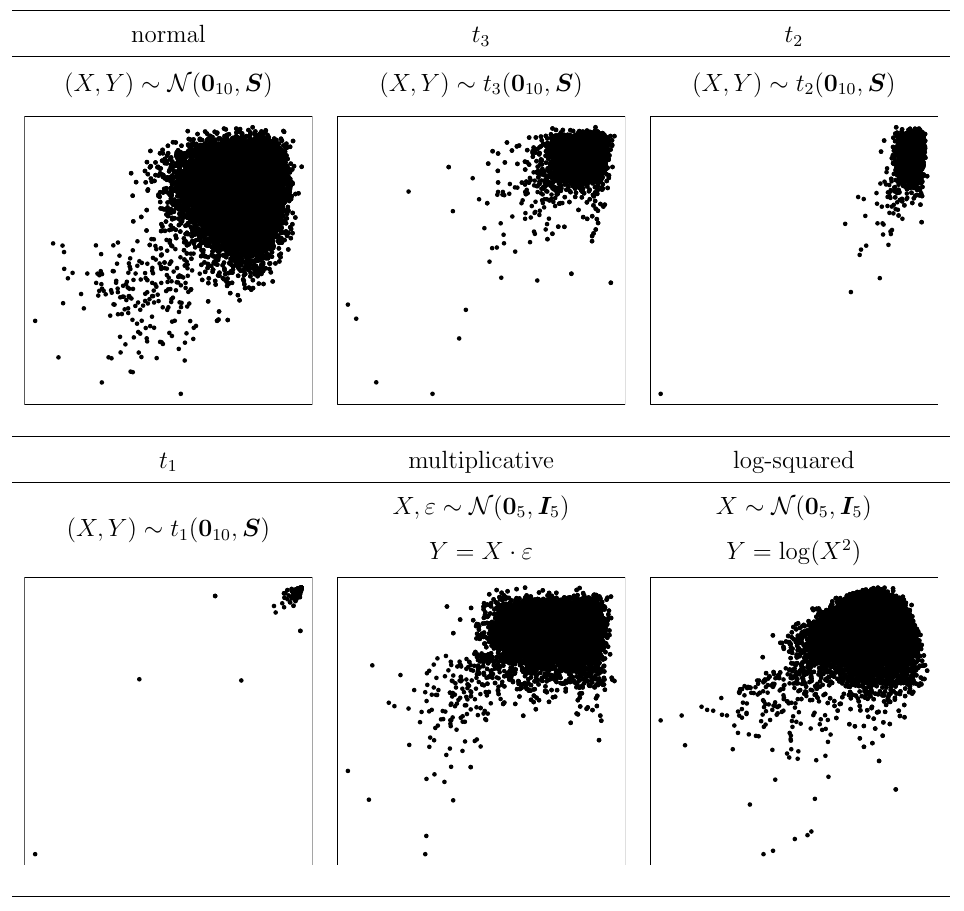}
\caption{\centering Plots of doubly centered distances 
   of $Y$ versus those of $X$ for the settings of the power
   simulation, where $\bm{S} = \begin{bsmallmatrix}
           \bm{I}_5 & \bm{1}_5 \bm{1}_5^{\intercal}/10\\
           \bm{1}_5\bm{1}_5^{\intercal}/10 & \bm{I}_5
       \end{bsmallmatrix}$.} 
\label{fig:simsettings_multivariate}
\end{figure}

\begin{figure}[!ht]
\begin{subfigure}{0.245\textwidth}
  \includegraphics[width=\linewidth]
  {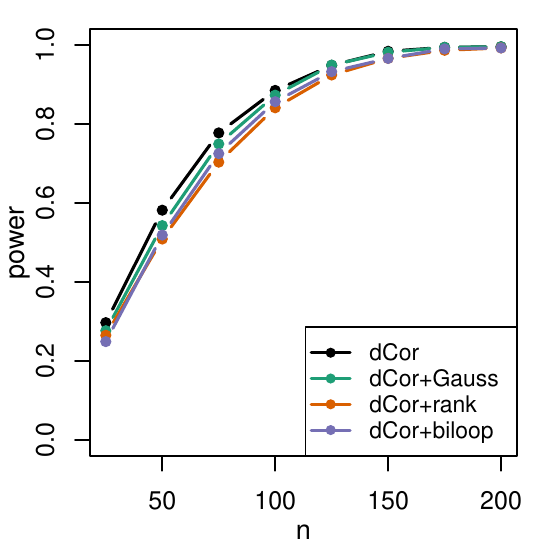}
  \caption{\centering normal}
\end{subfigure}
\begin{subfigure}{0.245\textwidth}
  \includegraphics[width=\linewidth]
  {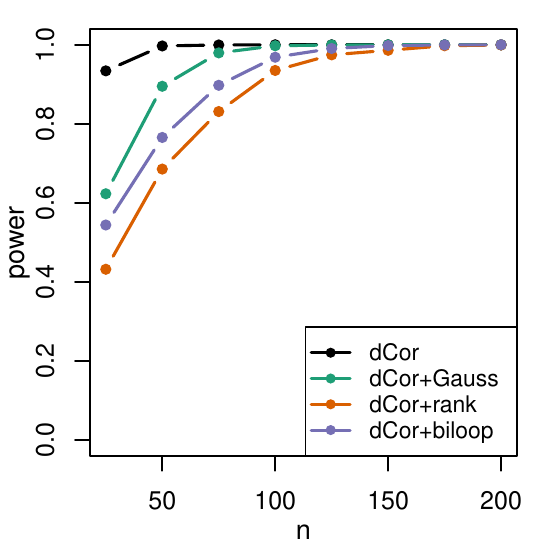}
  \caption{\centering $t_3$}
\end{subfigure}
\begin{subfigure}{0.245\textwidth}
  \includegraphics[width=\linewidth]
  {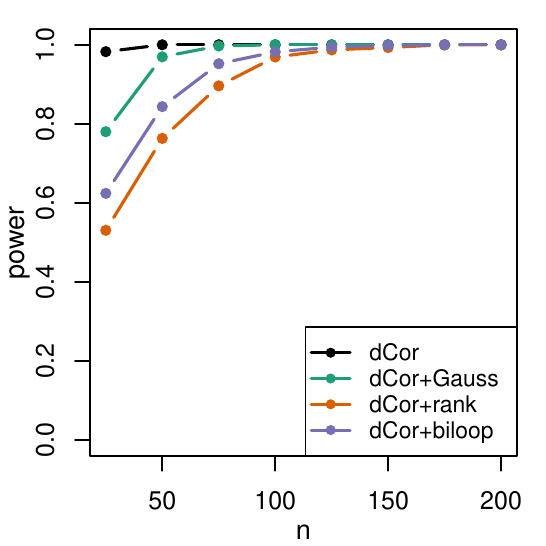}
  \caption{\centering $t_2$}
\end{subfigure}
\begin{subfigure}{0.245\textwidth}
  \includegraphics[width=\linewidth]
  {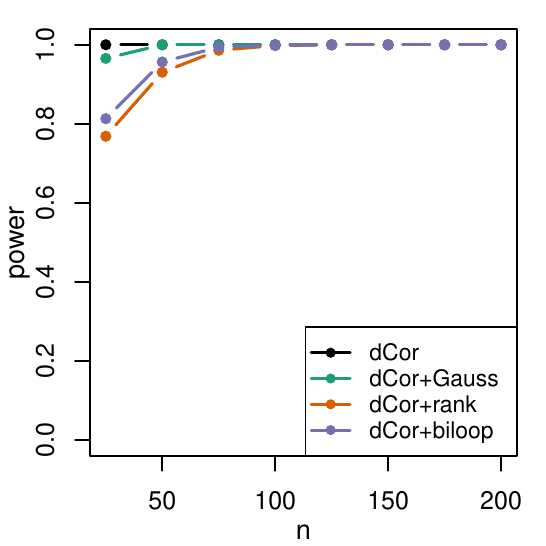}
  \caption{\centering $t_1$}
\end{subfigure}
\caption{\centering Power simulation in the multivariate 
         normal, $t_3$, $t_2$ and $t_1$ settings.} 
\label{fig:powersim_multivariate1}
\end{figure}

Figure~\ref{fig:simsettings_multivariate} summarizes
the multivariate settings. 
The power for the first four of these, with Gaussian 
data and various multivariate $t$-distributions,
is plotted in Figure~\ref{fig:powersim_multivariate1}.

For all methods, heavier tails led to higher power. 
This is because most methods give more weight to points further away from the center, which makes them more sensitive to dependence in the tails. This explanation is confirmed by the fact that $\dCor$ typically performs somewhat better than the other alternatives here. Its non-robustness causes it to easily pick up the dependence in the extreme tails of the distributions. Normal scores is next, and is also sensitive to dependence in the tails due to the unboundedness of the applied transformation. The biloop transform is ranked third here, and still performs well even though its robustness properties make it less dependent on the tails of the distribution. The $\dCor$ after rank transform performed only slightly lower.

\begin{figure}[!ht]
\centering
\begin{subfigure}{0.25\textwidth}
  \includegraphics[width=\linewidth]
  {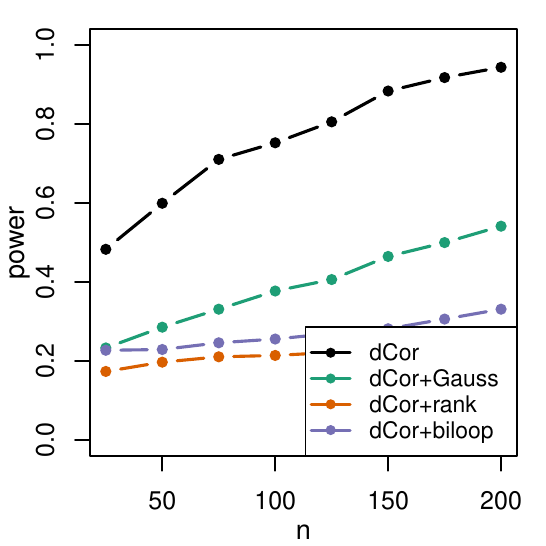}
  \caption{\centering multiplicative}
\end{subfigure}
\hspace{5mm}
\begin{subfigure}{0.25\textwidth}
  \includegraphics[width=\linewidth]
  {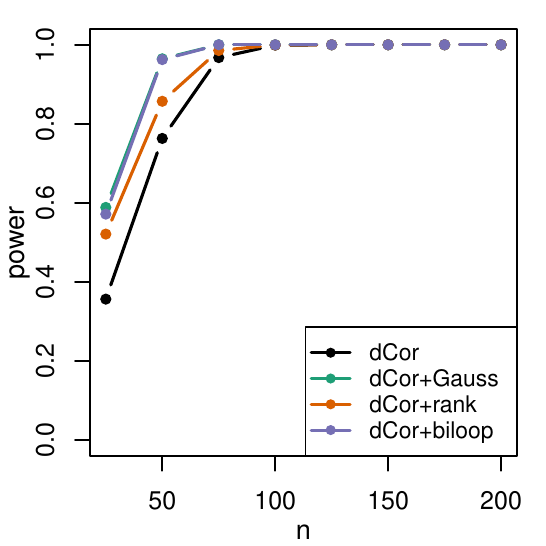}
  \caption{\centering log-squared}
\end{subfigure}
\caption{\centering Power simulation in the 
   multivariate settings
   multiplicative and log-squared.}
\label{fig:powersim_multivariate2}
\end{figure}

Figure \ref{fig:powersim_multivariate2} plots the power of the permutation tests on the multiplicative and log-squared settings. In the multiplicative setting $\dCor$ performed best, followed by the normal scores, biloop, and rank transforms. Note that his is again a setup where the dependence is most pronounced in the tails. In the log-squared setting we have more dependence in the center of the distribution. Here the biloop together with normal scores performed best. The rank transform still performed quite well, followed by the classical $\dCor$.

\subsection{Robustness simulation}
\label{sec:robsim}

We now explore the robustness of the various 
dependence measures to contamination. The 
clean data are samples of size $n=500$ from
$(X,Y)\sim \mathcal{N}_2\left(
\begin{bsmallmatrix} 0 \\ 0 \end{bsmallmatrix}\,, 
\begin{bsmallmatrix} 1 & \rho \\ \rho & 1 
\end{bsmallmatrix}\right)$,
with $\rho$ either 0 or 0.6. We then replace a 
fraction $\varepsilon$ of the generated points 
by outliers.

We first consider contamination generated as a 
small cloud of $\varepsilon n$ points following 
the distribution $\mathcal{N}_2\left(\mu_{c}, 
0.25\bm{I}_2\right)$. For $\rho = 0$ we 
take $\mu_c = [6 \; 6]^{\intercal}$, whereas for 
$\rho = 0.6$ we consider 
$\mu_c = [6 \; 6]^{\intercal}$ and 
$\mu_c =  [-6 \; 6]^{\intercal}$.

\begin{figure}[!ht]
\renewcommand*\thesubfigure{\arabic{subfigure}} 
    \centering 
\begin{subfigure}{0.25\textwidth}
  \includegraphics[width=\linewidth]
     {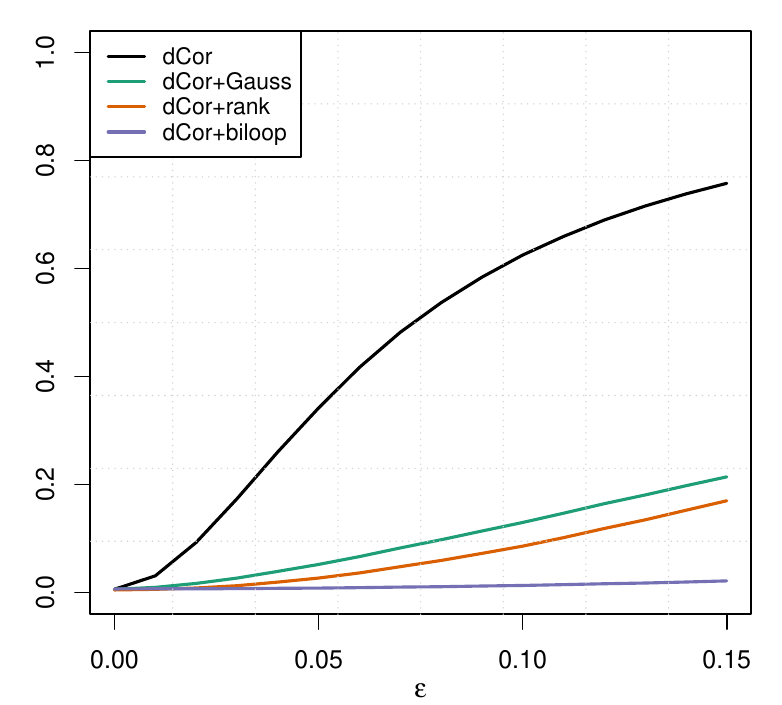}
  \caption{\centering $\rho = 0$ and $\mu_c = [6 \; 6]^{\intercal}$.}
\end{subfigure}
\hspace{5mm}
\begin{subfigure}{0.25\textwidth}
  \includegraphics[width=\linewidth]
     {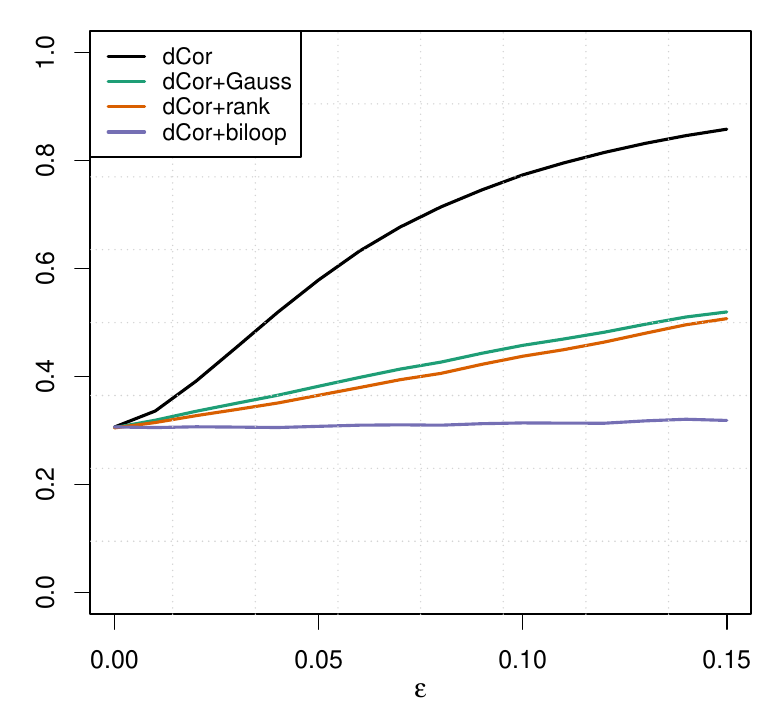}
  \caption{\centering $\rho = 0.6$ and $\mu_c = [6 \; 6]^{\intercal}$.}
\end{subfigure} 
\hspace{5mm}
\begin{subfigure}{0.25\textwidth}
  \includegraphics[width=\linewidth]
     {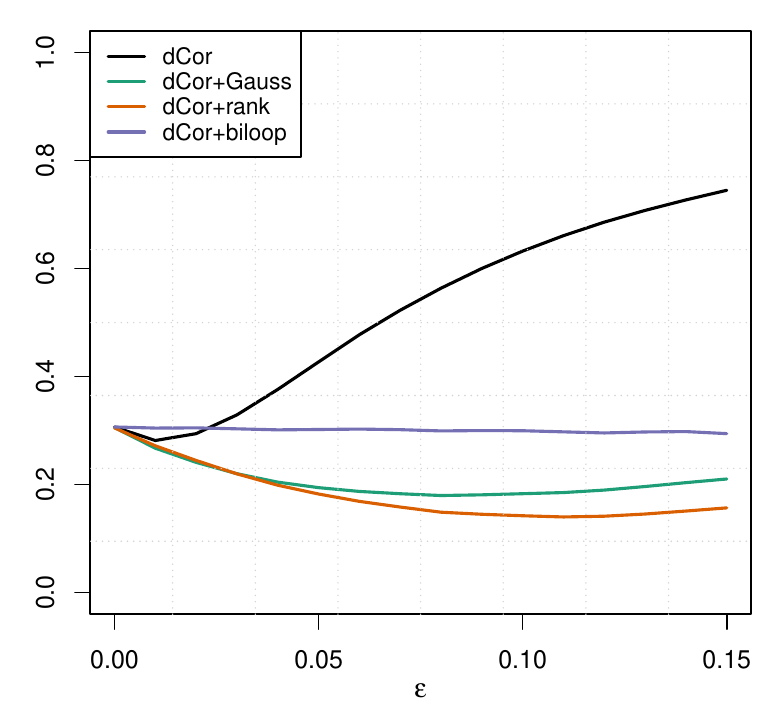}
  \caption{\centering $\rho = 0.6$ and $\mu_c = [-6 \; 6]^{\intercal}$.}
\end{subfigure}
\caption{\centering Robustness 
against $100\varepsilon\%$ of outliers.}
\label{fig:simresults_robustness1}
\end{figure}

Figure \ref{fig:simresults_robustness1} shows the average values of the different dependence measures for increasing levels of contamination. Note that we correct the dependence measures such that they estimate the same population quantity on the clean distribution by multiplying with the consistency factor $c_{\psi} = \dCor(X,Y)/\dCor(\psi(X),\psi(Y))$.
The pattern is quite similar in all three situations. The original $\dCor$ is most strongly affected by the contamination. If $\dCor$ is applied after normal scores or the rank transform, it is less affected by the contamination. The biloop $\dCor$ is barely affected.
 
\begin{figure}[!ht]
\renewcommand*\thesubfigure{\arabic{subfigure}} 
    \centering 
\begin{subfigure}{0.25\textwidth}
  \includegraphics[width=\linewidth]{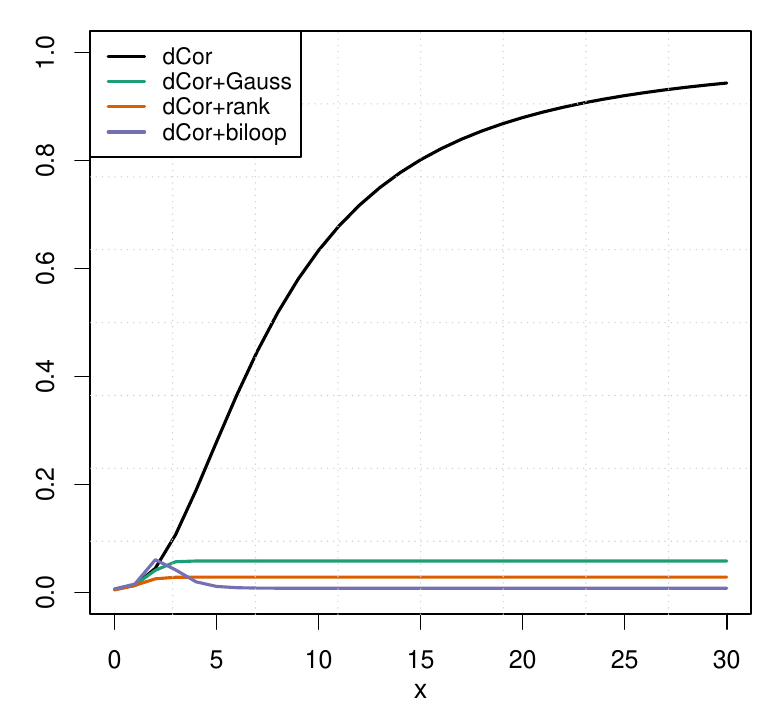}
  \caption{\centering $\rho = 0$ and 
  $\mu_c = [x \; x]^{\intercal}$.}
\end{subfigure}
\hspace{5mm}
\begin{subfigure}{0.25\textwidth}
  \includegraphics[width=\linewidth]{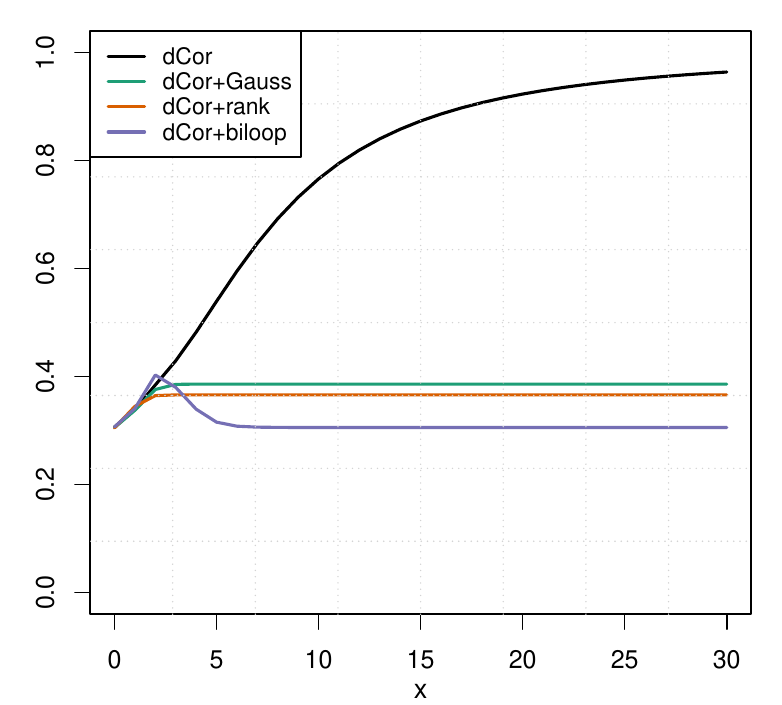}
  \caption{\centering $\rho = 0.6$ and 
  $\mu_c = [x \; x]^{\intercal}$.}
\end{subfigure}
\hspace{5mm}
\begin{subfigure}{0.25\textwidth}
  \includegraphics[width=\linewidth]{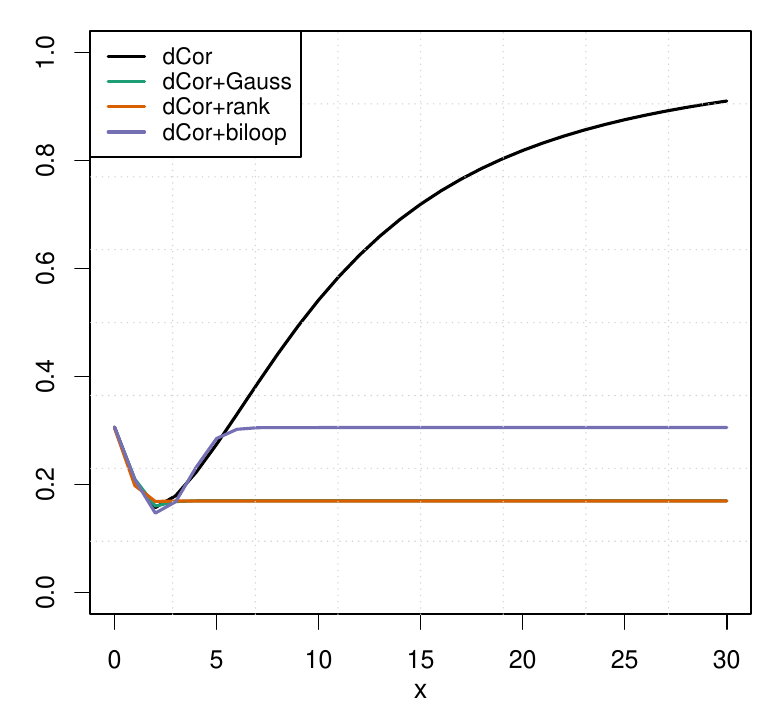}
  \caption{\centering $\rho = 0.6$ and 
  $\mu_c = [x \; -x]^{\intercal}$.}
\end{subfigure}
\caption{\centering Robustness against 5\% of outliers
         for varying degrees of outlyingness.}
\label{fig:simresults_robustness2}
\end{figure}

Next we consider the same distributions for the clean data, but this time we add 5\% outliers and vary their distance from the origin. We take the values of the outliers to be $\mu_c = [x \; x]^{\intercal}$ for $\rho= 0$, and $\mu_c = [x \; x]^{\intercal}$ or $\mu_c = [x \; -x]^{\intercal}$ for $\rho = 0.6$. Figure \ref{fig:simresults_robustness2} shows the results of this setup as a function of $x$. These results confirm what we expected from Section \ref{sec:bdv}. The original $\dCor$ has an unbounded sensitivity curve, which converges to its bounded influence function for growing sample size. In this simulation the sample size is fixed, so the curve deviates as the contamination moves further away from the center. The other three dependence measures are much less affected.

\subsection{Rejection of independence under contamination}

We now investigate the effect of contamination 
when the clean distribution has independence. 
More precisely, we investigate how often the 
null hypothesis of independence is falsely 
rejected when adding contamination to the clean 
data. The clean data is generated similarly to the setting used in Table 1 of \cite{szekely2007}. We sample ($X,Y$) from the independent bivariate standard normal, $t_3$, $t_2$, and $t_1$ distributions.
We perform permutation tests as in 
Section~\ref{sec:powersim}, at a significance level 
of 0.1 and with $\lfloor 5000/n + 200\rfloor$ 
permutations.

\begin{figure}[!ht]
\renewcommand*\thesubfigure{\arabic{subfigure}} 
    \centering 
\begin{subfigure}{0.245\textwidth}
  \includegraphics[width=\linewidth]
  {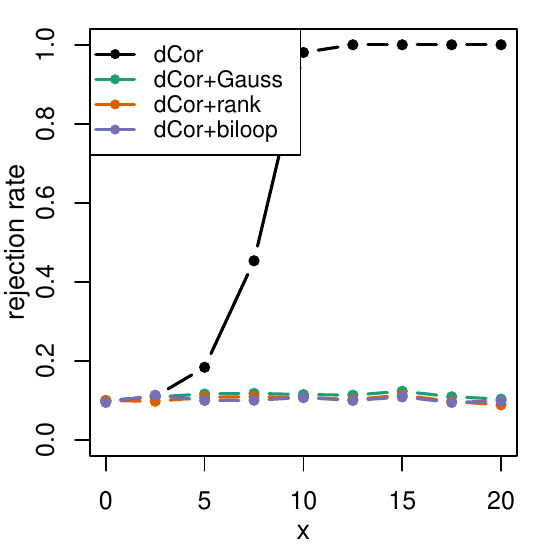}
  \caption{\centering normal}
\end{subfigure} 
\begin{subfigure}{0.245\textwidth}
  \includegraphics[width=\linewidth]
  {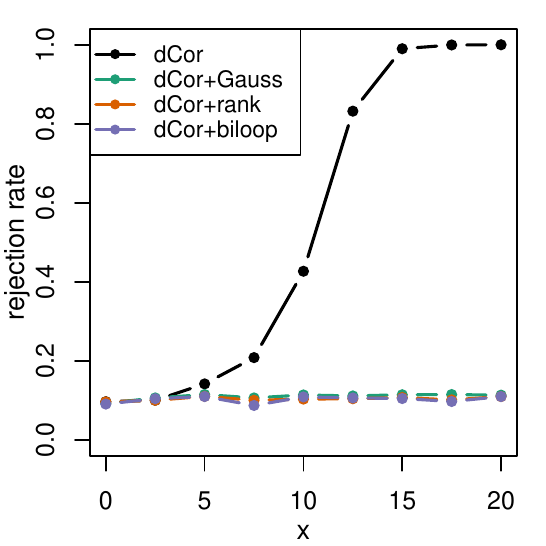}
  \caption{\centering $t_3$}
\end{subfigure}
\begin{subfigure}{0.245\textwidth}
  \includegraphics[width=\linewidth]
  {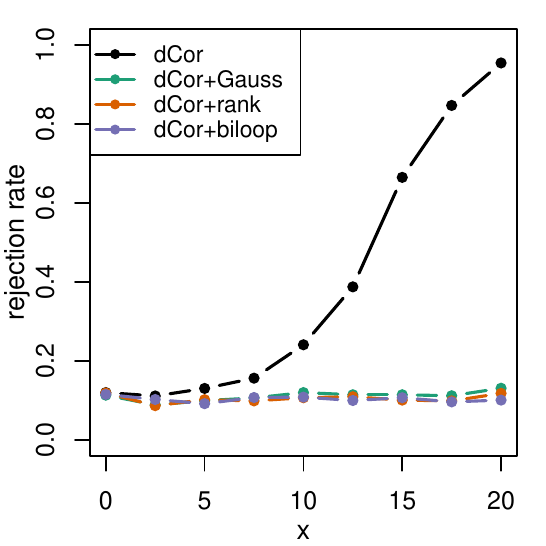}
  \caption{\centering $t_2$}
\end{subfigure}
\begin{subfigure}{0.245\textwidth}
  \includegraphics[width=\linewidth]
  {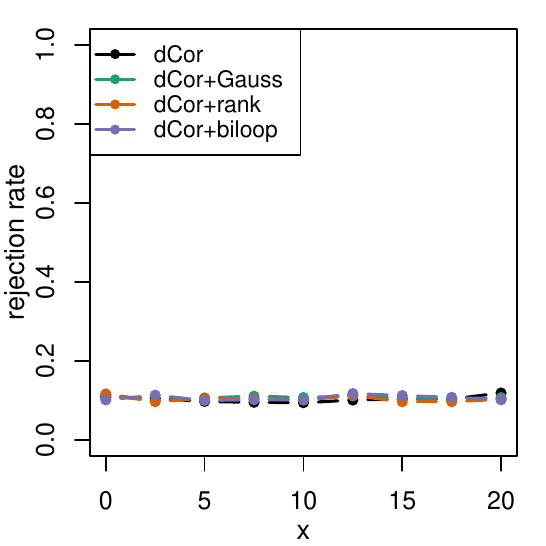}
  \caption{\centering $t_1$}
\end{subfigure}
\caption{\centering Rejection rates of the permutation tests when one outlier is added in $[x \; x]^{\intercal}$.}
\label{fig:type1error_2}
\end{figure}

First, we consider clean distributions plus a lone outlier $[x \; x]^{\intercal}$. The sample size is fixed at $n=200$. In Figure \ref{fig:type1error_2} the single outlier has a strong effect on the classical $\dCor$, whereas the other independence measures stay almost unaltered. The distortion is most pronounced for the lighter tailed normal distribution. The effect dies out as we move towards the $t_1$ distribution, as the size of the outlier becomes small relative to the clean observations in the tails, so its influence diminishes.

\begin{figure}[!ht]
\renewcommand*\thesubfigure{\arabic{subfigure}} 
    \centering 
\begin{subfigure}{0.245\textwidth}
  \includegraphics[width=\linewidth]{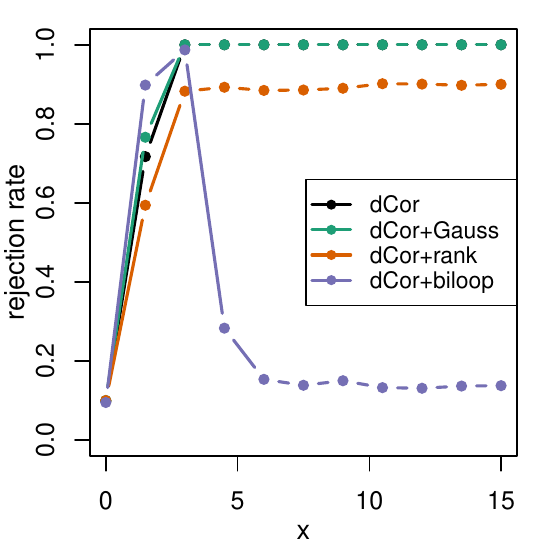}
  \caption{\centering normal}
\end{subfigure} 
\begin{subfigure}{0.245\textwidth}
  \includegraphics[width=\linewidth]
  {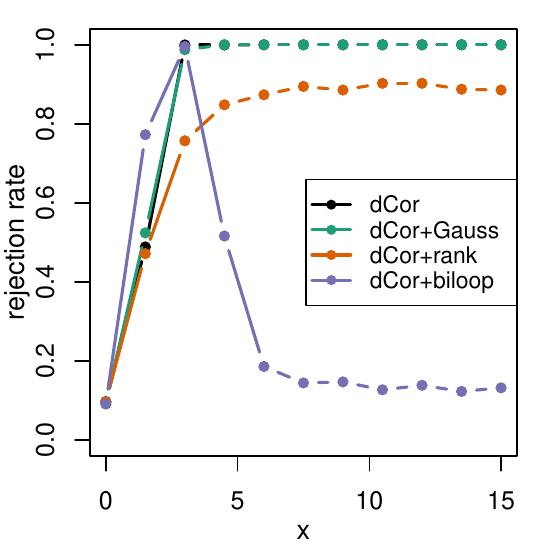}
  \caption{\centering $t_3$}
\end{subfigure}
\begin{subfigure}{0.245\textwidth}
  \includegraphics[width=\linewidth]
  {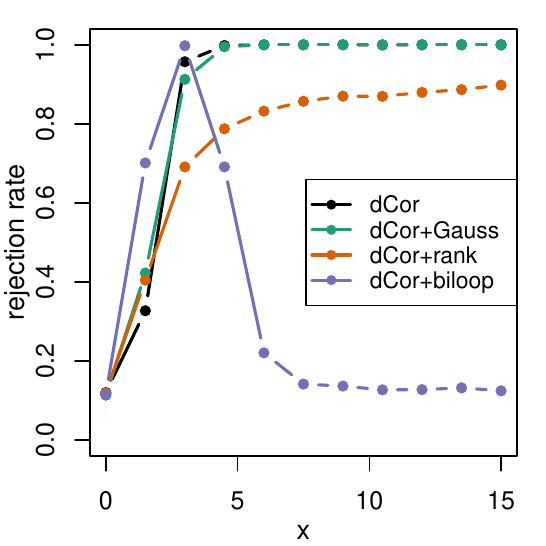}
  \caption{\centering $t_2$}
\end{subfigure}
\begin{subfigure}{0.245\textwidth}
  \includegraphics[width=\linewidth]
  {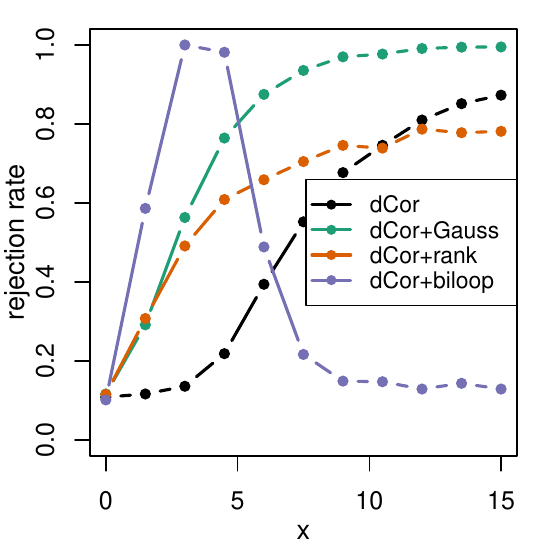}
  \caption{\centering $t_1$}
\end{subfigure}
\caption{\centering Rejection rates for $5\%$
of outliers in $[x \; x]^{\intercal}$.}
\label{fig:type1error_3}
\end{figure}

Figure \ref{fig:type1error_3} shows results for the same setup, but now with 5\% of contamination in $[x \; x]^{\intercal}$. This time the tests are affected more quickly when $x$ starts to grow. The effect is delayed for heavier-tailed distributions, for the same reason as before. The biloop $\dCor$ shows its unique redescending nature, as the rejection rate returns to approximately $10\%$ for very large $x$. Interestingly, for the $t_1$ distribution it is the classical $\dCor$ whose rejection rate grows the slowest when $x$ increases. This is again explained by the fact that, among the measures considered, the classical $\dCor$ assigns the most weight to the tails, and the tails of the $t_1$ distribution dominate the size of the outliers.

\begin{figure}[!ht]
\renewcommand*\thesubfigure{\arabic{subfigure}} 
    \centering 
\begin{subfigure}{0.245\textwidth}
  \includegraphics[width=\linewidth]{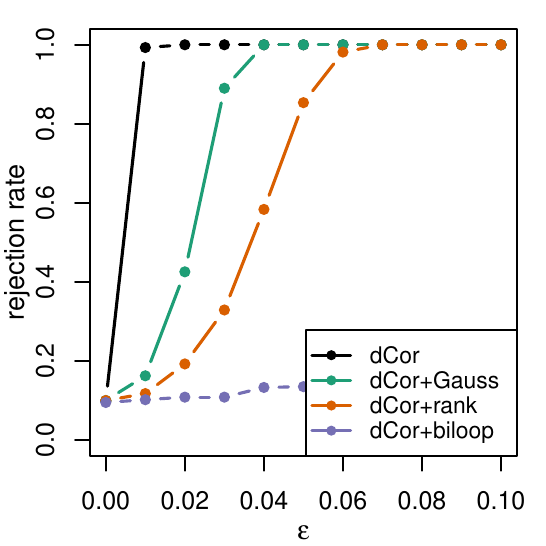}
  \caption{\centering normal}
\end{subfigure} 
\begin{subfigure}{0.245\textwidth}
  \includegraphics[width=\linewidth]{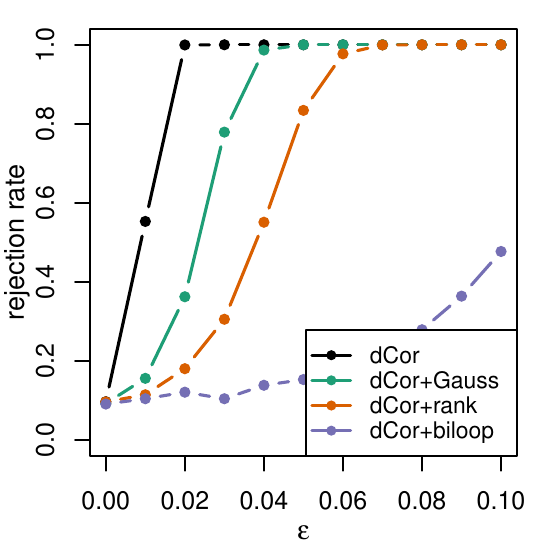}
  \caption{\centering $t_3$}
\end{subfigure}
\begin{subfigure}{0.245\textwidth}
  \includegraphics[width=\linewidth]{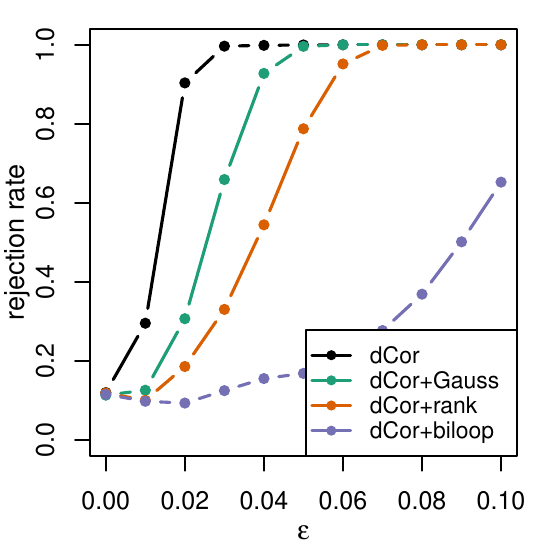}
  \caption{\centering $t_2$}
\end{subfigure}
\begin{subfigure}{0.245\textwidth}
  \includegraphics[width=\linewidth]{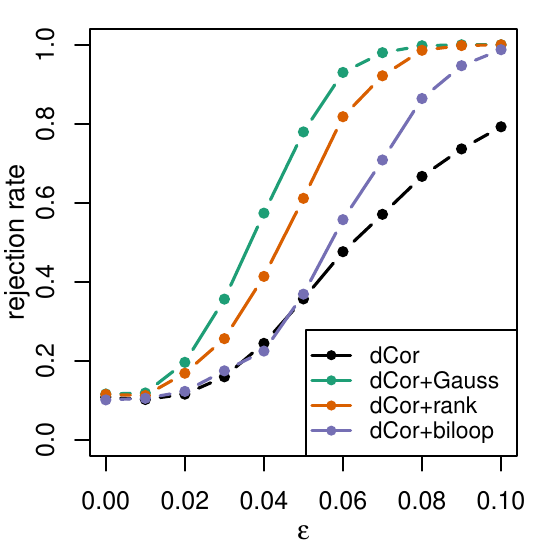}
  \caption{\centering $t_1$}
\end{subfigure}
\caption{\centering Rejection rates for 
  $100\varepsilon\%$ of outliers.}
\label{fig:type1error_4}
\end{figure}

Instead of focusing on the effect of increasing outlyingness, we now focus on the effect of increasing contamination level.
We consider the same normal, $t_3$, $t_2$, and $t_1$ distributions for the clean data, with $n=200$. We now contaminate by a fraction $\varepsilon$ of outliers sampled from $\mathcal{N}_2([6 \;6 ]^{\intercal},0.25\,\bm{I}_2)$.
Figure \ref{fig:type1error_4} shows the rejection rates of the different permutation tests as a function of $\varepsilon$. The curves clearly indicate that the measures assign a different importance to the tails. The original $\dCor$ is the most sensitive to them (except at the long-tailed $t_1$ distribution), followed by the normal scores and the rank transform. The biloop $\dCor$ is the least sensitive to the tails.

\section{Real data example} \label{sec:appl}

As an illustration we consider a data set originating
from \cite{Golub1999}, who aimed to distinguish between 
two types of acute leukemia on the basis of microarray
evidence. The input data is a matrix $\bX$ whose $n=38$
rows correspond to the patients, and with $p=7,129$ 
columns corresponding to the genes. 
The binary response vector $Y$ takes on the value 0 
for the 27 patients with leukemia type ALL and 1 for
the 11 patients with type AML. The data was previously 
analyzed by \cite{Hall2009}. Here our purpose is to
study dependence of general type between the 
variables (genes) $X^j$ and the response $Y$.

\begin{figure}[!ht]
\centering
\includegraphics[width=.48\textwidth]
   {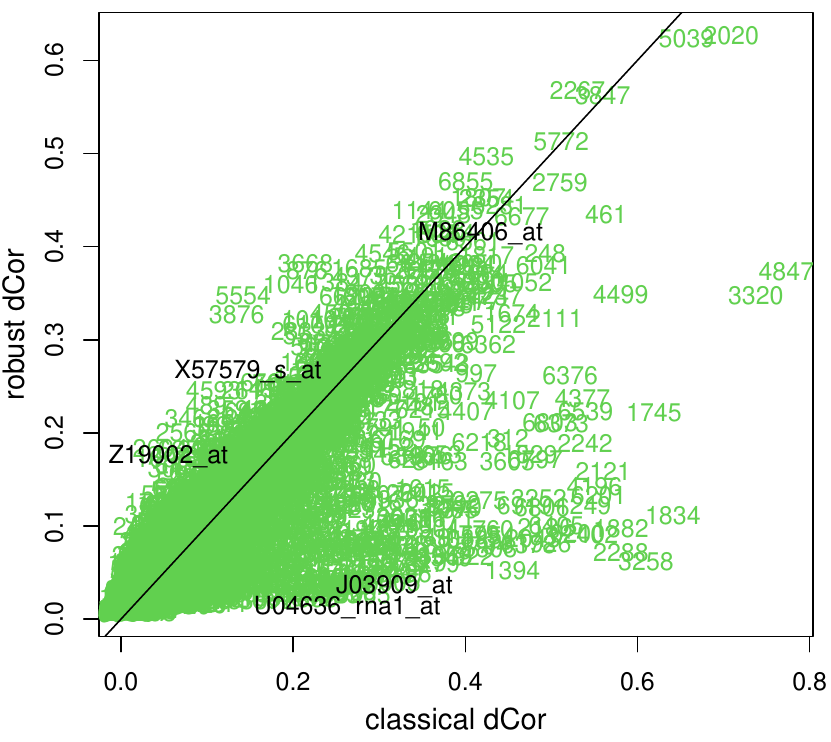} 
\caption{\centering Leukemia data: plot of robust 
   dCor versus classical dCor for all genes.}
\label{fig:CCplot}
\end{figure}

Figure~\ref{fig:CCplot} plots the 
biloop $\dCor$ versus the classical $\dCor$.
The light colored numbers inside the plot are the
values of $j$, the number of the gene. Most of the
points are concentrated around the main diagonal, so 
for them both versions of $\dCor$ are close together.

\begin{figure}[!ht]
\centering
\vspace{2mm}
\includegraphics[width=0.99\textwidth]
   {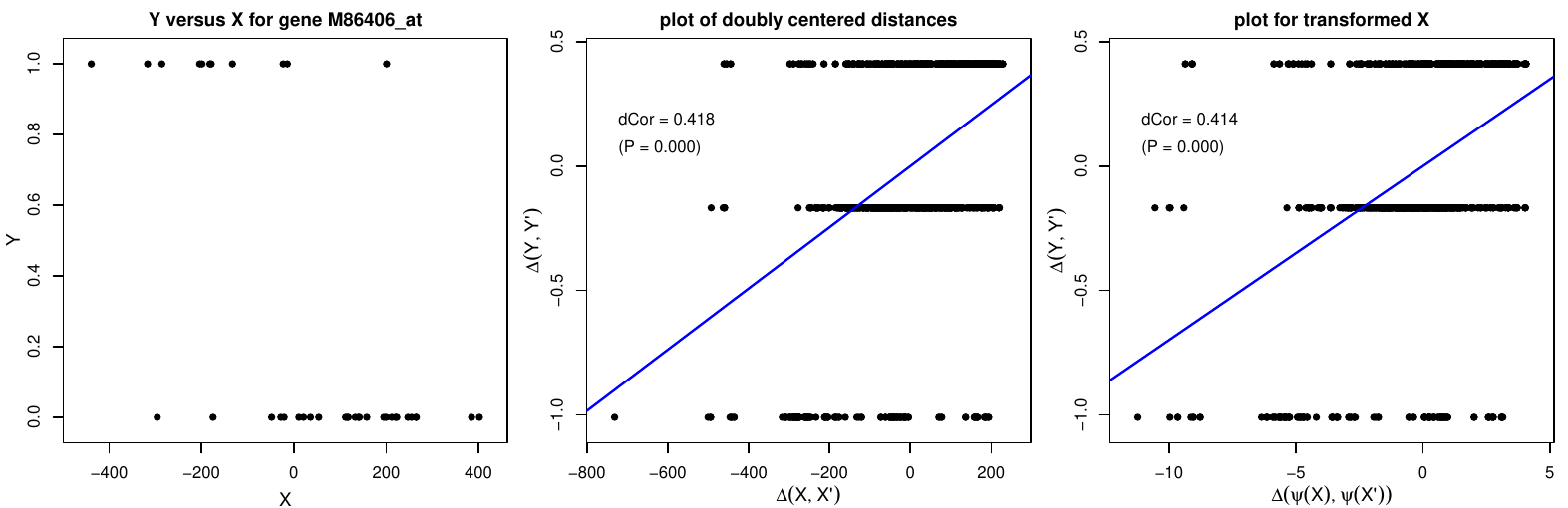}
\caption{\centering Gene \texttt{M86406\_at} 
  of the leukemia data, with $j=2301$. 
  Left: plot of $Y$ versus $X^j$. Middle: 
  doubly centered distances $\Delta(Y,Y')$ 
  of $Y$ versus those of $X^j$\,. Right: 
  versus those of $\psi_{\biloop}(X^j)$.}
  \label{fig:2301}
\end{figure}

As an example we look at gene \texttt{M86406\_at} 
corresponding to $j=2301$, which has high values 
for both as seen in Figure~\ref{fig:CCplot}. The 
leftmost panel of Figure \ref{fig:2301} is simply
a plot of $Y$ versus $X^j$. This indicates a kind
of decreasing relation, as would also be detected 
by a rank correlation. The middle 
panel is a plot of the doubly centered (DC)
distances $\Delta(Y,Y')$ of $Y$ versus the 
DC distances $\Delta(X^j,(X^j)')$ of $X^j$. 
This plot contains $38^2/2 = 722$ points instead 
of $38$. We note that the DC distances of $Y$ take 
only three values. This is because $Y$ itself is 
binary. It can easily be verified that the pairs 
$(y_j,y_k)$ with $y_j = 1 = y_k$ yield the lowest
of the three values, and the pairs with 
$y_j=0=y_k$ yield the middle value. The pairs 
for which $y_j \neq y_k$ obtain the top value. 
If the response $Y$ had an equal number of zeroes 
and ones, the pairs $(0,0)$ and $(1,1)$ would all
obtain the same value $-0.5$, and the pairs 
with differing response would obtain $0.5$.

\vspace{2mm}
The distance correlation of 0.418 equals
the Pearson correlation between the DC distances
of $X^j$ and those of $Y$. It is highly significant,
with $p$-value computed as 0.000 by the permutation 
test using 10,000 permutations.
The panel also contains the least squares regression 
line computed from these points, which is also
related to the dCor, and visualizes the strength of 
the relation. Note that the DC distances of $Y$ 
and $X^j$ do not follow 
a linear model, but that is not necessary since the
Pearson correlation arises here for a different
mathematical reason.

The right hand panel of Figure \ref{fig:2301}
plots the DC distances of $Y$ versus those of the
transformed variable $\psi_{\biloop}(X^j)$. 
This yields the
robustified version of dCor, given as 0.414 which
is again highly significant. The pattern in this
plot is quite similar to that in the previous plot,
since the variable $X^j$ did not contain outliers.
Note that the scale of the DC distances 
$\Delta(\psi_{\biloop}(X^j),(\psi_{\biloop}(X^j))')$ 
on the horizontal axis is totally different from 
that of the original $\Delta(X^j,(X^j)')$, because 
the transformation $\psi_{\biloop}$ contains a 
standardization. This does not matter, because the 
Pearson correlation is invariant to rescaling, so 
dCor is too. Also note that we did not transform 
$Y$ in the same way. That is because 
$\psi_{\biloop}(Y)$ would take only two values,
so its DC distances would simply be a rescaled 
version of those of $Y$ itself, and therefore yield
exactly the same dCor. 

\begin{figure}[!ht]
\centering
\includegraphics[width=0.80\textwidth]
   {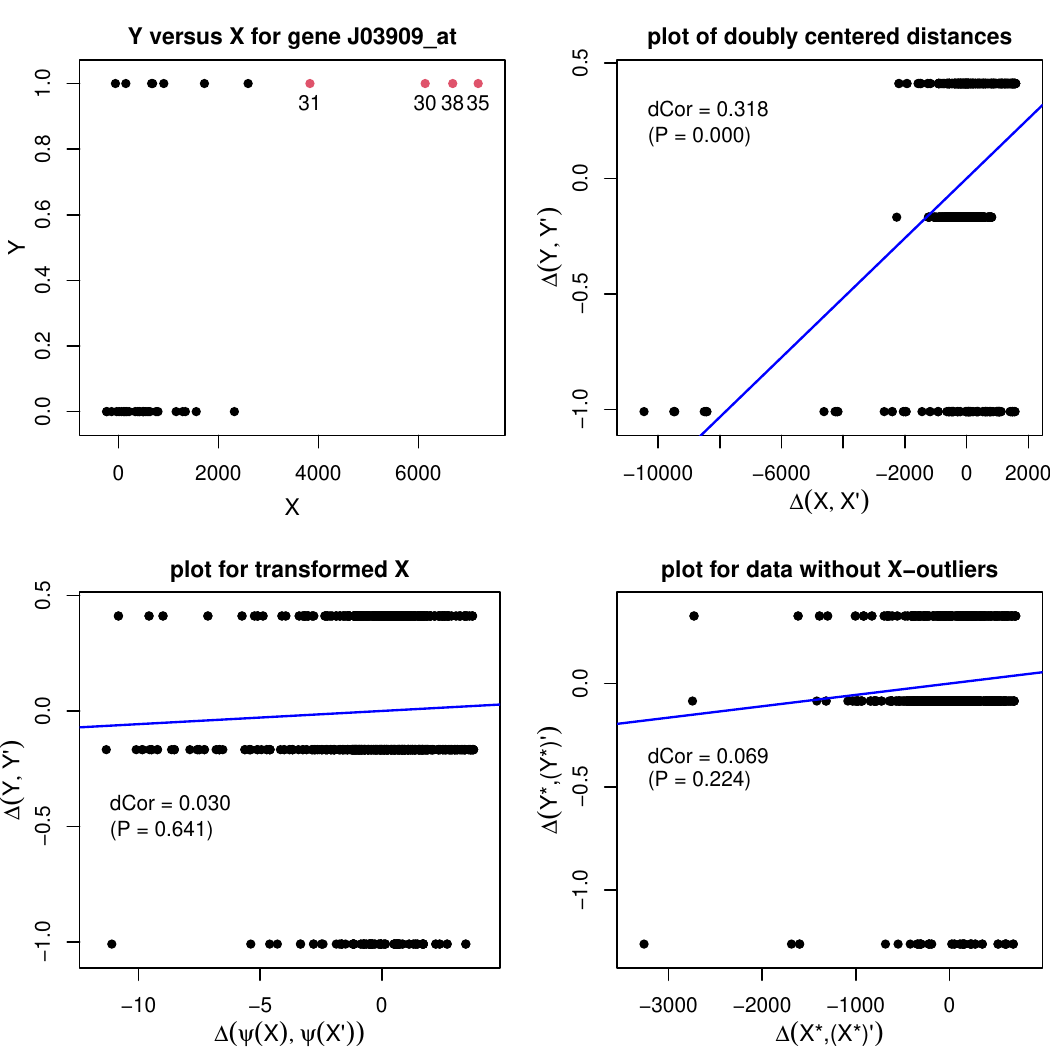}
\caption{\centering Gene \texttt{J03909\_at} of 
  the leukemia data ($j = 1092$).
  Top left: plot of $Y$ versus $X^j$, with the 
  four outliers. Top right: doubly 
  centered distances $\Delta(Y,Y')$ of $Y$ versus 
  those of $X^j$. Bottom left: $\Delta(Y,Y')$ 
  versus doubly centered distances of 
  $\psi_{\biloop}(X^j)$. Bottom right: plot 
  of doubly centered distances
  of the data $((X^j)^*,Y^*)$ without the  
  outliers.} \label{fig:1092}
\end{figure}

We are also interested in genes for which the 
classical and the robustified dCor are quite
different. We first look at an example where
the robustified dCor is much lower than the
classical one. For this purpose we restricted
attention to the genes with robust dCor
below 0.05, and among those we looked for the
largest value of classical minus robust.
This occurs for $j=1092$, which is the gene
\texttt{J03909\_at} indicated in 
Figure~\ref{fig:CCplot}.
The top left panel of Figure~\ref{fig:1092}
plots $Y$ versus $X^j$. We immediately note 
that there are several points with outlying
values of $X^j$ on the right. Their $x_{ij}$
lie far away relative to the scale of the
remaining $x_{ik}$. These patients all belong 
to the same class with $y_i = 1$.

\vspace{2mm}
The top right panel of Figure~\ref{fig:1092} 
plots the DC distances of $Y$ versus those of
$X^j$. We see that $\Delta(X^j,(X^j)')$ has
far outliers on the left hand side, which are 
due to the outliers in $X^j$. Those outliers
were on the right hand side, but that direction 
does not matter as a mirror image of $X^j$ 
yields the same DC distances since 
$\Delta(-X^j,(-X^j)')=\Delta(X^j,(X^j)')$.
The distance correlation of 0.318 is highly 
significant with $p$-value around 0.000.
The regression line is indeed quite steep,
its slope being determined mainly by the 
leverage of the outliers on the left.

In the bottom left panel we see the 
corresponding plot with the transformed variable 
$\psi_{\biloop}(X)$. The\linebreak 
$\Delta(\psi_{\biloop}(X^j),(\psi_{\biloop}(X^j))')$ 
still have a longer tail on the left than on the 
right, but there are no extreme outliers as in 
the previous panel. Now dCor is very low, with 
an insignificant p-value.

Finally, the bottom right panel shows what happens 
if we remove the four points with outlying 
$x$-values, yielding a reduced dataset denoted 
as $(X^*,Y^*)$ in the plot. 
The dCor is again quite low and 
insignificant. Indeed, if we remove the labeled
points in the top left panel there appears to be 
little structure left. Applying the robust dCor 
or taking out the outliers gives similar results 
here. This is an example where most of the information 
about the dependence between $X^j$ and $Y$ is
in the tails of the data, so it is reflected in the
classical dCor and not in its more robust version.
The difference between the classical and the robust 
dCor thus points our attention to the fact that the
conclusion rests heavily on these four cases.
Afterward it is up to the user to find out whether 
these x-values are correct or may be due to errors. 
In this dataset, measured by sophisticated 
equipment, they may well be correct.

The gene with the second largest difference 
classical - robust has $j = 5376$. 
It is analyzed in Section~F of 
the Supplementary Material.

\begin{figure}[!ht]
\centering
\includegraphics[width=0.80\textwidth]
   {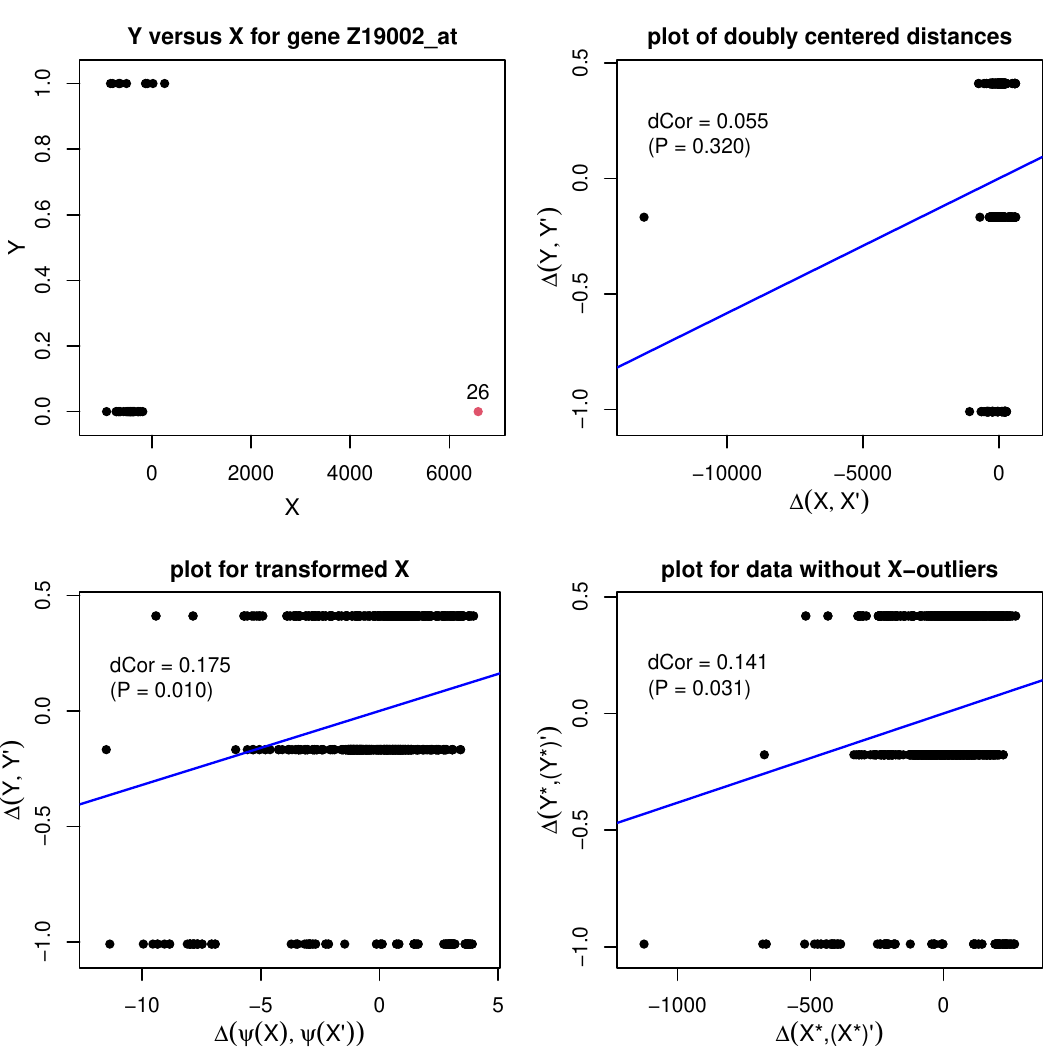}
\caption{\centering The panels are as in 
  Figure~\ref{fig:1092}, but for gene 
  \texttt{Z19002\_at} ($j=5071$).}
\label{fig:5071}
\end{figure}

Finally, we also want to look at an example where 
the robust dCor is higher than the classical one.
We looked at some genes where the difference
robust - classical was high, and for illustration
purposes we selected a simple one with only a 
single outlier. Figure~\ref{fig:5071} shows the
analysis for gene \texttt{Z19002\_at} with $j=5071$.

In the top left panel of Figure~\ref{fig:5071}
we see that patient 26 has a very outlying
value of $X^j$. This outlier yields a single
far outlier in the top right panel, because the
other DC distances involving patient 26 turn
out to be not nearly as far away. It may seem
strange that there is only one DC distance
from case 26 standing out, but this agrees
nicely with the proof of the breakdown value
in Section~C.1 of the Supplementary Material,
indicating that the DC distance of the 
outlier to itself
(denoted there as $\Delta_{11}$) 
dominates all others. The low distance
correlation of 0.055 is not significant at all.
But in the bottom left plot we see that the
use of the transformation $\psi_{\biloop}$ has 
brought the outlier much closer, thereby greatly 
reducing its effect. The robust dCor increases
to 0.175 and becomes significant. Also, taking
out the outlier entirely yields the bottom 
right plot which has a similar pattern.
So removing the single outlier brings the
classical dCor closer to the robust one.
(If one applies the robust dCor to the reduced
data set, it barely changes.)

\section{Outlook} \label{sec:outlook}

The distance correlation is increasingly being adopted 
by different scientific disciplines. 
\cite{martinez2014distance} identified new nonlinear 
dependencies between astrophysical variables by  
dCor, and found it worked better than the maximal 
correlation coefficient. 
In biology, \cite{lin2022linear} computed dCor 
between microbe taxa, using sparse and compositional data.
Among medical applications,  \cite{kong2012using} 
computed dCor between four variables: 
lifestyle, pedigree, disease, and mortality, some of which
were multivariate so they can be seen as groups of 
variables. \cite{geerligs2016functional} identified 
functional connectivity between regions of the human brain
by applying $\dCor$ to higher-dimensional variables.
In genomics, \cite{guo2014inferring} searched for nonlinear 
relationships in gene expression data in order to detect 
gene regulation, and found that $\dCor$ was better suited 
for this purpose than mutual information.
\cite{broadaway2016statistical} used dCov to detect genetic 
variations (genotypes) that affect several traits 
(phenotypes) simultaneously.

While distance correlation has become a popular tool in
statistical practice, its behavior in the presence of 
outliers has received relatively little attention.
One of the reasons may be that it is not obvious how to 
measure dependence robustly, which is the issue
that we address here. Therefore, we hope that our work 
will lead to a better understanding of the behavior of
dCov and dCor under contamination, and that the version
using the biloop transform will help to detect 
additional dependencies, as well as point out some
dependencies are concentrated in the tails.

Our work may also lead to new methodological development. 
The distance correlation, and independence measures in 
general, have seen a surge in research attention over 
the last two decades. 
For instance, various connections were found between
distance covariance and Brownian motion 
\citep{szekelyrizzo2009}, as well as with the 
Hilbert-Schmidt Independence Criterion (HSIC) of
\cite{gretton2005kernel} \citep{sejdinovic2013}
and with so-called global tests \citep{Edelmann2022}. 
Much of the current interest in measures of general
dependence is driven by novel and impactful 
methodological advances which rely on them. We present 
a brief overview of these methodologies here.

Feature selection is a statistical challenge for
which general dependence measures have been leveraged 
successfully. With the increasing popularity of 
nonlinear models in both statistics and machine learning, 
selecting features based on linear dependence measures 
is insufficient. Capturing general dependence allows for 
feature selection which caters to the use of 
more complex models in a second step. 
\cite{li2012feature} used dCor for this, as did we in 
our real data example, whereas \cite{song2012feature} 
applied the HSIC. \cite{kong2015using} applied dCov 
to feature selection with a high number of regressors 
by means of a sure independence screening approach, 
with applications to genetic data. 
\cite{kong2017interaction} used dCor to search for 
interaction terms. \cite{Vepakomma2018} employed dCor  
to construct a small set of latent regressors.

Independent component analysis (ICA) aims to construct 
a set of independent sources that together explain the
observed nongaussian multivariate dataset. It is
frequently used, e.g. in blind source separation.
\cite{matteson2017independent} developed a new
ICA algorithm based on dCov. We are currently working
on an extension based on the more robust variant
of dCov proposed here.

A key area where measuring independence is a 
fundamental building block is causal inference, and 
causal discovery in particular. In causal discovery, 
the goal is to identify a structural equation model 
and the underlying directed acyclic graph (DAG) from 
observational data. Under appropriate assumptions on 
the model classes and noise distributions, this can 
be done. However, inferring which DAGs can give rise 
to a given observational distribution can only be 
done by relying on general dependence measures.
For this purpose \cite{Zhang2011} constructed a
conditional independence test by means of an
extension of the HSIC. \cite{pfister2018kernel}
extended HSIC to test for joint independence of
variables. \cite{runge2019detecting} employed dCor,
and \cite{TSLINGAM} used both HSIC and dCor.

The machine learning community has adopted general 
dependence measures in several other methods. 
Measuring general dependence is often used to quantify 
a (dis)similarity between distributions or 
representations, without relying on particular 
distributional assumptions or linearity. An important
task is measuring the similarity of networks. 
Measuring similarities between 
different deep neural network (NN) representations is 
an important step in understanding the inner workings 
of these networks. \cite{kornblith2019similarity} 
introduce independence-based measures of NN similarity 
called centered kernel alignment (CKA), which reveal 
connections between network width and the sensitivity 
to training initialization. 

In transfer component 
analysis \citep{pan2010domain}, the goal is to use 
data in a source domain to build models that perform 
well in a different but related target domain. 
Typically, the strategy is to find a representation of 
the data that is similar in both domains. As in CKA, 
no distributional or linearity assumptions are made, 
which is why general independence measures are 
preferred to quantify similarity 
\citep{xiao2014feature,yan2017learning}. Other machine 
learning applications include clustering
\citep{niu2010multiple,cao2015diversity,WANG201950}, 
convolutional neural networks \citep{wang2020gcn}, 
and learning de-biased representations 
\citep{bahng2020learning}.

\section{Conclusions} 
\label{sec:concl}

The distance covariance $\dCov$ of 
\citet{szekely2007} is a popular measure of 
dependence between real random variables,
because it addresses all forms of dependence
rather than only linear or monotone
relations. Its simple definition is an
added benefit. Various claims have been made 
in the literature about its robustness to 
outliers, but this aspect was not yet studied
in detail.

In this paper we have investigated the 
robustness properties of $\dCov$, as well as
the distance variance and the distance 
correlation, by deriving 
influence functions and breakdown values.
This led to the surprising result that the 
influence function of the usual distance 
covariance is bounded, but that its breakdown
value is zero. The unbounded sensitivity function
converges to the bounded influence function for
increasing sample size. The robustness of 
$\dCor$ is thus not quite as high as expected. 
This led us to the construction of a more robust 
version based on the biloop, a novel data 
transformation.
The new version held its own in simulations
and was illustrated on a genetic dataset,
where comparing the classical $\dCor$ with its 
more robust version provided additional 
insight in the data.
In the outlook section we noted that 
there are many application areas of $\dCov$ 
and $\dCor$ in which the robust version could 
be useful.\\

\noindent{\bf Acknowledgment.} Thanks go to the
Editor and reviewers for helpful comments that
improved the presentation.\\

\noindent{\bf Software availability.} The dataset 
of the example and an R script reproducing the 
results are available at\linebreak 
\url{https://wis.kuleuven.be/statdatascience/robust/software}\,.\\

\clearpage

\pagenumbering{arabic}
\appendix

\setcounter{figure}{21} 
   
\begin{center}
\phantom{abc}\\
\vspace{10mm}

\LARGE{\bf Supplementary Material}\\
\end{center}
\vspace{3mm}

\section{Proofs of the influence functions of 
Section 2}\label{suppmat:IF}

\subsection{Proofs of Section 2.1}

\begin{proof}[\bf Proof of Proposition 1]
For the sake of simplifying notation, we exclude the 
exponent $\alpha$ since the proof remains unchanged.
We proceed as follows. Let $\left(X_{\varepsilon}, Y_{\varepsilon}\right) \sim F_{X, Y}^{\varepsilon}=(1-\varepsilon) F_{X, Y}+\varepsilon \Delta_{(s, t)}$. Then
$$
\dCov (X_{\varepsilon}, Y_{\varepsilon})=\underbrace{\mathbb{E}\left[\left|X_{\varepsilon}-X_{\varepsilon}^{\prime}\right| \;\left|Y_{\varepsilon}-Y_{\varepsilon}{ }^{\prime}\right|\right]}_{(1)}+\underbrace{\mathbb{E}\left[\left|X_{\varepsilon}-X_{\varepsilon}^{\prime}\right|\right] \; \mathbb{E}\left[\left|Y_{\varepsilon}-Y_{\varepsilon}^{\prime}\right|\right]}_{\text {(2) }}-2 \underbrace{\mathbb{E}\left[\left|X_{\varepsilon}-X_{\varepsilon}^{\prime}\right| \;\left|Y_{\varepsilon}-Y_{\varepsilon}^{\prime\prime}\right|\right]}_{(3)}.
$$
\noindent For the first term, we obtain:
\begin{align*}
\mathbb{E}[|X_{\varepsilon}-X_{\varepsilon}^{\prime}| \;|Y_{\varepsilon}-Y_{\varepsilon}^{\prime}|]=&\int|x-x^{\prime}||y-y^{\prime}| \underbrace{f_{X Y}^{\varepsilon}(x, y) f_{X Y}^{\varepsilon}(x^{\prime}, y^{\prime})}_{\substack{=(1-\varepsilon)^2 f_{X Y}(x, y) f_{X Y}(x^{\prime}, y^{\prime})+\varepsilon^2 \Delta_{s, t}\Delta_{s, t}  \\ +\varepsilon(1-\varepsilon) f_{X Y}(x, y) \Delta_{s, t}+(1-\varepsilon) \varepsilon \Delta_{s,t} f_{X Y}(x^{\prime}, y^{\prime})}} dx dx^{\prime} dy dy^{\prime} \\
=&(1-\varepsilon)^2 \mathbb{E}[|X-X^{\prime}||Y-Y^{\prime}|]+\varepsilon^2(|s-s||t-t|)\\
&+\varepsilon(1-\varepsilon) \mathbb{E}[|X-s||Y-t|]+(1-\varepsilon) \varepsilon \mathbb{E}[|s-X^{\prime}||t-Y^{\prime}|] \\
 =&(1-\varepsilon)^2 \mathbb{E}[|X-X^{\prime}||Y-Y^{\prime}|]+2 \varepsilon(1-\varepsilon) \mathbb{E}[|X-s||Y-t|] \;.
\end{align*}

\noindent The second term yields:
\begin{align*}
    \mathbb{E}[|X_{\varepsilon}-X_{\varepsilon}^{\prime}|] \; \mathbb{E}[|Y_{\varepsilon}-Y_{\varepsilon}^{\prime}|]=&\left((1-\varepsilon)^2 \mathbb{E}[|X-X^{\prime}|]+2 \varepsilon(1-\varepsilon) \mathbb{E}[|X-s| ]\right)\\
 &\left((1-\varepsilon)^2 \mathbb{E}[|Y-Y^{\prime}|]+2 \varepsilon(1-\varepsilon) \mathbb{E}[|Y-t|] \right) ,
\end{align*}
\noindent as a consequence of
\begin{align*}
 \mathbb{E}[|X_{\varepsilon}-X_{\varepsilon}^{\prime}|] &= \int|x-x^{\prime}| f_X^{\varepsilon}(x) f_X^{\varepsilon}(x^{\prime}) dx dx^{\prime} \\
& =(1-\varepsilon)^2 \mathbb{E}[|X-X^{\prime}|]+\varepsilon^2(|s-s|)+(1-\varepsilon) \varepsilon \mathbb{E}[|X-s|]+\varepsilon(1-\varepsilon) \mathbb{E}[| s-X^{\prime}|] \\
& =(1-\varepsilon)^2 \mathbb{E}[|X-X^{\prime}|]+2 \varepsilon(1-\varepsilon) \mathbb{E}[|X-s|] \;.
\end{align*}

\noindent Finally, for the third term we obtain
\begin{align*}
&\mathbb{E}[|X_{\varepsilon}-X_{\varepsilon}^{\prime}| \; |Y_{\varepsilon}-Y_{\varepsilon}^{\prime \prime}|]\\
= &\int|x-x^{\prime}||y-y^{\prime \prime}| f_{XY}^{\varepsilon}(x,y) f_X^{\varepsilon}(x^{\prime}) f_Y^{\varepsilon}(y^{\prime \prime}) dx dx^{\prime} dy dy^{\prime \prime}\\
= &\; (1-\varepsilon)^3 \mathbb{E}[|X-X^{\prime}| \;|Y-Y^{\prime\prime}|]\\
&\;+ \varepsilon(1-\varepsilon)^2(\mathbb{E}[|s-X^{\prime}||t-Y^{\prime\prime}|]+\mathbb{E}[|X-s||Y-Y^{\prime\prime}|]+\mathbb{E}[|X-X^{\prime}||Y-t|]) \\
&\;+ \varepsilon^2(1-\varepsilon) \;(\mathbb{E}[|s-s||t-Y^{\prime\prime}|]+\mathbb{E}[|s-X^{\prime}||t-t|]+\mathbb{E}[|X-s||Y-t|])+\varepsilon^3(|s-s| |t-t|)\\
=&\; (1-\varepsilon)^3 \mathbb{E}[|X-X^{\prime}| \;|Y-Y^{\prime\prime}|]\\
&\;+ \varepsilon(1-\varepsilon)^2(\mathbb{E}[|s-X^{\prime}||t-Y^{\prime\prime}|]+\mathbb{E}[|X-s||Y-Y^{\prime\prime}|]+\mathbb{E}[|X-X^{\prime}||Y-t|]) \\
&\;+ \varepsilon^2(1-\varepsilon) \;\mathbb{E}[|X-s||Y-t|]\;.
\end{align*}
\noindent
Putting everything together, we obtain:
\begin{align*}
 \dCov (X_{\varepsilon},Y_{\varepsilon}) &= \mbox{(term 1) + (term 2) - 2*(term 3)}\\
 &= (1-\varepsilon)^2 \mathbb{E}[|X-X^{\prime}||Y-Y^{\prime}|]+(1-\varepsilon)^4 \mathbb{E}[|X-X^{\prime}|] \mathbb{E}[| Y-Y^{\prime} |]\\
 &\qquad -2(1-\varepsilon)^3 \mathbb{E}[|X-X^{\prime}||Y-Y^{\prime \prime}|] +2 \varepsilon(1-\varepsilon)^2 \mathbb{E}[|X-s| |Y-t|]\\
 &\qquad +2 \varepsilon(1-\varepsilon)^3 \mathbb{E}[|X-s|] \mathbb{E}[|Y-Y^{\prime}|]
  +2 \varepsilon(1-\varepsilon)^3 \mathbb{E}[|X-X^{\prime}|] \mathbb{E}[|Y-t|] \\ 
 &\qquad +2 \varepsilon(1-\varepsilon)^2(2 \varepsilon-1) \mathbb{E}[|X-s|] \mathbb{E}[|Y-t|] 
 -2 \varepsilon(1-\varepsilon)^2 \mathbb{E}[|X-s||Y-Y^{\prime}|]\\
 &\qquad -2 \varepsilon(1-\varepsilon)^2 \mathbb{E}[|X-X^{\prime}||Y-t|]\;.
\end{align*}
\noindent
This implies that
\begin{equation*}
    \IF((s,t),\dCov,F) = \lim_{\varepsilon \to 0} \frac{\dCov(X_{\varepsilon}, Y_{\varepsilon})-\dCov(X, Y)}{\varepsilon}=\lim_{\varepsilon \to 0} \frac{a \varepsilon^0+b \varepsilon^1+O(\varepsilon^2)}{\varepsilon},
\end{equation*}
\noindent
where $a = 0$ and
\begin{align*}
b= & -2 \mathbb{E}[|X-X^{\prime}||Y-Y^{\prime}|]-4 \mathbb{E}[|X-X^{\prime}|] \mathbb{E}[|Y-Y^{\prime}|]+6 \mathbb{E}[|X-X^{\prime}||Y-Y^{\prime \prime}|] \\
& +2 \mathbb{E}[|X-s||Y-t|]+2 \mathbb{E}[|X-s|] \mathbb{E}[|Y-Y^{\prime}|]+2 \mathbb{E}[|X-X^{\prime}|] \mathbb{E}[|Y-t|] \\
& -2 \mathbb{E}[|X-s|] \mathbb{E}[| Y-t |]-2 \mathbb{E}[|X-s||Y-Y^{\prime}|]-2 \mathbb{E}[|X-X^{\prime}||Y-t|]\;.
\end{align*}
\noindent
Hence we obtain that $\IF((s,t),\dCov,F) = b$. This can be further simplified as:
\begin{align*}
b = &\; {-2 \mathbb{E}[|X-X^{\prime}||Y-Y^{\prime}|]-4 \mathbb{E}[|X-X^{\prime}|] \mathbb{E}[|Y-Y^{\prime}|]+6 \mathbb{E}[|X-X^{\prime}||Y-Y^{\prime \prime}|]} \\
& +2 \mathbb{E}[|X-s||Y-t|]+2 \mathbb{E}[|X-s|] \mathbb{E}[|Y-Y^{\prime}|]+2 \mathbb{E}[|X-X^{\prime}|] \mathbb{E}[|Y-t|] \\
& -2 \mathbb{E}[|X-s|] \mathbb{E}[| Y-t |]-2 \mathbb{E}[|X-s||Y-Y^{\prime}|]-2 \mathbb{E}[|X-X^{\prime}||Y-t|]\\
= &\; {-2 \dCov(X,Y)-2 \mathbb{E}[|X-X^{\prime}|] \mathbb{E}[|Y-Y^{\prime}|]+2 \mathbb{E}[|X-X^{\prime}||Y-Y^{\prime \prime}|]}\\
& +2 \mathbb{E}[|X-s||Y-t|]+2 \mathbb{E}[|X-s|] \mathbb{E}[|Y-Y^{\prime}|]+2 \mathbb{E}[|X-X^{\prime}|] \mathbb{E}[|Y-t|] \\
& -2 \mathbb{E}[|X-s|] \mathbb{E}[| Y-t |]-2 \mathbb{E}[|X-s||Y-Y^{\prime}|]-2 \mathbb{E}[|X-X^{\prime}||Y-t|]\\
= &\; -2 \dCov(X,Y) - {2 (\mathbb{E}[|X-s|] - \mathbb{E}[|X-X^{\prime}|])(\mathbb{E}[|Y-t|] - \mathbb{E}[|Y-Y^{\prime\prime}|])}\\ &+ {2 \mathbb{E}[(|X-s|-|X-X^{\prime}|)(|Y-t|-|Y-Y^{\prime\prime}|)]}\\
= &\; -2 \dCov(X,Y) + 2 \cov(|X-s|-|X-X^{\prime}|,|Y-t| - |Y-Y^{\prime \prime}|)\;.
\end{align*}
\end{proof}
\begin{proof}[\bf Proof of Proposition 2]
For the (un)boundedness of the IF, we study the behavior of 
\begin{align*}
&\cov(|X-s|^{\alpha}-|X-X^{\prime}|^{\alpha},|Y-t|^{\alpha}-|Y-Y^{\prime\prime}|^{\alpha})\\
&\; = \cov(|X-s|^{\alpha},|Y-t|^{\alpha})
  - \cov(|X-s|^{\alpha}, |Y-Y^{\prime\prime}|^{\alpha})\\
&\qquad - \cov( |X-X^{\prime}|^{\alpha}, |Y-t|^{\alpha})
  + \cov( |X-X^{\prime}|^{\alpha}, 
  |Y-Y^{\prime\prime}|^{\alpha}) \;.
\end{align*}
Hence we are interested in the first 3 terms. We start with $\alpha \leq 1$ using Cauchy-Schwarz:
\begin{align*}
    |\cov(|X-s|^{\alpha},|Y-t|^{\alpha})| &\leq \sqrt{\var(|X-s|^{\alpha})\;\var(|Y-t|^{\alpha})}\\
     |\cov(|X-s|^{\alpha},|Y-Y^{\prime\prime}|^{\alpha})|  &\leq \sqrt{\var(|X-s|^{\alpha})\;\var(|Y-Y^{\prime\prime}|^{\alpha})}\\
      |\cov(|X-X^{\prime}|^{\alpha},|Y-t|^{\alpha})|  &\leq \sqrt{\var(|X-X^{\prime}|^{\alpha})\;\var(|Y-t|^{\alpha})}\;.
\end{align*}
Next we show that $\var(|X-s|^{\alpha})$ and $\var(|Y-t|^{\alpha})$ are bounded for $\alpha \leq 1$:
\begin{align*}
    2 \var[|X-s|^{\alpha}] &= \E[(|X-s|^{\alpha} -|X^{\prime}-s|^{\alpha})^2]\\
    &\leq \E[(|X-X^{\prime}|^{2\alpha})]\\
    &\leq 1 + \E[|X-X^{\prime}|^2]\\
    &= 1 + 2\var[X]\;.
\end{align*}
Here the first inequality holds because it holds for any $(x, x^{\prime})$ when $0\leq \alpha \leq 1$. As the last quantity does not depend on $s$, this yields an upper bound and hence the $\alpha$-distance covariance has a bounded IF for $\alpha \leq 1$. For $\alpha <1$ the variance (and covariances) of interest are even redescending, see the proof of Proposition 3.
\\
Second we discuss $\alpha > 1$. The unboundedness of $\cov(|X-s|^{\alpha},|Y-t|^{\alpha})$ depends on the dependence between $X$ and $Y$. For example, if $X$ and $Y$ are independent, then $|X-s|^{\alpha}$ and $|Y-t|^{\alpha}$ are also independent and the term equals zero, just as the other two terms. However, when $X=Y$ and $t=s$, we obtain the IF of $\alpha$-distance variance which we show to be unbounded for $ \alpha > 1$ in the proof of Proposition 3.
\end{proof}

\subsection{Proofs of Section 2.2}

\begin{proof}[\bf Proof of Corollary 1]
For the first statement we use
\begin{align*}
 \IF(s,\dVar(X;\alpha),F_X) 
 &= \IF((s,s),\dCov(X,X;\alpha),F_X)\\
 &= -2\dVar(X;\alpha)
   + 2 \eta(s, s, X,X, \alpha))\,.
\end{align*}
For the second statement we compute
\begin{align*}
    \IF(s,\dVar(X;\alpha),F_X) &= \frac{\partial}{\partial \varepsilon} \left( \dVar(X_{\varepsilon};\alpha) \right) \vert_{\varepsilon = 0}\\
    &=\frac{\partial}{\partial \varepsilon} \left( \dStd(X_{\varepsilon};\alpha)^2 \right) \vert_{\varepsilon = 0}\\
    &= 2 \dStd(X;\alpha) \; \frac{\partial}{\partial \varepsilon} \left( \dStd(X_{\varepsilon};\alpha) \right) \vert_{\varepsilon = 0}\\
    &= 2 \dStd(X;\alpha) \; \IF(s,\dStd(X;\alpha),F)\;.
\end{align*}  
\end{proof}
  
\begin{proof}[\bf Proof of Proposition 3]
To investigate the boundedness of the IF of the $\alpha$-distance variance, we study the behavior of $$\var[|X-s|^\alpha] -2 \cov[|X-s|^{\alpha}, |X-X^{\prime}|^\alpha].$$
For $\alpha \leq 1$, we have already proven the boundedness of these terms in the proof of Proposition 2. 

Next, we consider $\alpha > 1$. For the variance term, we assume that $X$ is not degenerate, 
so there exists an $\varepsilon > 0$ such that the probability 
$p \coloneqq \Pr(|X-X^{\prime}| \geq \varepsilon)$ is strictly positive. 
Also denote $A_s =\{(X-s) (X'-s) > 0\}$, i.e. the event that $(X-s)$ and $(X'-s)$ have the same sign. Clearly, $\lim_{s \to \infty} \Pr(A_s) = 1$. We now have 
\begin{align*}
    2 \var[|X-s|^{\alpha}] &= \E[(|X-s|^{\alpha} -|X^{\prime}-s|^{\alpha})^2]\\
    &\geq  \E[(|X-s|^{\alpha} -|X^{\prime}-s|^{\alpha})^2\; \big| \;|X-X^{\prime}| \geq \varepsilon] \Pr(|X-X^{\prime}| \geq \varepsilon)\\
    &= p\;  \E[(|X-s|^{\alpha} -|X^{\prime}-s|^{\alpha})^2\; \big| \;|X-X^{\prime}| \geq \varepsilon]\\
     &\geq p\;  \E[(|X-s|^{\alpha} -|X^{\prime}-s|^{\alpha})^2\; \big| \;|X-X^{\prime}| \geq \varepsilon \cap A_s] \Pr(A_s)\\     
    &\geq p \alpha^2 \;  \E[ \min\{|X-s|, |X^{\prime}-s|\}^{2(\alpha - 1)} |X-X^{\prime}|^2\; \big| \;|X-X^{\prime}| \geq \varepsilon  \cap A_s ]\Pr(A_s)\\
    &\geq p \alpha^{2}  \varepsilon^{2}\;  \E[\min\{|X-s|, |X^{\prime}-s|\}^{{2}(\alpha - 1)} \; \big| \;|X-X^{\prime}| \geq \varepsilon  \cap A_s]\Pr(A_s) \;,
\end{align*}
where we have used the convexity of $x\to x^{\alpha}$ for $\alpha > 1$ in the second inequality (i.e. the graph lies above its tangents). The last term blows up as $s \to \infty$, because ${2}(\alpha - 1)> 0$ and $\lim_{s \to \infty} \Pr(A_s) = 1$. We thus obtain that $\var[|X-s|^{\alpha}]$ is unbounded.

 This gives us
\begin{align*}
    &\var[|X-s|^\alpha] -2 \cov[|X-s|^{\alpha}, |X-X^{\prime}|^\alpha]\\
    \geq &\var[|X-s|^\alpha] -2 | \cov[|X-s|^{\alpha}, |X-X^{\prime}|^\alpha] | \\
    \geq &\var[|X-s|^\alpha] -2 \sqrt{ \var[|X-s|^{\alpha}] \var[|X-X^{\prime}|^\alpha] } \\
    = & \var[|X-s|^\alpha] -2 \; C \; \sqrt{ \var[|X-s|^{\alpha}]} \; ,
\end{align*}
by using Cauchy-Schwarz. This explodes because 
$\var[|X-s|^\alpha]$ explodes for $s \to \infty$
when $\alpha > 1$. Therefore 
$\var[|X-s|^\alpha] -2 \cov[|X-s|^{\alpha}, 
|X-X^{\prime}|^\alpha]$ is unbounded for $\alpha > 1$.\\

\noindent
Lastly, we consider $  2 \var[|X-s|^{\alpha}] = \E[(|X-s|^{\alpha} -|X^{\prime}-s|^{\alpha})^2]$ with $\alpha < 1$.
Note that, due to the concavity of $x\to x^{\alpha}$ for $\alpha< 1$, we have
$$||X-s|^{\alpha} -|X^{\prime}-s|^{\alpha}| \leq \alpha \min\{|s-X|, |s-X'|\}^{\alpha-1} |X-X'|.$$
Unlike in the previous case for $\alpha > 1$, we do not need to consider $A_s$ here, since the above is true even if $(s-X)$ and $(s-X')$ have a different sign.  
\begin{align*}
    2 \var[|X-s|^{\alpha}] &= \E[(|X-s|^{\alpha} -|X^{\prime}-s|^{\alpha})^2]\\
    &\leq  \alpha^2 \E[\min\{|s-X|, |s-X'|\}^{2(\alpha-1)} |X-X'|^2]
\end{align*}
given that $2(\alpha-1) < 0$ and $\E[|X-X'|^2] < \infty$, this converges to 0. Therefore $\var[|X-s|^{\alpha}]$ converges to 0 for $s \to \infty$ if $\alpha<1$.

\end{proof}

\newpage
\subsection{Proofs of Section 2.3}

\begin{proof}[\bf Proof of Corollary 2]
We omit $\alpha$ to ease the notational burden, 
but the computation is the same. 
We compute
\begin{align*}
\IF&((s,t),\dCor,F) = \frac{\partial}{\partial\varepsilon} \left( \frac{\dCov(X_{\varepsilon},Y_{\varepsilon})}{\sqrt{\dVar(X_{\varepsilon})\dVar(Y_{\varepsilon})}}\right) \Bigg\vert_{\varepsilon=0}\\
&=\Big[\IF(\dCov(X,Y))\sqrt{\dVar(X)\dVar(Y)}-\frac{1}{2\sqrt{\dVar(X)\dVar(Y)}} (\IF(\dVar(X))\dVar(Y)\\
&\qquad + \dVar(X)\IF(\dVar(Y)))\dCov(X,Y)\Big]/
(\dVar(X)\dVar(Y))\\
&= \frac{\IF(\dCov(X,Y))}{\dStd(X)\dStd(Y)} - \frac{\IF(\dVar(X))\dVar(Y)+\dVar(X)\IF(\dVar(Y))}{2(\dVar(X)\dVar(Y))}\;\frac{\dCov(X,Y)}{\sqrt{\dVar(X)\dVar(Y)}}\\
&= \frac{\IF(\dCov(X,Y))}{\dStd(X)\dStd(Y)} - \left(\frac{\IF(\dVar(X))}{2\dVar(X)} + \frac{\IF(\dVar(Y))}{2\dVar(Y)}\right)\; \dCor(X,Y)\\
&= \frac{-2 \dCov(X,Y) + 2 \cov(|X-s|-|X-X^{\prime}|,|Y-t| - |Y-Y^{\prime \prime}|)}{\dStd(X)\dStd(Y)} \\
&\qquad- \Bigg(\frac{-2 \dVar(X) + 2 \cov (|X-s| - |X-X^{\prime}|,|X-s|-|X-X^{\prime \prime}|)}{2\dVar(X)} \\
    &\qquad \qquad + \frac{-2 \dVar(Y) + 2 \cov (|Y-t| - |Y-Y^{\prime}|,|Y-t|-|Y-Y^{\prime \prime}|)}{2\dVar(Y)}\Bigg)\; \dCor(X,Y)\\
&= -2\dCor(X,Y) + \frac{2 \cov(|X-s|-|X-X^{\prime}|,|Y-t| - |Y-Y^{\prime \prime}|)}{\dStd(X)\dStd(Y)} \quad -\\
&\qquad \Bigg(-2 + \frac{ \cov (|X-s| - |X-X^{\prime}|,|X-s|-|X-X^{\prime \prime}|)}{\dVar(X)}\\
&\qquad + \frac{\cov (|Y-t| - |Y-Y^{\prime}|,|Y-t|-|Y-Y^{\prime \prime}|)}{\dVar(Y)}\Bigg) \; \dCor(X,Y)\\
&= \frac{2 \cov(|X-s|-|X-X^{\prime}|,|Y-t| - |Y-Y^{\prime \prime}|)}{\dStd(X)\dStd(Y)}\\ 
&\qquad - \Bigg(\frac{ \cov (|X-s| - |X-X^{\prime}|,|X-s|-|X-X^{\prime \prime}|)}{\dVar(X)}\\ 
&\qquad + \frac{\cov (|Y-t| - |Y-Y^{\prime}|,|Y-t|-|Y-Y^{\prime \prime}|)}{\dVar(Y)}\Bigg) \; \dCor(X,Y)\;.
\end{align*}
\end{proof}

\clearpage

\begin{proof}[\bf Proof of Proposition 4]
For the IF of the $\alpha$-distance correlation we had the following expression:
{\small
    \begin{equation*}
        \frac{2 \eta(s,t,X,Y,\alpha)}{\dStd(X;\alpha)\dStd(Y;\alpha)}
        \; - \Bigg(\frac{ \eta(s,s,X,X,\alpha)}{\dVar(X;\alpha)}
        \; + \frac{\eta(t,t,Y,Y,\alpha)}{\dVar(Y;\alpha)}\Bigg) 
        \dCor(X,Y;\alpha)\;.
    \end{equation*} }
    
\noindent As shown in its derivation, this can also be written as $$\frac{\IF(\dCov(X,Y;\alpha))}{\dStd(X;\alpha)\dStd(Y;\alpha)} - \left(\frac{\IF(\dVar(X;\alpha))}{2\dVar(X;\alpha)} + \frac{\IF(\dVar(Y;\alpha))}{2\dVar(Y;\alpha)}\right) \dCor(X,Y;\alpha)\,.$$ 
Hence we immediately have that the IF of $\dCor(X,Y;\alpha)$ is bounded (and redescending) for $\alpha \leq 1$ ($\alpha <1$) as it is a combination of bounded (and redescending) terms. For $\alpha > 1$, the (un)boundedness of the IF again depends on the distribution of $(X,Y)$.
\end{proof}

\newpage
\section{Distance standard deviation as a scale estimator}
\label{suppmat:dStd}

In this part of the Supplementary Material, we undertake a more detailed analysis of the distance standard deviation as a scale estimator. More precisely, we study its theoretical efficiency and perform a small simulation to assess its robustness. 

First, we calculate its asymptotic variance (ASV) to derive its efficiency (eff). For this, we study the consistent estimator $\sqrt{v_\alpha c}\, \dStd(X;\alpha)^{1/\alpha}$ at the normal model $\mathcal{N}(0,\sigma^2)$.
 \begin{align*}
     \IF(s,\sqrt{v_\alpha c}\,\dStd(X;\alpha)^{1/\alpha},\Phi) &= \frac{\partial}{\partial \varepsilon} \left( \sqrt{v_\alpha c}\,\dStd(X_{\varepsilon};\alpha)^{1/\alpha} \right) \vert_{\varepsilon = 0}\\
     &= \sqrt{v_\alpha c}\, \frac{1}{\alpha} \left(\dStd(X;\alpha)\right)^{\frac{1}{\alpha} - 1} \frac{\partial}{\partial \varepsilon} \left( \dStd(X_{\varepsilon};\alpha)\right) \vert_{\varepsilon = 0}\\
     &= \frac{1}{\alpha} \left(\dStd(X;\alpha)\right)^{-1} \IF(s,\dStd(X;\alpha),\Phi)\\
     &= \frac{\left(\dStd(X;\alpha)\right)^{-1}}{\alpha} \frac{\IF(s,\dVar(X;\alpha),\Phi)}{2 \, \dStd(X;\alpha)}\\
     &=  \frac{\IF(s,\dVar(X;\alpha),\Phi)}{2 \alpha \dVar(X;\alpha)}
\end{align*}
\begin{align*}
  &\implies \text{ASV}(\sqrt{v_\alpha c}\,\dStd(X;\alpha)^{1/\alpha},\Phi) = \mathbb{E}_{\Phi}[\IF(s,\sqrt{v_\alpha c}\,\dStd(X;\alpha)^{1/\alpha},\Phi)^2]\\ 
  &\implies \text{eff}(\sqrt{v_\alpha c}\,\dStd(X;\alpha)^{1/\alpha},\Phi) 
     = \frac{1}{2\;\text{ASV}(\sqrt{v_\alpha c}\,\dStd(X;\alpha)^{1/\alpha},\Phi)}
\end{align*} 
\vskip2mm
 
\noindent where $2$ is the Fisher information of $\sigma$ 
at the scale model $\mathcal{N}(0,\sigma^2)$. 
Calculating the integrals numerically in Mathematica 
we obtain the following efficiencies per $\alpha$:

\begin{center}
\vskip-4mm
\begin{tabular}{ |c||c|c|c|c|c|c|c|c|c|c| } 
 \hline
 $\alpha$ & 0.6 & 0.7 & 0.8 & 0.9 & 1 & 1.1 & 1.2 & 1.3 & 1.4\\ 
 \hline 
 efficiency & 0.5793 & 0.6380 & 0.6911 & 0.7396 & 0.7839 & 0.8244 & 0.8610 & 0.8936 & 0.9220\\
 \hline
\end{tabular}
\end{center}
\vskip4mm

Naturally, the efficiency increases towards 1 
when $\alpha \to 2$ because then the 
estimate converges to the classical standard 
deviation. 
To study the finite-sample efficiency of 
$\sqrt{v_\alpha c}\,\dStd(X;\alpha)$, we 
compare it to scale M-estimators with the same 
influence function per $\alpha$.
\begin{figure}[!ht]
\centering
\includegraphics[width=.49\textwidth]
{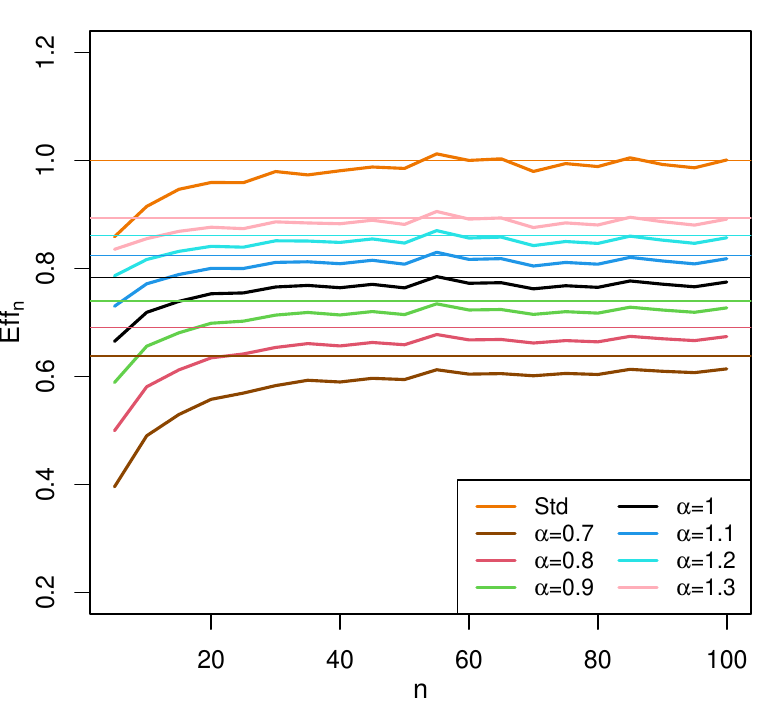}
\includegraphics[width=.49\textwidth]
{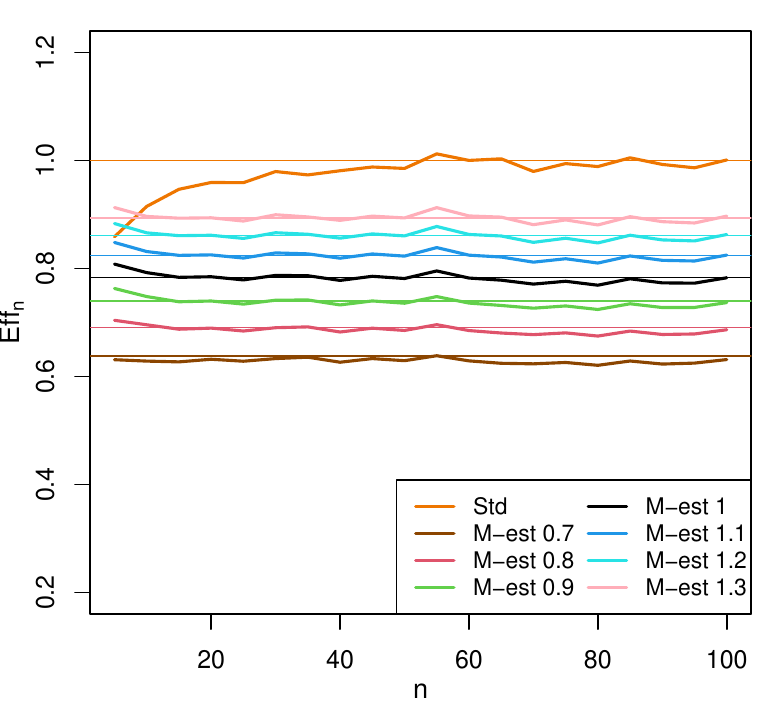}
\caption{\centering Finite-sample efficiencies for different sample sizes for: on the left the consistent $\alpha$-distance standard deviation for different $\alpha$, and on the right the scale M-estimators with the same influence functions as the estimators in the left figure.}
\label{fig:finsampeff}
\end{figure}

In Figure \ref{fig:finsampeff} the M-estimators 
converge faster to the theoretical 
efficiencies than the $\alpha$-distance standard 
deviation, but both estimators converge rather fast. 
The convergence is also faster for higher $\alpha$, 
when the estimator gets closer to the classical 
standard deviation.

\begin{figure}[!ht]
\renewcommand*\thesubfigure{\arabic{subfigure}} 
    \centering 
\begin{subfigure}{0.49\textwidth}
  \includegraphics[width=\linewidth]{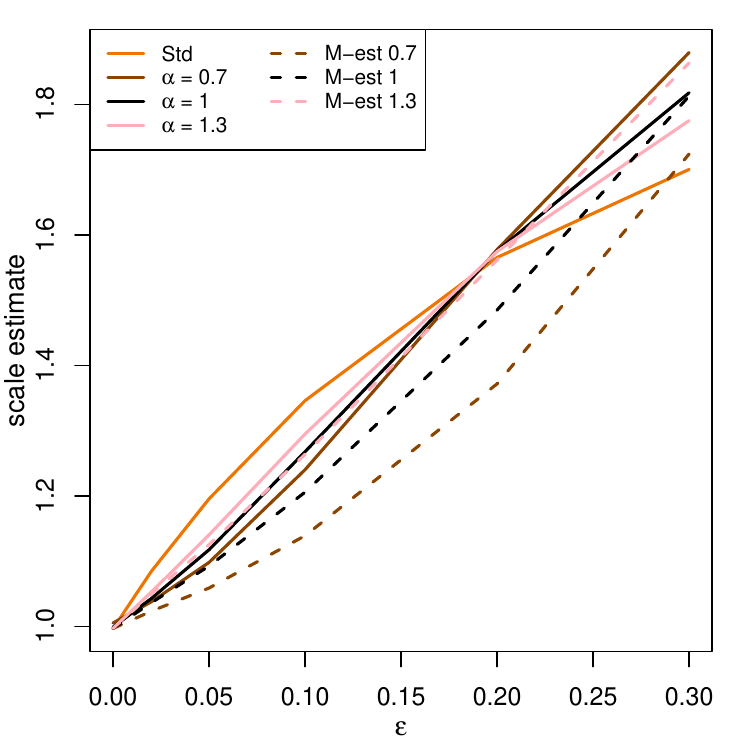}
  \caption{\centering contaminated by $\mathcal{N}(3,1)$.}
\end{subfigure} 
\begin{subfigure}{0.49\textwidth}
  \includegraphics[width=\linewidth]{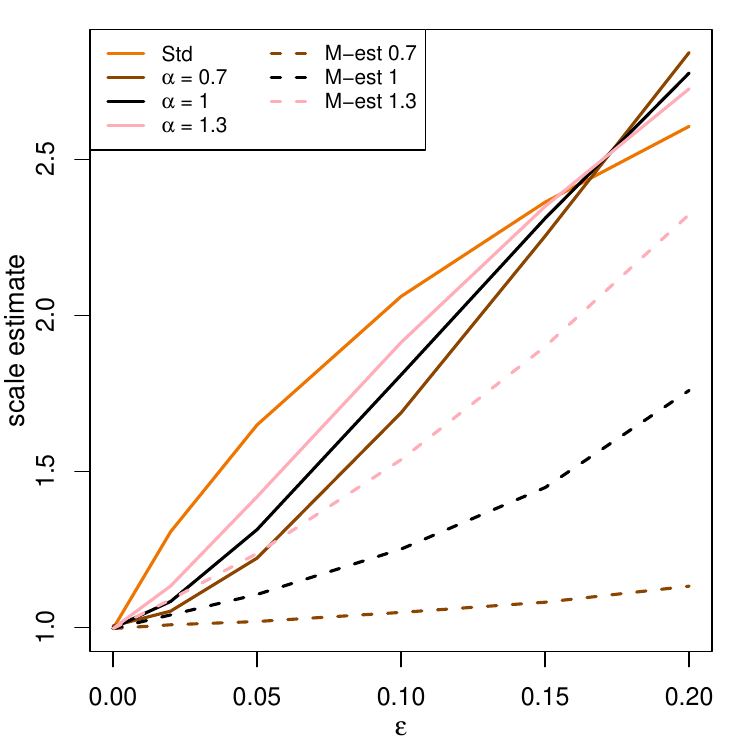}
  \caption{\centering contaminated by $\mathcal{N}(6,1)$.}
\end{subfigure}
\begin{subfigure}{0.49\textwidth}
  \includegraphics[width=\linewidth]{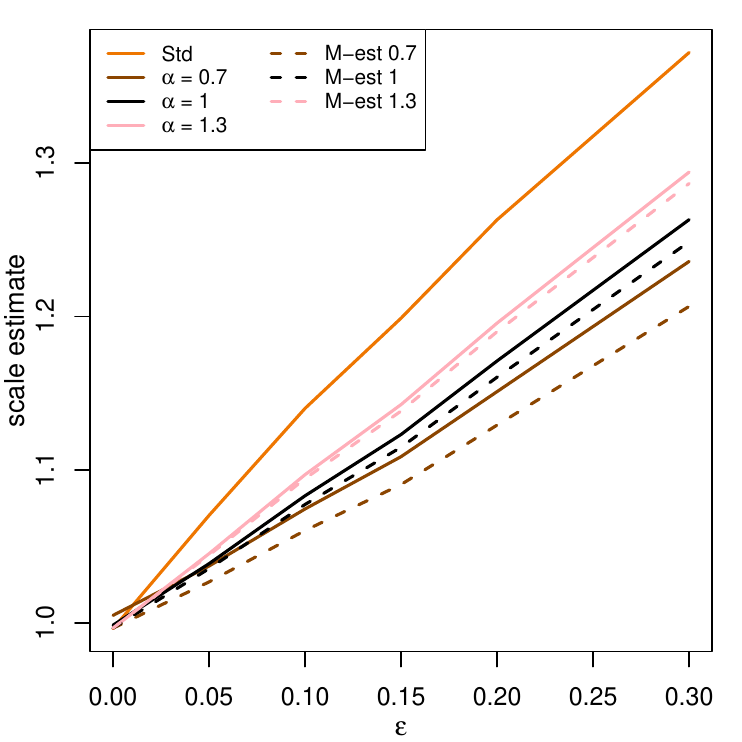}
  \caption{\centering contaminated by $\mathcal{N}(0,4)$.}
\end{subfigure}
\begin{subfigure}{0.49\textwidth}
  \includegraphics[width=\linewidth]{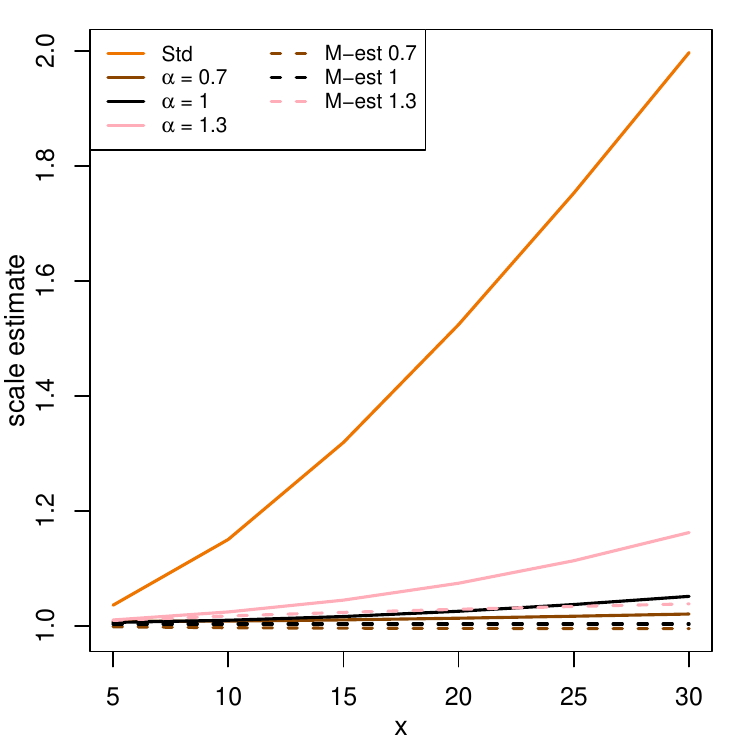}
  \caption{\centering contaminated by one outlier in $x$.}
\end{subfigure}
\caption{\centering Average value of the $\alpha$-distance 
standard deviation, the corresponding M-estimator, and the 
classical standard deviation, in the 4 settings.}
\label{fig:robustness_scaleest}
\end{figure}

Next, we compare the robustness of $\sqrt{v_\alpha c} \, \dStd(X;\alpha)^{1/\alpha}$ per $\alpha$ with the robustness of the corresponding M-estimators and the classical standard deviation. We do this for the following 4 standard normal settings containing $100\varepsilon\%$ contamination:
\begin{enumerate}
    \item $(1-\varepsilon)\; \mathcal{N}(0,1) + \varepsilon\; \mathcal{N}(3,1)$
    \item $(1-\varepsilon)\; \mathcal{N}(0,1) + \varepsilon\; \mathcal{N}(6,1)$
    \item $(1-\varepsilon)\; \mathcal{N}(0,1) + \varepsilon\; \mathcal{N}(0,4)$
    \item $\mathcal{N}(0,1)$ with one outlier added in $x$\,.
\end{enumerate}
In each setting we draw 1000 samples of size 
$n=300$ and compute the scale estimators. 
Their average values are shown in 
Figure~\ref{fig:robustness_scaleest}.

In these graphs the contamination affects every 
scale estimator. Among them, the M-estimator with 
$\alpha=0.7$ appears to be the most resistant. 
While the $\alpha$-distance standard deviations 
are less robust, they are still superior to the 
classical measure. We also observe that the 
$\alpha$-distance standard deviation is more
robust for low $\alpha$ values, which is in 
agreement with the influence functions in the 
main text.

\clearpage
\section{Proofs of the breakdown values in Section 3}
\label{suppmat:bdv}

\subsection{Finite-sample breakdown values}\label{sec:finitebdv}

From the expression 
$$d_{1j} = |s-x_j|^\alpha = u^2 + O(u)$$   
we derive
$$\overline{d_{1.}} = \frac{1}{n} \sum_{j \neq 1} 
  (u^2 + O(u)) = \frac{n-1}{n} u^2 + O(u)\;.$$
For $i \neq 1$ we obtain
\begin{align*}
    \overline{d_{i.}} = \frac{1}{n} \sum_{j=1}^n d_{ij} = 
    \frac{1}{n} \left( d_{i1}+\sum_{j=2}^n d_{ij} \right) 
    = \frac{1}{n} \left(u^2 + O(u) + o(u) \right) = \frac{u^2}{n} + O(u)
\end{align*}
and for $j \neq 1$ it holds by symmetry that
$$\overline{d_{.j}} = \frac{u^2}{n} + O(u)\,.$$
Finally
\begin{align*}
      \overline{d_{..}} 
      &= \frac{1}{n} \sum_{i=1}^n \overline{d_{i.}}
      = \frac{1}{n} \left(\overline{d_{1.}} 
        + \sum_{i=2}^n \overline{d_{i.}}\right)\\
      &= \frac{1}{n} \left( \frac{n-1}{n} u^2 + O(u) 
       + (n-1)\left(\frac{u^2}{n} + O(u)\right) \right)
      = \frac{2(n-1)}{n^2} u^2 +O(u)\,.
\end{align*}
Next we find for all $j \neq 1$ that
\begin{align*}
 \Delta_{1j} &= d_{1j} - \overline{d_{1.}}
    - \overline{d_{.j}} + \overline{d_{..}}\\
  &= u^2 - \frac{n-1}{n} u^2
  - \frac{1}{n} u^2 + \frac{2(n-1)}{n^2} u^2 + O(u)\\
  &= \frac{2(n-1)}{n^2} u^2 + O(u)
\end{align*}
and
\begin{align*}
 \Delta_{11} &= d_{11} - \overline{d_{1.}}
    - \overline{d_{.1}} + \overline{d_{..}}\\
  &= 0 - \frac{n-1}{n} u^2
  - \frac{n-1}{n} u^2 + \frac{2(n-1)}{n^2} u^2 + O(u)\\
  &= \frac{-2(n-1)}{n} u^2 + \frac{2(n-1)}{n^2} u^2 + O(u)\\
  &= \frac{-2(n-1)^2}{n^2} u^2 +O(u)\;.
\end{align*}
For the remaining $(i,j)$ with $i \neq 1 \neq j$ we find
$$\Delta_{ij} = -\frac{1}{n} u^2 
  - \frac{1}{n} u^2 + \frac{2(n-1)}{n^2} u^2 + O(u)
  = \frac{-2}{n^2} u^2 + O(u)\;.$$
The overall distance variance of the dataset $X$ then yields
\begin{align*}
  \dVar(X) &= \;\frac{1}{n^2} \sum_{i,j=1}^n 
     \Delta_{ij}^2\\
  &= \frac{1}{n^2} \left[\Delta_{11}^2 
  + \sum_{j=2}^n \Delta_{1j}^2
  + \sum_{i=2}^n \Delta_{i1}^2
  + \sum_{i,j=2}^n \Delta_{ij}^2\right]\\
  &= \frac{1}{n^2} \left[\left(\frac{-2(n-1)^2}{n^2} u^2\right)^2
  + 2(n-1)\left(\frac{2(n-1)^2}{n^2} u^2\right)^2
  + (n-1)^2\left(\frac{-2}{n^2} u^2\right)^2 + n^2 O(u^3)\right]\\
  &= \frac{4}{n^6} u^4[(n-1)^4 + 2(n-1)^3 + (n-1)^2]+ O(u^3)\\
  &= 4\frac{(n-1)^2}{n^4}u^4 + O(u^3)\;.
\end{align*}
For the finite-sample breakdown value of $\dCov$ we obtain 
analogously
\begin{align*}
  \dCov(Z) &= \;\frac{1}{n^2} \sum_{i,j=1}^n 
     \Delta_{X,i,j}\Delta_{Y,i,j}\\
  &= \frac{1}{n^2} \left[
     \Delta_{X,1,1} \Delta_{Y,1,1} 
  + \sum_{j=2}^n \Delta_{X,1,j}\Delta_{Y,1,j}
  + \sum_{i=2}^n \Delta_{X,i,1}\Delta_{Y,i,1}
  + \sum_{i,j=2}^n \Delta_{X,i,j}\Delta_{Y,i,j}
  \right]\\
  &=\; ... = 4\frac{(n-1)^2}{n^4} s^{\alpha} t^{\alpha}
  + O(s^{3\alpha/4}t^{3\alpha/4})\\
  &\approx \frac{4}{n^2} s^\alpha t^\alpha
  + O(s^{3\alpha/4}t^{3\alpha/4})\,.
\end{align*}

\subsection{Asymptotic breakdown value of dVar} 
\label{sec:dVar+}

Take $H = \Delta_s$ and consider $F_{\eps} = (1-\eps)F + \eps H$. 
Then the three terms of the distance variance become
$$E[||\Xe -\Xe'||^2] = (1-\eps)^2E[||X - X'||^2] + 2 \eps (1-\eps) E[||X - s||^2], $$
$$E[||\Xe -\Xe'||]^2 = ((1-\eps)^2E[||X - X'||] + 2 \eps (1-\eps) E[||X - s||])^2, $$
\begin{align*}
    E[||\Xe -\Xe'|| ||\Xe -\Xe''||] =&\;(1-\eps)^3 E[||X - X'|| ||X - X''||]\\
    &\; +2 \eps (1-\eps)^2 E[||X-X'|| ||X - s||]\\
    &\; + \eps (1-\eps)^2 E[||s-X'|| ||s - X''||]\\
    &\; + \eps^2 (1-\eps) E[||X-s||^2] \;.
\end{align*}
If we reconstruct $\dVar$ from these terms (summing the first two and subtracting twice the third term), we obtain
\begin{align*}
    \dVar(F, \eps, s) =&\; \mbox{constant}\\
    &\;+ \left\{2 \eps (1-\eps) - 2\eps^2(1-\eps) \right\} E[||X-s||^2]\\
    &\;+ \left\{ 4\eps^2(1-\eps)^2 - 2(1-\eps)^2\eps \right\} E[||X-s||]^2\\
    &\;+ 4\eps(1-\eps)^3 E[||X-X'||]E[||X-s||]\\
    &\;-4 \eps (1-\eps)^2 E[||X-X'|| ||X-s||] \;.
\end{align*}
This can be simplified to
\begin{align*}
    \dVar(F, \eps, s) =&\;  2\eps (1-\eps)^2 E[||X-s||^2]\\
    &\;+ 2 \eps (1-\eps)^2 (2\eps-1) E[||X-s||]^2\\
    &\;+ 4\eps(1-\eps)^3 E[||X-X'||]E[||X-s||]\\
    &\;- 4 \eps (1-\eps)^2 E[||X-X'|| \; ||X-s||]\\
    &\;+ \mbox{constant}.
\end{align*}
Note that the second term is negative. Now note that for the first and second term, we can bound them (using the fact that the variance is non-negative) from below by 
\begin{align*}
    &2\eps (1-\eps)^2 E[||X-s||^2] + 2 \eps (1-\eps)^2 (2\eps-1) E[||X-s||]^2\\
    &\; \geq 2\eps (1-\eps)^2 E[||X-s||^2] + 2 \eps (1-\eps)^2 (2\eps-1) E[||X-s||^2]\\
    &\;= 4\eps^2 (1-\eps)^2 E[||X-s||^2] \;.
\end{align*}
Dropping the positive third term, we obtain :
\begin{align*}
  \dVar(F, \eps, s) 
      \geq &\; \mbox{constant}\\
    &\:+  4\eps^2 (1-\eps)^2 E[||X-s||^2]\\
    &\;- 4 \eps (1-\eps)^2 E[||X-X'|| \; ||X-s||]\\
      \geq &\; \mbox{constant}\\
    &\; +  4\eps^2 (1-\eps)^2 E[||X-s||^2]\\
    &\: -4 \eps (1-\eps)^2 \sqrt{E[||X-X'||^2] E[||X-s||^2]}\\
      \geq &\; \mbox{constant}\\ 
    &\; + \left\{2\eps (1-\eps) \sqrt{E[||X-s||^2]} - (1-\eps) \sqrt{E[||X-X'||^2]}\right\}^2 \\
    &\; -(1-\eps)^2 E[||X-X'||^2]\\
      \geq &\; \mbox{constant'}\\
    &\; + \left\{2\eps (1-\eps) \sqrt{E[||X-s||^2]} - (1-\eps) \sqrt{E[||X-X'||^2]}\right\}^2,
\end{align*}
where we have used Cauchy-Schwarz for the second inequality and completed the square for the second-to-last equality. For any fixed $\eps$, the last term explodes as $s$ goes to $\infty$.

\subsection{Asymptotic breakdown value of dCov} 
\label{sec:dCov}

Take $H = \Delta_{(s,s)}$ and consider 
$F_{\eps} = (1-\eps)F + \eps H$. We will show that 
adding a point mass at $(s,s)$ can make $\dCov(F_\eps)$
arbitrarily large. 
The three terms of the distance covariance become
$$E[||\Xe -\Xe'||||\Ye -\Ye'||] = (1-\eps)^2E[||X - X'|| \; 
  ||Y - Y'||] + 2 \eps (1-\eps) E[||X - s|| ||Y-s||] $$
\begin{align*}
    E[||\Xe -\Xe'||] E[||\Ye -\Ye'||] =&\; 
      (1-\eps)^4 E[||X - X'||]E[||Y - Y'||]\\
    &\; + 2 \eps (1-\eps)^3 \left\{E[||X - s||] E[||Y - Y'||]
        + E[||X - X'||] E[||Y - s||]  \right\}\\
    &\; + 4\eps^2(1-\eps)^2 E[||X - s||]E[||Y - s||]
\end{align*}
\begin{align*}
    E[||\Xe -\Xe'|| ||\Ye -\Ye''||] =&\;
      (1-\eps)^3 E[||X - X'|| ||Y - Y''||]\\
    &\; + \eps (1-\eps)^2 \{ E[||X-X'|| ||Y - s||]
        + E[||X-s||] E[||Y - s||]\\
    &\; + E[||X-s|| ||Y - Y''||] \} \\
    &\; + \eps^2 (1-\eps) E[||X-s|| ||Y-s||]\;.
\end{align*}
Then we obtain:
\begin{align*}
    \dCov(F_{\eps}) =&\; \mbox{constant}\\
    &\;+E[||X-s||||Y-s||] \left\{ 2\eps(1-\eps) - 2 \eps^2(1-\eps)\right\}\\
    &\;+E[||X-s||]E[||Y-s||] \left\{4\eps^2(1-\eps)^2 - 2\eps (1-\eps)^2 \right\}\\
    &\;+E[||X-X'||]E[||Y-s||] \left\{ 2\eps (1-\eps)^3\right\}\\
    &\;+E[||X-X'| ||Y-s||] \left\{ -2(1-\eps)^2\eps\right\}\\
    &\;+E[||X-s||]E[||Y-Y'||] \left\{ 2\eps (1-\eps)^3\right\}\\
    &\;+E[||X-s||||Y-Y'||] \left\{ -2(1-\eps)^2\eps \right\}\;.
\end{align*}
Simplifying:
\begin{align*}
    \dCov(F_{\eps}) =&\;E[||X-s||||Y-s||] \left\{ 2\eps(1-\eps)^2\right\}\\
    &\;+E[||X-s||]E[||Y-s||] \left\{-2\eps(1-\eps)^2(1-2\eps) \right\}\\
    &\;+E[||X-X'||]E[||Y-s||] \left\{ 2\eps (1-\eps)^3\right\}\\
    &\;+E[||X-X'| ||Y-s||] \left\{ -2(1-\eps)^2\eps\right\}\\
    &\;+E[||X-s||]E[||Y-Y'||] \left\{ 2\eps (1-\eps)^3\right\}\\
    &\;+E[||X-s||||Y-Y'||] \left\{ -2(1-\eps)^2\eps \right\}\\
    &+ \mbox{constant}.
\end{align*}
Now we need to show that this explodes for $s \to \infty$. 
Consider the first two terms only, which are the only ones
that are ``quadratic'' in $s$. Consider the first term 
$E[||X-s||||Y-s||]$. By Cauchy-Schwarz, we have\\ 
$$ E[||X-s||||Y-s||] \geq E[||X-s||]E[||Y-s||] - 
\sqrt{\var[||X-s||]\var[||Y-s||]}\;.$$\\
For $s$ large enough, we have 
$\sqrt{\var[||X-s||]\var[||Y-s||]} 
\approx \sqrt{\var[X]\var[Y]}$\;.
Using this, we can lower bound the first two terms (for $s$ large enough) by:
\begin{align*}
   & E[||X-s||||Y-s||] \left\{ 2\eps(1-\eps)^2\right\}+E[||X-s||]E[||Y-s||] \left\{-2\eps(1-\eps)^2(1-2\eps) \right\}\\
    \geq &\;  4 \eps^2 (1-\eps)^2 E[||X-s||||Y-s||] - \left\{2\eps(1-\eps)^2(1 - 2\eps) \right\} \sqrt{\var[X]\var[Y]}\;.
\end{align*}
Therefore we have 
\begin{align*}
    \dCov(F_{\eps}) \geq &\; \mbox{constant}\\
    &\;+4 \eps^2 (1-\eps)^2 E[||X-s||||Y-s||]\\
    &\;- \left\{2\eps(1-\eps)^2(1 - 2\eps) \right\} \sqrt{\var[X]\var[Y]}\\
    &\;+E[||X-X'||]E[||Y-s||] \left\{ 2\eps (1-\eps)^3\right\}\\
    &\;+E[||X-X'| ||Y-s||] \left\{ -2(1-\eps)^2\eps\right\}\\
    &\;+E[||X-s||]E[||Y-Y'||] \left\{ 2\eps (1-\eps)^3\right\}\\
    &\;+E[||X-s||||Y-Y'||] \left\{ -2(1-\eps)^2\eps \right\}\;.
\end{align*}
We can drop the fourth and sixth terms as they are 
positive, and absorb the third term in the constant:
\begin{align*}
    \dCov(F_{\eps}) \geq&\; \mbox{constant}\\
    &\;+4 \eps^2 (1-\eps)^2 E[||X-s||||Y-s||]\\
    &\;+E[||X-X'| ||Y-s||] \left\{ -2(1-\eps)^2\eps\right\}\\
    &\;+E[||X-s||||Y-Y'||] \left\{ -2(1-\eps)^2\eps \right\}\;.
\end{align*}
It now becomes clearer that we have something quadratic 
in $s$, and some terms which are linear in $s$.
It suffices to show that the expression
\begin{align*}
&2\eps^2 (1-\eps)^2 E[||X-s||||Y-s||] - 2(1-\eps)^2\eps E[||X-X'| ||Y-s||]\\
&\;= 2 \eps (1-\eps)^2 E[\left(\eps ||X-s|| - ||X-X'||\right) ||Y-s||] \;.
\end{align*}
explodes for $s \to \infty$. We can write
\begin{align*}
  &E[\left(\eps ||X-s|| - ||X-X'||\right) ||Y-s||]\\
  &\; \geq E[\left(\eps ||s|| - \eps||X|| - ||X|| -||X'||\right) ||Y-s||]\\
  &\; = E[\eps ||s||  ||Y-s||] - E[\left((1+\eps)||X|| +||X'||\right) ||Y-s||]\\
  &\; \geq \eps E[||s||  \left(||s|| - ||Y||\right)] 
   - E[\left((1+\eps)||X|| +||X'||\right) (||Y|| + ||s||)]\\
  &\; \coloneqq a ||s||^2 - b ||s|| - c
\end{align*}
for some $a, b, c > 0$. This indeed explodes 
as $||s|| \to \infty$.

\clearpage
\section{Distance correlation after rank transforms}
\label{suppmat:gaussianrank}

\begin{proposition}
    The influence function of $\dCov(F_X(X),F_Y(Y))$ is given by
\begin{align*}
   \IF((s,t),\dCov(F_X(X),F_Y(Y)),F) = -4 &\dCov(F_X(X),F_Y(Y))\\
   &+ 2\,\eta(F_X(s),F_Y(t),F_X(X),F_Y(Y),1) \\
   &+\dCov(I(X \geq s),F_Y(Y)) \\
   &+\dCov(I(Y \geq t),F_X(X)).
\end{align*}
The IF of $\dCov(\Phi^{-1}(F_X(X)),\Phi^{-1}(F_Y(Y)))$ is given by
\begin{align*}
   \IF&((s,t),\dCov(\Phi^{-1}(F_X(X)),\Phi^{-1}(F_Y(Y))),F)\\
   =& -2 \dCov(X,Y) + 2\,\eta(s,t,X,Y,1) \\
   & \quad + 2\mathbb{E} \Big[ \text{sign}(X-X^{\prime})
   \left( \frac{I(X \geq s)- \Phi(X)}{\phi(X)}\right)\\
   &\qquad \qquad \left(|Y-Y^{\prime}|+ \mathbb{E}[|Y-Y^{\prime}|]-
   |Y-Y^{\prime\prime}|-
   |Y^{\prime}-Y^{\prime\prime}|\right)\Big] \\
   & \quad +2\mathbb{E} \Big[ \text{sign}(Y-Y^{\prime}) 
   \left( \frac{I(Y \geq t)-\Phi(Y)}{\phi(Y)} \right)\\
   &\qquad \qquad (|X-X^{\prime}|+\mathbb{E}[|X-X^{\prime}|]
   -|X-X^{\prime\prime}|-|X^{\prime}-X^{\prime\prime}|)\Big]
\end{align*}
for bivariate normal $(X,Y)$.
\end{proposition}

From these we can easily derive the corresponding 
influence functions for the $\dCor$ using
Corollary 2. 
The proof of this proposition follows the proof of 
Proposition 1, but takes into account 
the dependence of the rank transforms on the 
contamination.

\newpage
\section{Additional simulation results}\label{suppmat:simul}

Figure \ref{fig:powersim_univaraiate_additional} below 
presents the results of the power simulation for
the bivariate settings which were not shown in the main 
text. The performance differences between the various methods are smaller here.\\

\begin{figure}[!ht]
\renewcommand*\thesubfigure{\arabic{subfigure}} 
    \centering 
\begin{subfigure}{0.3\textwidth}
  \includegraphics[width=\linewidth]{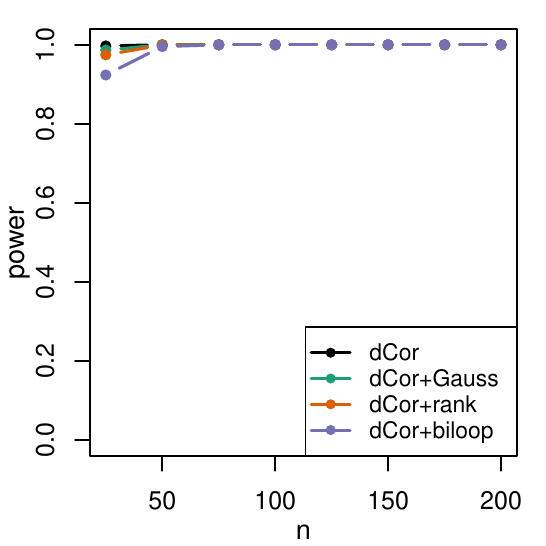}
  \caption{cubic}
\end{subfigure}
\begin{subfigure}{0.3\textwidth}
  \includegraphics[width=\linewidth]{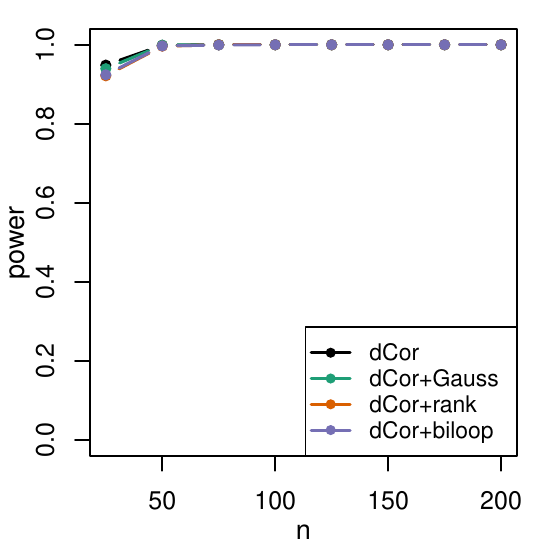}
  \caption{logarithm}
\end{subfigure}\hfil 
\begin{subfigure}{0.3\textwidth}
  \includegraphics[width=\linewidth]{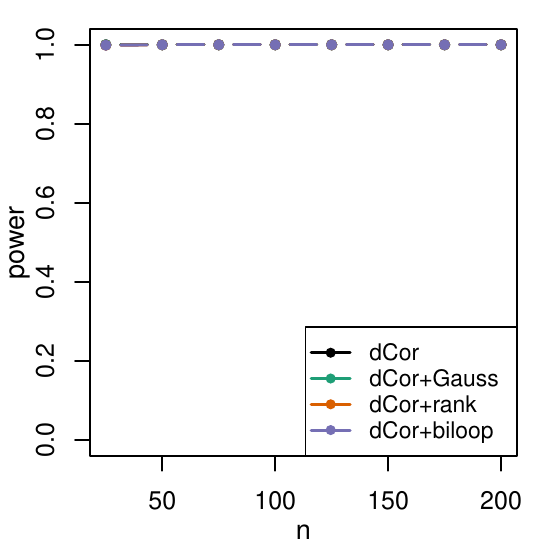}
  \caption{exponential}
\end{subfigure}\hfil 
\begin{subfigure}{0.3\textwidth}
  \includegraphics[width=\linewidth]{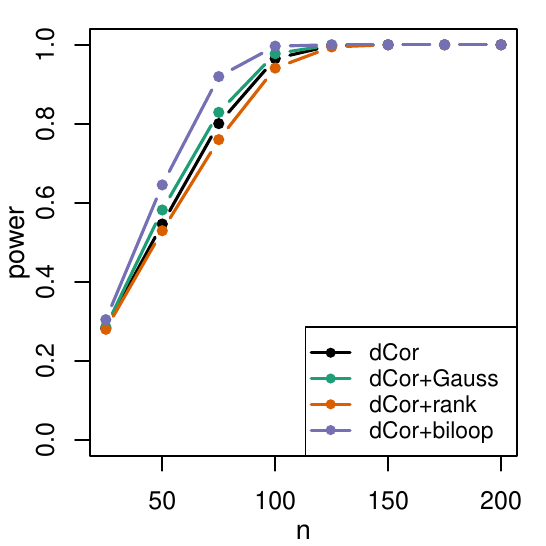}
  \caption{sine}
\end{subfigure}\hfil 
\begin{subfigure}{0.3\textwidth}
  \includegraphics[width=\linewidth]{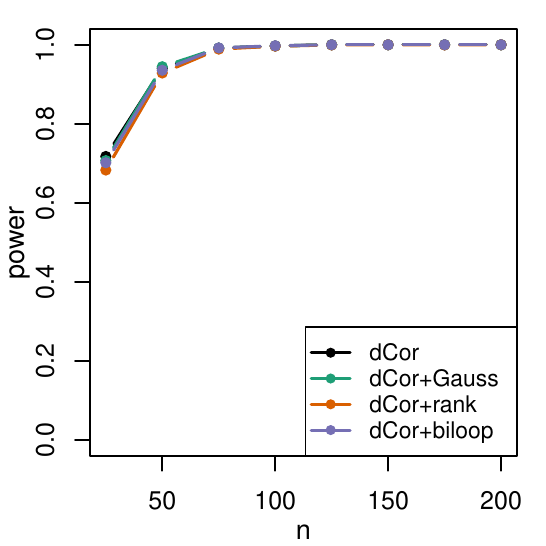}
  \caption{$4^{th}$ root}
\end{subfigure}\hfil 
\begin{subfigure}{0.3\textwidth}
  \includegraphics[width=\linewidth]{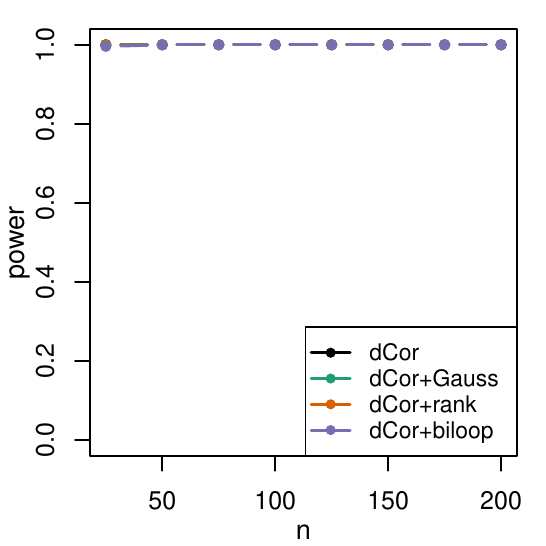}
  \caption{step}
\end{subfigure}
\caption{\centering Results of the power simulation per sample size $n$.}
\label{fig:powersim_univaraiate_additional}
\end{figure}

\clearpage

\clearpage
\section{More on the real data example of Section 6}
\label{suppmat:app}

\begin{figure}[!ht]
\centering
\includegraphics[width=.80\textwidth]
   {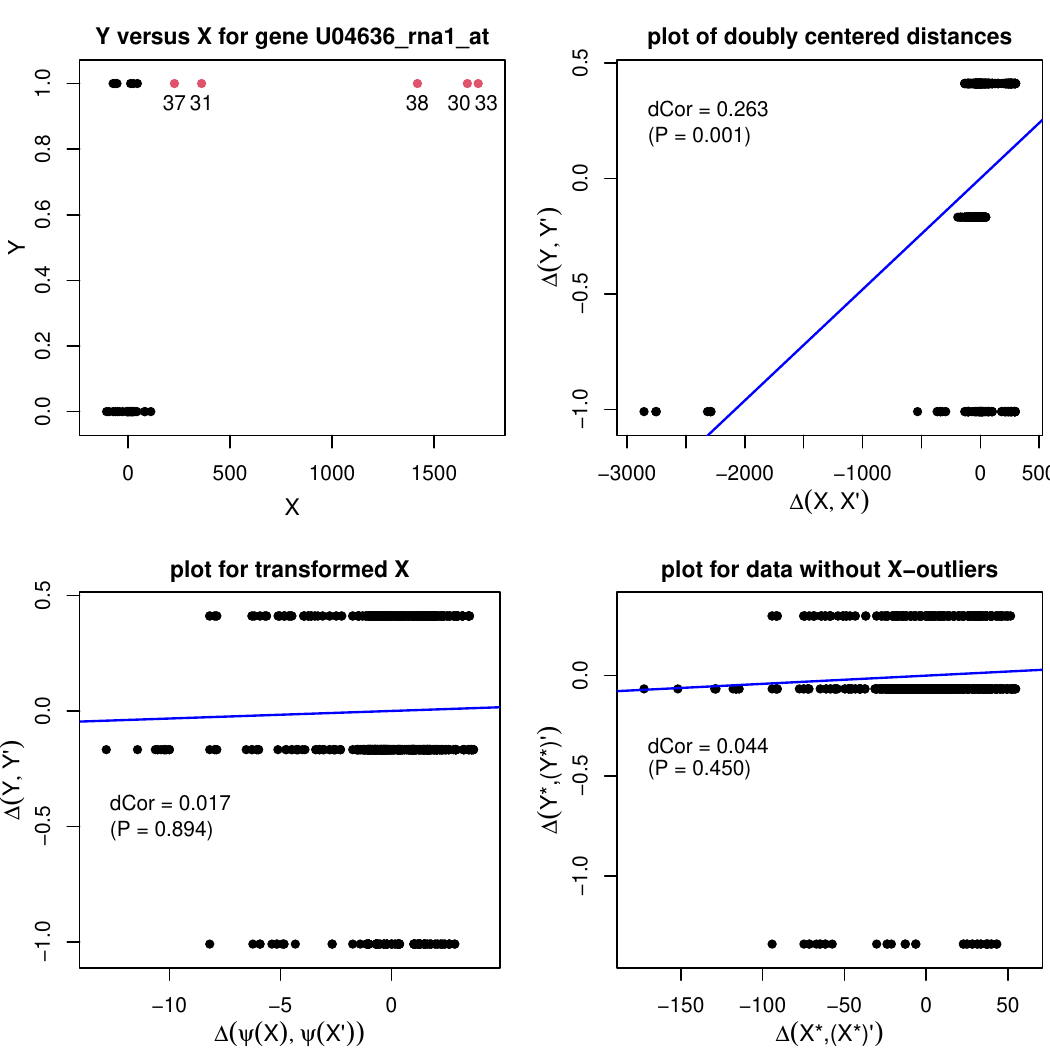}
\vspace{2mm}
\caption{Gene U04636\_rna1\_at of the leukemia 
  data ($j=5376$).
  Top left: plot of $Y$ versus $X^j$, with the five
  outliers marked in red. Top right: doubly 
  centered distances $\Delta(Y,Y')$ of $Y$ versus 
  those of $X^j$. Bottom left: $\Delta(Y,Y')$ 
  versus doubly centered distances of
  $\psi_{\biloop}(X^j)$. Bottom right: plot of 
  doubly centered distances for the data 
  $((X^j)^*,Y^*)$ without the five outlying 
  points.} 
\label{fig:5376}
\end{figure}

Among the genes for which the robust dCor with
$Y$ is below 0.05, the one with the second largest 
difference\linebreak classical $-$ robust has
$j = 5376$, corresponding to gene 
U04636\_rna1\_at\,. Figure \ref{fig:5376} is 
similar to Figure 20 in the main
text, except that now five outliers are 
marked instead of four. The analysis is
analogous.

In the main text we analyzed gene Z19002\_at 
with $j=5071$, where the robust dCor is higher 
than the classical one. 
In Figure 21 we saw that it has
a single outlier in $X^j$, which does not obey 
the increasing trend of the remaining data 
points. Here we look instead at gene 
X57579\_s\_at ($j=5972$). Now the outlier
(patient 38) has $y_i = 1$, which is in 
agreement with the increasing trend of the 
inliers. In a logistic regression, the fit
would not change much because of this point.
But here it still creates a far leverage
point in the top right panel, which makes the
Pearson correlation lower than when the
point is removed, as seen in the bottom
right panel. Another way to interpret this
is that the leverage point has reduced the 
slope of the regression line.
Still, although the classical dCor is lower
with the outlier than without it, here it 
remains significant in both situations.

\begin{figure}[!ht]
\centering
\includegraphics[width=.80\textwidth]
   {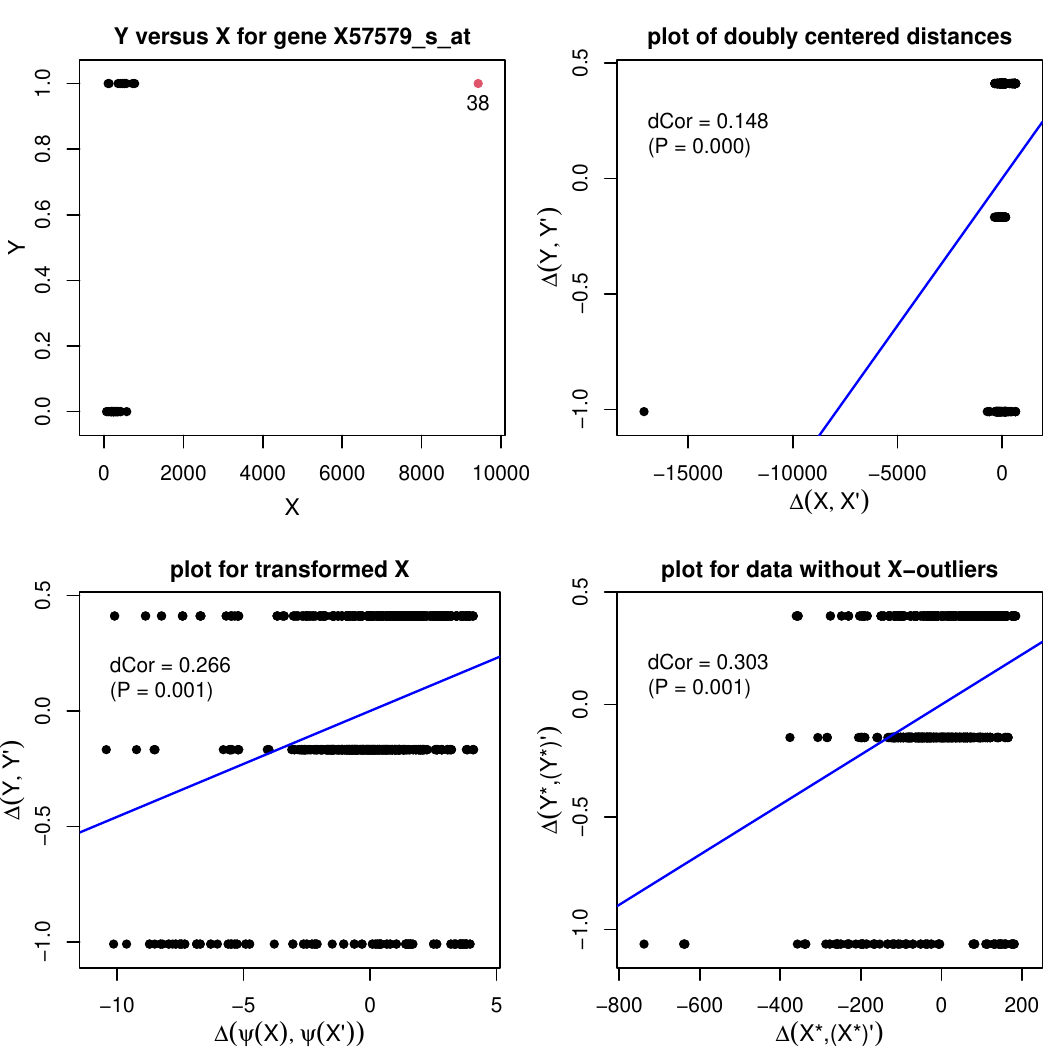}
\vspace{2mm}
\caption{The panels are as in Figure~\ref{fig:5376}, 
  but for gene X57579\_s\_at ($j=5972$).}
\label{fig:5972}
\end{figure}

\clearpage
\section{Supplementary Material for the rejoinder}

In our reply to the invited comment by
Klaus Nordhausen and Una Radoji\u{c}i\'c 
we gave a counterexample to the subadditivity 
of $\dVar$ using Bernoulli distributions.
Consider $X \sim \mbox{Bernoulli}(p)$ with 
success probability $0 < p < 1$ and $X', X''$  
independent copies of $X$. We used that
\begin{equation*}
  \dVar(X) = \E[|X-X'|^2] + \E[|X-X'|]^2 
   - 2 \E[|X-X'||X-X''|] = 4p^2(1-p)^2\;.
\end{equation*}
This is found as follows. We have that 
$$\dVar(X) = \E[|X-X'|^2] + \E[|X-X'|]^2 - 
2 \E[|X-X'||X-X''|].$$
Now $\E[|X-X'|] = 2p(1-p)$, 
$\E[|X-X'|^2] = 2p(1-p)$, and 
$\E[|X-X'||X-X''|] =p(1-p)^2 + (1-p)p^2$ 
(note that $|X-X'||X-X''| \neq 0$ only if 
$(X,X',X'') = (1, 0, 0)$ or 
$(X, X', X'') = (0, 1, 1)$, which happens with probabilities $p(1-p)^2$ and $(1-p)p^2$).
So we have $\dVar(X) = 2p(1-p) + (2p(1-p))^2 
- 2(p(1-p)^2 + (1-p)p^2) = 4p^2(1-p)^2 $.

Now consider $Z \coloneqq X+X'$. Then 
\begin{equation*}
    Z = \begin{cases}
        0 &\mbox{ with probability  }\; P_0 \coloneqq (1-p)^2\\
        1 &\mbox{ with probability  }\; P_1 \coloneqq 2p(1-p)\\
        2 &\mbox{ with probability  }\; P_2 \coloneqq p^2\;.
    \end{cases}
\end{equation*}
As a result, we have 
\begin{align*}
    \E[|Z-Z'|] =&\, 2 * P(|Z-Z'| = 2) +  P(|Z-Z'| = 1)\\
    =&\, 2 (2P_0P_2) + (2P_0P_1 + 2P_1P_2)
\end{align*}
and similarly
\begin{align*}
    \E[|Z-Z'|^2] =&\, 4 * P(|Z-Z'| = 2) +  P(|Z-Z'| = 1)\\
    =&\, 4 (2P_0P_2) + (2P_0P_1 + 2P_1P_2)\;.
\end{align*}

The third term is the most cumbersome. The joint distribution of $(Z, Z', Z'')$ is given by the
following table:

\clearpage
\begin{table}[!ht]
\begin{tabular}{|c|c|c|c|c|}
\hline
$Z$ & $Z'$ & $Z''$ & $|Z - Z'| \cdot |Z - Z''|$ & Probability \\
\hline
0 & 0 & 0 & 0 & $P_0^3$ \\
0 & 0 & 1 & 0 & $P_0^2 P_1$ \\
0 & 0 & 2 & 0 & $P_0^2 P_2$ \\
0 & 1 & 0 & 0 & $P_0 P_1 P_0$ \\
0 & 2 & 0 & 0 & $P_0 P_2 P_0$ \\
1 & 0 & 1 & 0 & $P_1 P_0 P_1$ \\
1 & 1 & 0 & 0 & $P_1^2 P_0$ \\
1 & 1 & 1 & 0 & $P_1^3$ \\
1 & 1 & 2 & 0 & $P_1^2 P_2$ \\
1 & 2 & 1 & 0 & $P_1 P_2 P_1$ \\
2 & 0 & 2 & 0 & $P_2 P_0 P_2$ \\
2 & 1 & 2 & 0 & $P_2 P_1 P_2$ \\
2 & 2 & 0 & 0 & $P_2^2 P_0$ \\
2 & 2 & 1 & 0 & $P_2^2 P_1$ \\
2 & 2 & 2 & 0 & $P_2^3$ \\
\hline
\hline
0 & 1 & 1 & 1 & $P_0 P_1^2$ \\
1 & 0 & 0 & 1 & $P_1 P_0^2$ \\
1 & 0 & 2 & 1 & $P_1 P_0 P_2$ \\
1 & 2 & 0 & 1 & $P_1 P_2 P_0$ \\
1 & 2 & 2 & 1 & $P_1 P_2^2$ \\
2 & 1 & 1 & 1 & $P_2 P_1^2$ \\
\hline
\hline
0 & 1 & 2 & 2 & $P_0 P_1 P_2$ \\
0 & 2 & 1 & 2 & $P_0 P_2 P_1$ \\
2 & 0 & 1 & 2 & $P_2 P_0 P_1$ \\
2 & 1 & 0 & 2 & $P_2 P_1 P_0$ \\
\hline
\hline
0 & 2 & 2 & 4 & $P_0 P_2^2$ \\
2 & 0 & 0 & 4 & $P_2 P_0^2$ \\
\hline
\end{tabular}
\end{table}

Therefore we can calculate the probabilities of 
$|Z-Z'||Z-Z''|$ taking on the values 1, 2 and 4 
by summing the corresponding probabilities in 
the table. We get
\begin{equation*}
    P(|Z-Z'||Z-Z''| = 1) =  P_0P_1(P_0 + P_1) + P_1P_2(2P_0+P_1+P_2)=
    P_1P_0(P_0+P_1+2P_2) + P_1P_2(P_1+P_2)\;,
\end{equation*}
\begin{equation*}
    P(|Z-Z'||Z-Z''| = 2) = 4 P_0 P_1 P_2\;,
\end{equation*}
\begin{equation*}
    P(|Z-Z'||Z-Z''| = 4) = P_0 P_2(P_0+P_2)\;.
\end{equation*}
\noindent Combining these results yields
\begin{align*}
    \dVar(Z) =&\; 4 (2P_0P_2) + (2P_0P_1 + 2P_1P_2) + (2 (2P_0P_2) + (2P_0P_1 + 2P_1P_2))^2\\
    - &\; 2(P_0P_1(P_0 + P_1) + P_1P_2(2P_0+P_1+P_2) + 8P_0 P_1 P_2 + 4 P_0 P_2(P_0+P_2))\\
    =&\;16 p^2 - 80 p^3 + 176 p^4 - 208 p^5 + 144 p^6 - 64 p^7 + 16 p^8 
\end{align*}
so finally
\begin{equation*}
    \dVar(X + X') = 16 p^2 - 80 p^3 + 176 p^4 - 208 p^5 + 144 p^6 - 64 p^7 + 16 p^8 \;.
\end{equation*}

\vspace{5mm}
\noindent The resulting figure can also be confirmed 
empirically by the following \textsf{R} code:

\begin{verbatim}
dvarbern = function(pgrid, n){
  out = matrix(NA, nrow = length(pgrid), ncol = 3)
  i = 0
  for (p in pgrid) {
    i = i+1
    out[i,1] = p
    set.seed(1)
    x = rbinom(n, 1, p) 
    y = rbinom(n, 1, p) 
    out[i,2] = dcov::dcov(x + y, x + y)
    out[i,3] = dcov::dcov(x, x) + dcov::dcov(y, y)
  } 
  out
}

pgrid = seq(0.01, 0.99, by = 0.01)
out = dvarbern(pgrid, n = 1e6)
outmat = rbind(c(0,0,0),out,c(1,0,0))
plot(outmat[,c(1,3)], type="l", col="red", xlab="p",
     ylab = "dVar(X+Y) and dVar(X)+dVar(Y)")
lines(outmat[,c(1,2)], col="blue")
abline(h=0)
abline(h=0.5, lty = 2, lwd = 1)
\end{verbatim}

\end{document}